\documentclass[11pt]{article}
\pdfoutput=1 
\usepackage{jheppub}
\usepackage{tabu}
\usepackage[vcentermath]{youngtab}
\usepackage[usenames,dvipsnames,table]{xcolor}
\usepackage{graphicx,amsmath,amssymb,amsthm,multirow,array,bm,bbm,esint}
\usepackage[mathscr]{eucal}
\usepackage[bbgreekl]{mathbbol}
\usepackage{epsf,amsfonts}
\usepackage{slashed}
\usepackage[numbers,sort&compress]{natbib}
\usepackage{minibox}
\usepackage{rotating}
\usepackage{pdflscape}

\usepackage{tikz, pgf}
\usepackage{amsmath}
\usetikzlibrary{shapes.misc}
\usetikzlibrary{shapes}
\usetikzlibrary{fit}
\usetikzlibrary{arrows}

\makeatletter
\newcommand{\textlabel}[2]{%
  \phantomsection
  #1\def\@currentlabel{\unexpanded{#1}}\label{#2}%
}
\makeatother

\usepackage{tikz}
\usetikzlibrary{shapes.geometric, arrows}
\tikzstyle{thread} = [rectangle, minimum width=.3\textwidth, minimum height=.6cm, text centered, text width=.3\textwidth, draw=black, fill=blue!10]
\tikzstyle{details} = [rectangle, minimum width=.3\textwidth, minimum height=.6cm, text centered, text width=.3\textwidth, draw=black, fill=green!5]
\tikzstyle{arrowT} = [very thick,->,>=stealth']
\tikzstyle{arrow2} = [very thick,dashed,->,>=stealth']
\tikzstyle{arrowB} = [ultra thick,->,>=stealth']

\newcommand{\qmax}{\lfloor\frac{n+1}{2}\rfloor}

\definecolor{rust}{rgb}{0.8,0.2,0.2}

\newcommand{\Tr}[1]{\hbox{Tr}\left(#1\right)}

\newcommand{\vev}[1]{\left\langle #1 \right\rangle}

\newcommand{\be}{\begin{equation}}
\newcommand{\ee}{\end{equation}}
\newcommand{\ba}{\begin{eqnarray}}
\newcommand{\ea}{\end{eqnarray}}
\newcommand{\bi}{\begin{itemize}}
\newcommand{\ei}{\end{itemize}}
\newcommand{\bse}{\begin{subequations}}
\newcommand{\ese}{\end{subequations}}

\newcommand{\rhoi}{\hat{\rho}_{\text{initial}}}
\newcommand{\rhoT}{\hat{\rho}_{_T}}

\newcommand{\bose}{\mathfrak{f}}
\newcommand{\cth}{{\sf N}}

\newcommand{\ad}[1]{\bm{#1}}

\newcommand{\comm}[2]{ [#1, #2 ] }
\newcommand{\anti}[2]{ \{ #1, #2 \}}
\newcommand{\fullcomm}[1]{{[\hspace{-.75mm}[} #1{]\hspace{-.75mm}]}}
\newcommand{\Ecomm}[3]{ [ #1, #2 ]_{\varepsilon_{#3}}}
\newcommand{\tcomm}[3]{ {}_{\varepsilon_{_{#3}}}\!\ad{[} #1, #2 \ad{]} }

\newcommand{\ta}[2]{ {}_{_+}\ad{[} #1, #2 \ad{]} }
\newcommand{\tc}[2]{ {}_{_-}\ad{[} #1, #2 \ad{]} }

\newcommand{\kmsfd}[2]{\mathcal{K}_{(#1\,,\,#2)}}
\newcommand{\dptp}{\delta^{\lhook}}

\newcommand{\Op}[1]{\mathbb{#1}}
\newcommand{\OpH}[1]{\widehat{\mathbb{#1}}}
\newcommand{\skR}{\text{\tiny R}}
\newcommand{\skL}{\text{\tiny L}}

\title{Thermal out-of-time-order correlators, KMS relations, and spectral functions}

\author[a]{Felix M. Haehl}
\author[b]{\!, R.\ Loganayagam}
\author[b]{\!, Prithvi Narayan}
\author[b]{\!, Amin A. Nizami}
\author[c]{, \\Mukund Rangamani}

\affiliation[\,a]{Department of Physics and Astronomy, University of British Columbia,\\
6224 Agricultural Road, Vancouver, B.C.\ V6T 1Z1, Canada.}
\affiliation[\,b]{International Centre for Theoretical Sciences (ICTS-TIFR), \\
Shivakote, Hesaraghatta Hobli, Bengaluru 560089, India.}
\affiliation[\,c]{
Center for Quantum Mathematics and Physics (QMAP),  \\
Department of Physics, University of California, Davis, CA 95616 USA.}

\emailAdd{f.m.haehl@gmail.com}
\emailAdd{nayagam@gmail.com}
\emailAdd{prithvi.narayan@gmail.com}
\emailAdd{amin@icts.res.in}
\emailAdd{mukund@physics.ucdavis.edu}

\abstract{
We describe general features of thermal correlation functions in quantum systems, with specific focus on the fluctuation-dissipation type relations implied by the KMS condition.   These end up relating correlation functions with different time ordering and thus should naturally be viewed in the larger context of out-of-time-ordered (OTO) observables. In particular, eschewing the standard formulation of KMS relations where thermal periodicity is combined with time-reversal to stay within the purview of Schwinger-Keldysh functional integrals, we show that there is a natural way to phrase them directly in terms of OTO correlators. We use these observations to construct a natural causal basis for thermal $n$-point functions in terms of fully nested commutators. We provide several general results which can be inferred from cyclic orbits of permutations, and exemplify the abstract results using a quantum oscillator as an explicit example.
}

\begin{document}
\maketitle
\flushbottom


\section{Introduction}
\label{sec:intro}

Understanding thermal equilibrium in quantum systems is a necessary precursor to developing a comprehensive physical picture of out-of-equilibrium dynamics. Fortunately, the simplicity of the thermal state, wherein the Gibbs density matrix is a simple functional of the quantum Hamiltonian, $\rhoT = e^{-\beta \,\Op{H}}$, allows one significant insight. By analytic continuation to Euclidean time one can study an equivalent classical statistical mechanical problem; a compact thermal circle allows decomposition into discrete Matsubara modes and thus one can analyze various equilibrium properties. 

However, it remains of interest to ask how deviations from equilibrium can be physically quantified. These are captured by response functions, which by virtue of causal ordering of events,  necessarily involve real-time ordering. A remarkable fact about thermal states is that the real-time response functions are related to fluctuations about equilibrium thanks to the fluctuation-dissipation relations, which in turn follow from the Kubo-Martin-Schwinger (KMS) relations \cite{Kubo:1957mj,Martin:1959jp}. These conditions have played a key role in not only in  thermal field theory  (cf., \cite{Chou:1984es} for a review), but have also been useful in the context of axiomatic formulations of QFT \cite{Haag:1967sg}. 

Traditionally, the fluctuation-dissipation relations (abbreviated FDT) are characterized by relating the retarded Green's function in thermal equilibrium (causal response), to the symmetrized two-point function (the fluctuation). This well-celebrated generalization of the Einstein relation clearly admits generalization to higher point functions.  Over the years many authors have attempted to construct a full set of relations, translating the information into a series of statements about higher point spectral functions in the field theory. Some preliminary attempts are for instance summarized in the excellent review \cite{Chou:1984es}; further attempts to understand these relations within the real-time Schwinger-Keldysh formalism can be found in \cite{Carrington:1996rx,Hou:1998yc,Wang:1998wg,Carrington:2000zn,Weldon:2005nr}. In addition, several groups have tried to ascertain the relation between the real-time and imaginary time formalisms, cf., \cite{Evans:1990hy,Evans:1990qh,Aurenche:1991hi,Evans:1991ky,Taylor:1992ib,Taylor:1993ub,Guerin:1993ik,Guerin:2001mx}. These works demonstrate that  an effective implementation of KMS relations can be very useful in simplifying  the finite temperature  Schwinger-Keldysh formalism as applied to $\phi^4$ theories as well as gauge theories.

Much of the aforementioned analysis was inspired by the need to better understand the connection between the imaginary time formalism (or statistical field theory) and the real time Schwinger-Keldysh formalism for finite temperature QFT. Another motivation was the need to compute thermal observables in hot QCD plasma. More recently, there is growing interest in quantum systems driven out of equilibrium such as e.g., quantum quenches, owing to our improved ability to experimentally simulate time-dependent many-body quantum Hamiltonians. 

The current work is aimed at synthesizing these developments by attempting to form a coherent picture that transcends the limitations of the Schwinger-Keldysh construction. To appreciate this perspective, let us first note that the general FDT arises from the fact that thermal correlators are trace class observables in the Gibbs density matrix. As such the cyclicity inherent in such observables descends directly onto the correlation functions. This is very clear if we think of the cyclic structure made explicit in the imaginary time formalism.  Whilst simple, this cyclicity does not play well with causal ordering -- cyclic transpositions of ordered operators are no longer ordered in a similar manner. In the Schwinger-Keldysh formalism we are able to compute correlators which have a restricted form of out-of-time ordering, but not the most general OTO correlator. Despite being reasonably intuitive, it was only recently argued in \cite{Haehl:2017qfl} (see also preliminary observations in \cite{Haehl:2016pec}), that of the set of real-time Wightman $n$-point functions,  characterized by $n!$ orderings,  only a small subset are computed by the real-time Schwinger-Keldysh contour. Most of the Wightman functions, are out-of-time-order, and are computed by more general functional integral contours, which involve many instances of forward/backward evolution. Such out-of-time order (OTO) functional integral contours, and correlation functions computed therefrom,  have been the focus of recent interest owing to the intricate connection between chaos, ergodicity, thermalization, and black hole physics \cite{Maldacena:2015waa}.\footnote{ Various authors have also explored the OTO correlators from a quantum information perspective, cf., \cite{Hosur:2015ylk,Roberts:2016hpo} for connections to quantum channels, and \cite{Halpern:2017abm} for connections to weak measurements and quasiprobabilities.}

Traditionally, the approach in the analysis of fluctuation-dissipation theorems has been to eschew the occurrence of OTO-correlators by suitably combining the KMS transformation with other discrete symmetries such as time-reversal or CPT, cf., \cite{Sieberer:2015hba}. This is not necessary, and indeed we will show that one can formulate the general set of KMS relations quite simply once we enlarge our considerations to include OTO correlation functions. One of the aims of the current discussion is to give a unified and general picture of such thermal relations from a real-time perspective and elucidate the simplicity gained by moving to a framework involving OTO correlators. For a special class of OTO $2k$-point functions, fluctuation-dissipation relations for suitably regulated ``bipartite'' correlators have been investigated in \cite{Tsuji:2016kep}.\footnote{ It was also argued in \cite{Halpern:2016zcm} that the $4$-point chaos correlator could be given an interpretation of a generalized FD relation along the lines of the Jarzynski relation \cite{Jarzynski:1997aa,Jarzynski:1997ab}. We however believe that the relation derived there is better viewed as a suitable writing of the generating function after fusing two operators into a single composite, which is somewhat different from the more standard notion of fluctuation-dissipation relations.}

\paragraph{Summary of results:} We can distill the essential features of thermal correlation functions into the following set of statements:
\begin{itemize}
\item The KMS relations and general fluctuation-dissipation theorems for higher-point functions can be formulated for any time-ordering of the operators. One does not need to restrict the relations to only involve correlation functions obtained from the Schwinger-Keldysh path integral contour. More specifically, the $n!$ Wightman functions can be partitioned into $(n-1)!$ equivalence classes after taking into account the KMS relations, which act by cyclic permutations and imaginary time shifts of the operator insertions.
\item Once one allows for this general perspective, one does not need to invoke any form of ${\mathbb Z}_2$ involution originating from time-reversal or CPT. The rationale for doing so in more traditional presentations of KMS relations originates from the desire to relate correlators computed within the Schwinger-Keldysh contour (which in our classification would be a $1$-OTO contour). There is no need for such a restriction when using the $k$-OTO path integral contours and the general perspective espoused herein.
\item The KMS relations end up relating proper $q$-OTO Wightman functions with proper $(q+1)$-OTO correlators. The proper OTO number refers to the minimal number of forward-backward evolutions necessary to account for the time-ordering in the correlation function. In particular, this set of relations has the effect of reducing the switchbacks in the path integral contours necessary for computing thermal Wightman functions.
\item  In particular, while it is well known that all time orderings in two-point functions are captured by the Schwinger-Keldysh 1-OTO contour, for three- and four-point functions we need to employ 2-OTO contours for generic initial states. However, in the thermal state the 2-OTO three-point functions can be related to 1-OTO 3-point functions via the KMS relations, thereby reducing the required proper OTO number. The first time we encounter a genuinely 2-OTO thermal correlation is for four-point functions, as for example illustrated by the now familiar chaos correlator \cite{Maldacena:2015waa}. 
\item More generally, an $n$-point thermal Wightman function can be computed using at most an $\lfloor \frac{n}{2}\rfloor$-OTO contour, whereas in the absence of the KMS relations one would have to go up to $\lfloor \frac{n+1}{2} \rfloor$-OTO contours. Consequently, one may be tempted to speculate, as we do in \S\ref{sec:discuss}, that even-point functions where we first encounter new OTO contour order (e.g., 2-OTO for four-point, 3-OTO for six-point etc.) may be the natural place to look for detailed features of how systems thermalize.
\end{itemize}

While the general set of statements above holds for any thermal Wightman functions, it is useful to express the relations in terms of nested commutators and anti-commutators. As is well known from the Keldysh construction, fully nested commutators capture causal response, and fully nested anti-commutators encode fluctuations. One expects based on the two-point FDT that there would be general relations between such objects and this is indeed borne out. We will in particular argue that {\it $n$-point function generalized FDTs} (which constrain the physics of thermal OTO correlators) can be described as follows:
\begin{itemize}
\item The set of nested commutators with the innermost operator held fixed provides a complete basis of thermal correlation functions (written down in \eqref{eq:causalb}). That is, there are $(n-1)!$ such correlators and they are not related by either generalized Jacobi or KMS relations. We refer to this set of correlators as the \emph{causal basis}.  
\item The KMS relations take a very simple form when expressed in terms of fully nested correlators. This statement can be argued for in a couple of different ways. The first involves writing down thermal Jacobi type operator identities, \emph{tJacobi relations}, which can be proven using thermally deformed commutators and anti-commutators (see Appendix \ref{app:tJacobi}).  Alternately, one can recurse the KMS relations taking into account the generalized Jacobi relations of  \cite{Haehl:2017qfl} to obtain a sequence of iterated KMS relations.
\item We solve the KMS relations explicitly by giving formulae that express classes of correlation functions (such as thermal Wightman functions, nested correlators, and advanced/retarded Green functions) in terms of our causal basis (see \S\ref{sec:Causal}).
\item All told, the causal basis captures the essence of the KMS relations most efficiently. We argue that it is a very useful alternative to a basis of Wightman functions, since it implements nice causal properties and conforms with various analyses of thermal spectral functions in the literature. 
\end{itemize}

The outline of the paper is as follows: In \S\ref{sec:Wkms} we explore general features of thermal Wightman functions exemplifying the cyclic structure and construct the master Wightman function which generates all the KMS relations.\footnote{ This master function is similar to the Floquet/Bloch wavefunction.} In \S\ref{sec:causal} we describe various features of nested commutators, construct the causal basis, and explore iterated KMS relations. \S\ref{sec:sho} uses the simple harmonic oscillator to illustrate the general features of our the discussion in a somewhat explicit manner. Finally, in \S\ref{sec:toto} we turn to the OTO characterization of the KMS relations. We have collected in the appendices several technical details which prove useful to verify the statements in the main text. In addition, for completeness we also check that the thermal identities we derive are consistent with the relations obtained in the thermal quantum field theory literature (Appendix \ref{app:thspec}).

\section{Wightman correlators and the KMS condition}
\label{sec:Wkms}

The Wightman correlation functions of interest are correlation functions of Heisenberg operators $\Op{O}(t) = U(t_0,t)^\dagger \,\Op{O} \,U(t_0,t)$ with no prescribed time-ordering.  We will consider generic $n$-point functions, and w.l.o.g.\ fix the temporal insertion points to be ordered $t_1 > t_2 > t_3 > \cdots > t_n$ and simply permute the operators of interest. The Wightman basis of $n$-point functions is then given by 
\begin{equation}
G_\sigma(t_1, t_2,\, \cdots\,, t_n) =
\vev{\Op{O}_{\sigma(1)} \, \Op{O}_{\sigma(2)}   \, \cdots \, \Op{O}_{\sigma(n)} } \,, \qquad \sigma \in S_n \,,
\label{eq:tobasis}
\end{equation}
where $S_n$ denotes the group of permutations of $n$ objects and $\Op{O}_{\sigma(i)} \equiv \Op{O}_{\sigma(i)} (t_{\sigma(i)})$. This accounts for the $n!$ possibilities of time-ordering of $n$-operators. 

In what follows we will find it convenient to simplify notation -- we refer to the operators by their temporal insertion points thereby abbreviating, $\Op{O}_j(t_j) \equiv j$, which helps declutter formulae below.

While Wightman correlation functions can be studied in any given state, in this paper we will focus exclusively on thermal correlation functions in equilibrium. We therefore take our quantum system to be in a thermal density matrix 
\begin{equation}
\rhoT = \frac{1}{\mathcal{Z}(\beta)} \, e^{-\beta \, H } \,.
\label{eq:thermaldensity}
\end{equation}	
The correlation functions of interest are then
\begin{equation}
G_\sigma^\beta(t_1, t_2,\, \cdots\,, t_n) = \Tr{\rhoT \, \Op{O}_{\sigma(1)} \, \Op{O}_{\sigma(2)}   \, \cdots \, \Op{O}_{\sigma(n)} } \,, \qquad \sigma \in S_n \,,
\label{eq:}
\end{equation}	

A special feature of the thermal density matrix is the fact that it involves the Hamiltonian, and can be interpreted as 
evolving the system by an imaginary amount $t = -i\, \beta$. This fact can be encoded in the Kubo-Martin-Schwinger (KMS) conditions \cite{Kubo:1957mj,Martin:1959jp}.  Formally, one may state the KMS conditions  in terms of the Schwinger functions of the Euclidean theory. 

Two-point  functions are required to be analytic in a strip in the complex time plane: with $t_\mathbb{C} = t_{\sigma(1)} - t_{\sigma(2)} - 
i \, t_\text{E}$ the KMS condition requires that the Green's functions are analytic in the strip $\{ t_\mathbb{C } \in \mathbb{C}: 0 < t_\text{E} < \beta \} $.  Hence for two Heisenberg operators $\Op{A}(t)$ and $\Op{B}(t)$ which are elements of the algebra of observables, the two-point thermal correlator  obeys the periodicity condition
\begin{equation}
\Tr{\rhoT\, \Op{A} (t -i\,\beta)\, \Op{B}(0) } = \Tr{\rhoT\, \Op{B}(0)\, \Op{A} (t) }\,,
\label{eq:therkms}
\end{equation}
for bosonic operators\footnote{  The generalization to fermionic operators is straightforward. We will exclusively work with bosonic operators for simplicity.} $\Op{A}$ and $\Op{B}$. 
We used here conjugation of $\Op{A}$ by the density matrix operator $\rhoT$ and cyclicity of the trace. This motivates us to define the KMS conjugate of an operator: 
\begin{align}
\tilde{\Op{A}}(t) \equiv \Op{A}(t-i\beta) = \rhoT^{-1} \, \Op{A}(t) \, \rhoT\,.
\label{eq:kmsconj}
\end{align}

We are now in a position to state the general KMS condition for $n$-point functions.
Consider a correlation function of the form
\begin{equation}
\begin{split}
\vev{ 1_{k_1  \beta} 2_{k_2 \beta} \cdots n_{k_n \beta} }_\beta &\equiv G^\beta_{\text{id}}( t_1 - i k_1 \beta,  t_2 - i k_2 \beta, \cdots ,  t_n - i k_n\beta) \\ 
& =\Tr{\rhoT \Op{O}_1(t_1-i\, k_1 \beta) \, \Op{O}_2(t_2- i\,k_2 \beta)  \cdots \Op{O}_n(t_n- i\,k_n \beta) }\,.
\end{split}
\label{eq:TimeCorelator1}
\end{equation}
We have allowed complex time shifts  by $k _i\, \beta$ being mindful of the analyticity of the Euclidean correlator to only continue the imaginary argument into the lower half plane. This analytic continuation is valid provided as we read from left to right in the above correlator, the  imaginary part of the time argument always decreases, viz., $k_n+1 \geq k_1\geq k_2\geq\ldots\geq k_n$, independent of the real time ordering, cf., \cite{Evans:1991ky}.\footnote{  This can be seen by inserting a complete set of states between operators and demanding the complex phase factors  thence generated are exponentially damped at high energies. }

Then the KMS conditions can be stated as a relation among cyclic shifts in the imaginary time argument of the correlation functions. For the particular ordering chosen above, we have 
\begin{equation}
\begin{split}
\vev{ 1_{k_1 \beta} 2_{k_2 \beta} \cdots n_{k_n \beta} }_\beta 
&= \vev{ n_{(k_n+1)\beta}1_{k_1 \beta} \cdots  (n-1)_{k_{n-1}  \beta}  }_\beta \\ 
& \qquad \qquad  \vdots  \\
& = \vev{  3_{(k_3+1) \beta} \cdots n_{(k_n +1)\beta} 1_{k_1 \beta} 2_{k_2 \beta} }_\beta \\
& = \vev{  2_{(k_2+1) \beta} \cdots n_{(k_n+1) \beta} 1_{k_1 \beta}  }_\beta \,,
\end{split}
\end{equation}
where we have assumed that all the time arguments fall into the admissible domain.

More generally, we can consider the correlation function defined by the permutation $\sigma \in S_n$ with a sequence of imaginary time excursions characterized by $k_i$ for each of the operators., viz.,
\begin{equation}
\begin{split}
\vev{  \prod_{j=1}^n \; \sigma(j)_{k_{\sigma(j)}\beta} }_\beta &\equiv G_\sigma^\beta( t_1 - i k_1 \beta,  t_2 - i k_2 \beta, \cdots ,  t_n - i k_n\beta) \\ 
& \hspace{-1cm}=\Tr{\rhoT \Op{O}_{\sigma(1)}(t_{\sigma(1)}-i\, k_{\sigma(1)} \beta) \, \Op{O}_{\sigma(2)}(t_{\sigma(2)}- i\,k_{\sigma(2)} \beta)  \cdots \Op{O}_{\sigma(n)}(t_{\sigma(n)}- i\,k_{\sigma(n)} \beta) } .
\end{split}
\label{eq:TimeCorrelator2}
\end{equation}
The KMS condition relates correlators within a cyclic orbit of the permutation $\sigma$ giving relations of the form:
\begin{equation}
\vev{  \prod_{j=1}^n \; \sigma(j)_{ k_{\sigma(j)} \beta} }_\beta = \vev{  \prod_{j=p+1}^n \; \sigma(j)_{ (k_{\sigma(j)} +1) \beta} 
\prod_{\ell=1}^p \, \sigma(\ell)_{ k_{\sigma(\ell)} \beta} }_\beta
 \,, \qquad\text{for}  \;\; p =1,\cdots , (n-1) .
\label{eq:kmst}
\end{equation}	

Thus given a particular time-ordering (specified by the permutation $\sigma$) the KMS conditions relate it to its cyclic permutations, with operators that pass through the density matrix being shifted in imaginary time by an extra unit along the thermal circle. This clearly, breaks up the $n!$ Wightman correlators into $(n-1)!$ equivalence class each comprising of $n$ correlators related by a KMS relation and imaginary time shifted arguments.\footnote{  This statement will be trivial to see in the frequency domain, cf., Eqs.~\eqref{eq:2okms}-\eqref{eq:4okms} below.}

To make these statements more explicit, we will give some examples below. To be concise, we will limit ourselves to the case where all the correlators we will consider have the complex time shifts either $0$ or $- i \beta$, i.e., $k_i \in \{0,1\}$. 

\subsection{KMS relations in time domain}
\label{sec:KMStime}
The KMS relations can be written out explicitly for various low-point Wightman functions. For instance, we have for $n=2$ and $n=3$ the following decomposition of the correlators:

\paragraph{Two-point functions:} At the two-point level there is the standard relation which we express as  
\begin{equation}
\vev{12}=\vev{2_\beta  1}=\vev{1_\beta 2_\beta  }
\label{eq:2tkms}
\end{equation}	
Note that this is the same as the relation quoted originally in \eqref{eq:therkms}, and in particular means that once we take into account the KMS condition there is a single two-point function in thermal equilibrium.

\paragraph{Three-point functions:} One can similarly carry out the exercise for three-point functions.
\begin{equation}
\begin{split}
&   \vev{  1 2_\beta 3}=\vev{ 3_\beta  12_\beta },\,\, 
     \vev{2 1_\beta3 }=\vev{ 3_\beta 2  1_\beta }   \\
&  \vev{  {2}3_\beta  1}=\vev{  1_\beta23_\beta},\,\,
\vev{32_\beta 1 }=\vev{ 1_\beta3 {2}_\beta } ,\,\, 
\vev{3 1_\beta2 }=\vev{ {2}_\beta3 1_\beta },\,\, 
\vev{ 13_\beta {2} }= \vev{2_\beta 13_\beta }             \\
  &   \vev{  {2}3 1}=\vev{  1_\beta 23  }= 
  \vev{3_\beta  1_\beta 2 } = \vev{  {2}_\beta  3_\beta  1_\beta  } ,\,\,  \vev{  13 {2}}=\vev{ 2_\beta 13 }=\vev{3_\beta 2_\beta  1}=\vev{ 1_\beta 3_\beta  {2}_\beta }  \\
  &  \vev{  123}=\vev{ 3_\beta 12  }= \vev{ {2}_\beta 3_\beta 1 }= \vev{  1_\beta  2_\beta 3_\beta  } ,  \,\,\,
                                 \vev{2  13}=\vev{ 3_\beta2 1  }= \vev{ 1_\beta 3_\beta  {2}  }= \vev{ 2_\beta  1_\beta  3_\beta }   \\
                                &      \vev{ 3 2 1}=\vev{  1_\beta3 {2}  }= \vev{2_\beta  1_\beta 3 }= \vev{  3_\beta 2_\beta  1_\beta  } ,   \,\,\,
                                     \vev{ 3 12}=\vev{  {2}_\beta3 1  }= \vev{  1_\beta 2_\beta 3 }= \vev{  3_\beta  1_\beta 2_\beta  }   
\end{split}
\label{eq:3tkms}
\end{equation}
%

\subsection{KMS relations in frequency domain}
\label{sec:KMSom}

In frequency space, KMS relations take a much simpler form since we can write:\footnote{  We will abuse notation and not distinguish between the time and frequency domain incarnation of the operators. By convention we take 
$\Op{O}(\omega) = \int \, dt\, e^{i\,\omega \,t}\, \Op{O}(t)$. }
\begin{equation}
\Op{O}_i(t_i - i\beta) \mapsto e^{-\beta \omega_i} \, \Op{O}_i(\omega_i)\,.
\label{eq:ft}
\end{equation}	
By virtue of the fact that the imaginary time shift factors out, we end up with simple forms for the KMS relations in frequency space:
\begin{equation}
\vev{  \prod_{j=1}^n \; \sigma(j) }_\beta = e^{-\beta \sum_{j=p+1}^n \, \omega_j} \vev{  \prod_{j=p+1}^n \; \sigma(j) 
\prod_{\ell=1}^p \, \sigma(\ell) }_\beta
 \,, \qquad\text{for}  \;\; p =1,\cdots ,(n-1)\,.
\label{eq:kmsfr}
\end{equation}	
This is self consistent since $\sum_{i=1}^n \omega_i=0$ owing to time translational invariance. These relations are again best illustrated with examples for low values of $n$.

\paragraph{Two-point functions:}  The $n=2$ KMS relations \eqref{eq:2tkms} take  the form:
\begin{equation}
\vev{12}= e^{-\beta \omega_2} \vev{ 21 } = e^{-\beta (\omega_1+  \omega_2)} \vev{ 12 }  
\label{eq:2okms}
\end{equation}	

\paragraph{Three-point functions:} The  $n=3$ KMS relations in frequency space can again be immediately written down from the earlier expressions in \eqref{eq:3tkms}. They take the much simpler compact form 
\begin{equation}
\begin{split}
\vev{ {1}23}&=e^{-\beta\omega_3}\vev{3 {1}2  }= e^{-\beta(\omega_2+\omega_3)}\vev{{2}3 {1} }  \\
\vev{2  {1}3}&=e^{-\beta\omega_3}\vev{32 {1}  }=e^{-\beta(\omega_1+\omega_3)} \vev{{1} 3  {2}  }
\end{split}
\label{eq:3okms}
\end{equation}
\paragraph{Four-point functions:} Finally, we record the  $n=4$ KMS relations in frequency space. To wit,  
\begin{equation}
\begin{split}
\vev{ \, 1\, \sigma(2)\sigma(3)\sigma(4)}
& =e^{-\beta\omega_{\sigma(4)}}\vev{ \sigma(4)\, 1\, \sigma(2)\sigma(3)  } \\ 
& =
e^{-\beta(\omega_{\sigma(3)}+\omega_{\sigma(4)})} \vev{\sigma(3)\sigma(4)\, 1\,\sigma(2)  }  \\
&= e^{-\beta(\omega_{\sigma(2)}+\omega_{\sigma(3)}+\omega_{\sigma(4)})}\vev{\sigma(2)\sigma(3)\sigma(4)\, 1\, }
\end{split}
\label{eq:4okms}
\end{equation}
where $\sigma \in S_3$ is any permutation acting on $2,3,4$.

\subsection{The generator of KMS relations}
\label{sec:masterKMS}

The cyclic symmetry inherent in the KMS condition is very clearly visible in the  frequency domain, where associated with a given cyclic permutation we have associated a prefactor involving the frequencies of operators that have been conjugated through the density matrix measured  in thermal units. This simple transformation law suggests that we can recover the KMS relations from a set of master functions that diagonalize the cyclic permutations. 

Consider a Wightman correlation function specified by a given permutation $\sigma \in S_n$. We consider the action of the cyclic group ${\mathbb Z}_n$ on the sequence $\sigma(1) \sigma (2) \cdots \sigma(n)$. Denoting by $\pi$ an element of ${\mathbb Z}_n$, we would end up with map
\begin{equation}
\sigma(1)\sigma(2)\cdots \sigma(n) \;\; \mapsto\;\; \pi\sigma(1) \pi\sigma(2) \cdots \pi\sigma(n) = 
\prod_{j=p+1}^n \, \sigma(j) \,\prod_{\ell=1}^p \, \sigma(\ell) \,.
\label{eq:szact}
\end{equation}	

Taking a weighted average of the cyclic permutations we construct the master function for the permutation $\sigma \in S_n$:
\begin{equation}
\Psi_\sigma(1,\; \cdots, n) \equiv 
 \frac{\sum_{\pi \in {\mathbb Z}_n} \vev{ \pi \sigma(1) \,  \pi \sigma(2) \,  \cdots \, \pi \sigma(n)}}{ \sum_{\pi \in {\mathbb Z}_n} e^{- \frac{\beta}{n} \sum_j \,  j\,\omega_{\pi \sigma(j)}   }} \,,
\end{equation} 
where $\pi \in {\mathbb Z}_n$ is a cyclic permutation. Then the KMS relations can be stated in terms of recovering a given Wightman correlator in frequency space directly from $\Psi_\sigma$. One finds:
\begin{equation}
\label{eq:master}
\vev{ \sigma(1)\,\cdots \,\sigma(n) }  = \Psi_\sigma(1,\,\cdots ,\,n)  \;   e^{-\frac{\beta}{n} \, \sum_j j\,\omega_{  \sigma(j)}} \,.
\end{equation}

Note that the master function $\Psi_\sigma$ is indexed by a permutation $\sigma \in S_n$. Since we are averaging over the ${\mathbb Z}_n$ orbit of $\sigma$, we only need to consider $(n-1)!$ elements of $S_n$. Combined with the action of $\pi$ we would end up covering all $n!$ permutations. 

There is a physically interesting way to write $\Psi_\sigma$. We claim that 
 \begin{equation}
\Psi_\sigma(1,\cdots , n) = \vev{(\rhoT)^{\frac{1}{n}} \sigma(1) \; (\rhoT)^{\frac{1}{n}} \sigma(2)  \; 
\cdots \; (\rhoT)^{\frac{1}{n}}\sigma(n)  } 
\end{equation}
which makes the cyclic invariance manifest.

For $n=2$ this spectral function is naturally interpreted in the thermofield double construction with factors of $\rhoT^{\frac{1}{2}}$ appearing between the forward and backward evolutions. Similarly, the chaos correlator in \cite{Maldacena:2015waa} was computed using an Euclidean regulator which inserts powers of $\rhoT^\frac{1}{4}$ between the operators in a $4$-point function. More generally, inspired by these developments, \cite{Tsuji:2016kep} considered arbitrary combinations of commutators and anti-commutators of two operators, $\comm{\Op{A}}{\Op{B}}$ and $\anti{\Op{A}}{\Op{B}}$, to derive a set of bipartite fluctuation-dissipation relations. Their results follow from the statements mentioned above as the reader can quickly infer. 

 The statement of \eqref{eq:master} is that $\Psi_\sigma$ is the natural object that symmetrizes the KMS relations amongst the $(n-1)!$ independent Wightman functions, and should be viewed as providing a suitable basis for the spectral functions.\footnote{ This is analogous to a Bloch, or perhaps more appropriately, Floquet, basis for wavefunctions.}

Once again it is helpful to view these relations through some examples of low point correlation functions.

\paragraph{Two-point functions:}
There being a single ${\mathbb Z}_2$ orbit since ${\mathbb Z}_2 \simeq S_2$, the unique  2-point cyclic  function $\Psi$ is
\begin{equation}
\Psi(1,2) = \frac{\vev {12 }  + \vev {21 }}{ e^{-\frac{\beta}{2} \omega_1} + e^{-\frac{\beta}{2} \omega_2} } \,.
\end{equation}
The KMS relations \eqref{eq:2okms} can then be re-expressed as  
\begin{equation}
\vev{ 12 }  =  \Psi(1,2) \; e^{\frac{\beta}{2} \, \omega_2} \,,  
\qquad 
\vev{ 21 }  =  \Psi(1,2) \; e^{\frac{\beta}{2} \, \omega_1}  \,.
\end{equation}

\paragraph{Three-point functions:} There are two non-trivial permutations of $S_3$ whose ${\mathbb Z}_3$ orbits help fill out the six elements. Thus we need to consider two master $\Psi$ functions indexed by the two independent orderings. Consider first $\sigma =\text{id}$
and 
\begin{equation}
\Psi_{\text{id}}(1,2,3) =
\Psi(1,2,3) = {\vev{ 1 2 3 }+ \vev{ 2 3 1 } +
\frac {\vev{ 3 1 2 } }{ e^{- \frac{\beta}{3} (\omega_3 -\omega_1) } + e^{- \frac{\beta}{3} (\omega_1 -\omega_2) }+ e^{- \frac{\beta}{3} (\omega_2 -\omega_3) }   }} \,.
\label{eq:P31}
\end{equation}	
In terms of this function we can write down three KMS relations in the ${\mathbb Z}_3$ orbit of $\text{id}$ as
\begin{equation}
\begin{split}
\vev{ 1 2 3 }  &=\Psi(1,2,3) e^{- \frac{\beta}{3} (\omega_3 -\omega_1) }  \\
\vev{ 2 3 1 }  &=\Psi(1,2,3) e^{- \frac{\beta}{3} (\omega_1 -\omega_2) }  \\
\vev{ 3 1 2 }  &=\Psi(1,2,3) e^{- \frac{\beta}{3} (\omega_2 -\omega_3) }  \,.
\end{split}
\end{equation}

We are then left with the second choice $\sigma = (23)$ (using the cycle notation for permutations), which results in 
\begin{equation}
\Psi_{(23)} (1,2,3) \equiv \Psi(1,3,2) = \frac{\vev{ 1 3 2 }+ \vev{ 3 2 1 } + \vev{ 2 1 3 } }{e^{- \frac{\beta}{3} (\omega_2 -\omega_1) } + e^{- \frac{\beta}{3} (\omega_1 -\omega_3) }+ e^{- \frac{\beta}{3} (\omega_3 -\omega_2) }   } \,,
\label{eq:P32}
\end{equation}	
which then gives us   three further relations
\begin{equation}
\begin{split}
\vev{ 1 3 2 }  &=\Psi(1,3,2)\; e^{- \frac{\beta}{3} (\omega_2 -\omega_1) }  \\
\vev{ 3 2 1 }  &=\Psi(1,3,2)\; e^{- \frac{\beta}{3} (\omega_1 -\omega_3) }  \\
\vev{ 2 1 3 }  &=\Psi(1,3,2)\; e^{- \frac{\beta}{3} (\omega_3 -\omega_2) }  \,.
\end{split}
\end{equation}
 
\section{Generalized fluctuation-dissipation relations}
\label{sec:causal}

While Wightman functions encode all the information inherent in the KMS relations quite succinctly, it is often useful in physical applications to consider correlation functions of various combinations involving nested sequence of commutators and anti-commutators. This basis was referred to as the nested basis in \cite{Haehl:2017qfl} and its physical utility is that causal response functions are determined in terms of nested commutators. 

Unlike the Wightman basis the set of nested correlators has redundancies.  For instance even without including imaginary time shifted operators, it was noted in \cite{Haehl:2017qfl} that one naively has $2^{n-2}\, n!$ nested correlators, since each of the $n!$ permutations allows for $(n-1)$ binary choices of either a commutator or an anti-commutator (and we only need to consider alternating permutations owing to an overall sign flip resulting from swapping order of operators in a commutator). In this section we resolve this redundancy, leading to a minimal set of $(n-1)!$ {\it spectral functions}. Generalized fluctuation-dissipation relations encode how to express other correlators (such as Wightman functions, or generic nested correlators) in terms of these spectral functions.

\tikzset{
     block/.style={rectangle, draw, text width=40em,
                   text centered, rounded corners, minimum height=3em},
     arrow/.style={-{Stealth[]}}
     }
     
\begin{figure}
\centering
\small{
\begin{tikzpicture}[node distance=1.2cm]
\node (wightman) [details] {{\bf Wightman correlators}\\ $\vev{\sigma(1) \cdots \sigma(n)}$\\ $\sigma \!\!\!\!\in\!\!\!\! S_n$\\\vspace{.1cm} (count: $n!$)};
\node (wightman2) [thread, below of = wightman, yshift = -4.5cm] {{\bf Reduced Wightman}\\$\vev{1\,\rho(2)\cdots \rho(n)}$\\$\rho \!\!\!\!\in\!\!\!\! S_{n-1}$\\\vspace{.1cm} (count: $(n-1)!$)};
\node (causal) [thread, right of = wightman2, xshift = 5.5cm] {{\bf Causal basis}\\$\vev{\fullcomm{1\,\rho(2)\cdots \rho(n)}}$\\$\rho \!\!\!\!\in\!\!\!\! S_{n-1}$\\\vspace{.1cm} (count: $(n-1)!$)};

\node (nested) [details, right of = wightman, xshift = 9cm] {{\bf Nested correlators}\\ $\vev{\fullcomm{ \sigma(1) \sigma(2)_{\varepsilon_2} \cdots \sigma(n)_{\varepsilon_n}}}$\\ $\sigma \!\!\in\!\! S_n^+,\; \varepsilon_i = \pm$\\\vspace{.1cm} (count: $2^{n-2} n!$)};
\node (Bn) [details, below of = nested, yshift = -1.3cm] {$\mathfrak{B}_n$ basis \eqref{eq:basisGeneral}\\\vspace{.1cm}
(count: n!)};

\draw [arrowT] (wightman) -- node [right] {KMS (\S\ref{sec:Wkms})} (wightman2);
\draw [arrowT] (wightman2) -- node [above] {Lemma \ref{lemma2}$\;\;$} (causal);
\draw [arrowT] (causal.185) -- node [below] {Lemma \ref{lemma5}$\;\;$} (wightman2.-5);
\draw [arrow2] (nested) -- node [right] {sJacobi (\S\ref{app:nestedGeneric})}  (Bn);
\draw [arrowT] (Bn) -- node [right] {$\;\;$KMS} (causal);
\draw [arrowB] (wightman.east) [bend left=30] to node [right=-.2cm] {${\Large\substack{{\mathbf n}\text{\bf -point}\\\text{\bf OTO FDTs}\\\;\\(\S\ref{sec:Causal},\,\S\ref{sec:relrel})\\\;\\\;\\\;\\\,}}$} (causal.150);
\draw [arrowB] (nested.west) [bend right=23] to (causal.125);
\end{tikzpicture}
}
\caption{A pictorial collection of the various classes of correlators discussed in this section. The green colored boxes denote classes that are useful for generic states. The KMS condition in thermal states introduces a further reduction to the objects collected in blue colored boxes. Generalized fluctuation-dissipation theorems (FDTs) describe these redundancies. For completeness, note also Appendix \ref{app:thspec}, which discusses the relation with standard thermal retarded-advanced Green's functions.}
\label{fig:flowchart}
\end{figure}
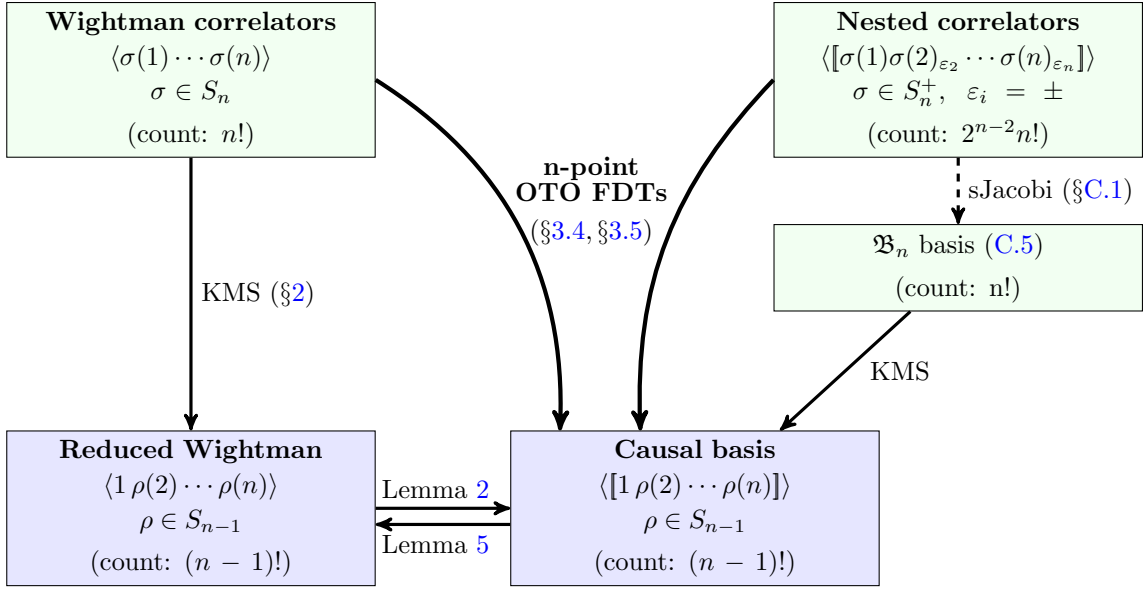

\subsection{Notation} 
To simplify some of the formulae that appear in the sequel, we  introduce some new notation which we collect here for quick reference:
\begin{itemize}	
\item We will be interested in nested correlators with both commutators and anti-commutators. To avoid clutter of brackets, we will write out a single unbroken string of operators with the innermost operator in the sequence being the left-most entry in the sequence. We specify anti-commutators by marking the outer-most entry with a `$+$' subscript, so that 
\begin{equation}
\fullcomm{ij\cdots k_{_+}l} \equiv \comm{\,\cdots
			\anti{\cdots
				\comm{\Op{O}_i}{\Op{O}_j}
				}
				{\cdots\, \Op{O}_k}}
		{\Op{O}_l} \,,
\label{eq:fcomm} 
\end{equation}	
where we remind the reader of our shortcut notation $ \Op{O}_j(t_j) \equiv j $.
For example for 3-point correlators we would have combinations of the form
\begin{equation}
\begin{split}
\fullcomm{123} = \comm{\comm{\Op{O}_1}{\Op{O}_2}}{\Op{O}_3} \,, 
\quad \fullcomm{12_{_+}3} = \comm{\anti{\Op{O}_1}{\Op{O}_2}}{\Op{O}_3}
\,, \quad 
\fullcomm{12_{_+}3_{_+}} = \anti{\anti{\Op{O}_1}{\Op{O}_2}}{\Op{O}_3} \,, 
\end{split}
\end{equation}
among others. In particular, note that a fully nested commutator will have no $+$ subscript markings, i.e., $\fullcomm{ij \cdots kl} \equiv \comm{\,\comm{\cdots\comm{i}{j}}{\cdots k}}{l}$.
\item Various thermal factors which appear in the formulae will also be abbreviated:
\begin{equation}
\begin{split}
&\text{Bose-Einstein factor}: \quad 
\bose_i \equiv \frac{1}{e^{\beta\, \omega_i} -1} \,, \qquad 
\bose_{i,\ldots,\,j} \equiv \frac{1}{e^{\beta\,( \omega_i + \ldots + \omega_j)} -1} \,, \\
&\text{Thermal factors}: \qquad 
 \cth_i = \coth\left(\frac{1}{2}\, \beta \omega_i\right) \,, \\
\end{split}
\label{eq:}
\end{equation}
The following relation between Bose-Einstein factors in correlators with $n$ insertions is very often useful to simplify various expressions:
\begin{equation} \label{eq:ThermalRel}
\begin{split}
  \bose_{A} = - (1+ \bose_{A^c}) \quad \text{ for } A \cup A^c = \{ 1 , \ldots , n \}\,,\;\; A \cap A^c = \emptyset\,.
\end{split}
\end{equation}
This is a consequence of energy conservation, $\omega_1 + \cdots + \omega_n = 0$. For example, in 4-point functions we have $\bose_{1,2,4} = - (1+ \bose_3)$ etc.. An analogous statement for thermal factors can be written as 
\begin{equation}\label{eq:ThermalRel2}
  \prod_{i=1}^n (\cth_i+1) = \prod_{i=1}^n (\cth_i-1) \,.
\end{equation}
\item In the context of fluctuation-dissipation relations, we will use a shortcut for the combination that appears in the $n=2$ FDT:
\begin{equation}
\kmsfd{\Op{X}}{\Op{A}} \equiv \vev{\fullcomm{\Op{X\, A_{_+}}}} + \cth_{\Op{A}} \, \vev{
\fullcomm{\Op{X\, A}}}
 \label{eq:masterFD}
\end{equation}	
\item When necessary we capture both commutators and anti-commutators by the definition 
\begin{equation}\label{eq:gradedcomm}
  \Ecomm{\Op{A}}{\Op{B}}{\Op{B}} = \Op{A} \,\Op{B} + \varepsilon_{\Op{B}}\, \Op{B} \,\Op{A} \,,\qquad
  \varepsilon_{\Op{B}} \in \{+,-\} \,.
\end{equation}
\item Inspired by various relations which we encounter, we also introduce a new bracket notation.  Define the thermally deformed commutator and anti-commutator as follows:
\begin{equation}
\tcomm{\Op{A}}{\Op{B}}{\Op{A}} = \Op{A}\,\Op{B} + \varepsilon_{\Op{A}} \, e^{\beta\, \omega_{_\Op{A}}}\, \Op{B}\, \Op{A} \,,\qquad \varepsilon_{\Op{A}} \in \{+,-\}\,.
\label{eq:tac}
\end{equation}	
Note that the definition singles out the inner-most operator in the (anti-)commutator and reduces to the standard commutator and anti-commutator when $\beta \to 0$. This thermal commutator is used in Appendix \ref{app:tJacobi} and provides an alternative route to deriving the fluctuation-dissipation relations for nested thermal correlators.
\end{itemize}

\subsection{Relations among nested  correlators: examples}
\label{sec:egNest}

It is helpful to first see some explicit examples (which are well known) to understand the rationale behind the use of the nested correlators.

\paragraph{Two-point functions:} The usual presentation of the KMS relation is in the form of a fluctuation-dissipation (FD) condition. Recall that response functions, which involve commutators, naturally diagnose dissipation (and transport), while the symmetrized Green's functions capture fluctuations \cite{Chou:1984es}. Starting with \eqref{eq:2okms} one can by taking suitable linear combinations arrive at the standard form of the FD relation, viz.,
\begin{equation}
\vev{\fullcomm{12_{_+}}}
\equiv \vev{ \{ 1, 2 \}} = \cth_1 \vev{[1,2]} \equiv \cth_1 \vev{ \fullcomm{12}}   .
\label{eq:FD2}
\end{equation}
We have chosen here to use the KMS relation to solve for the fluctuation measure in terms of the commutator. Whilst it is conventional to do so, this relation is singular when $\beta\, \omega_1 \to 0$ for $\cth_1$ diverges. What this really means is that in this limit the two operators simply commute, as would have been manifest had we solved the KMS relation to give instead
\begin{equation}
 \vev{ \fullcomm{12}}   = \frac{1}{\cth_1}\, \vev{\fullcomm{12_{_+}}}
\label{eq:FD2alt}
\end{equation}	
In what follows we will typically solve the KMS relations and present the results in terms of nested commutators. The reader should be aware then that the relations hold modulo \emph{contact-terms}, which could arise when the statistics factors $\cth_i$ diverge. 

\paragraph{Three-point functions:} The analogous story for 3-point functions is a bit more involved. We have 6 permutations of the operators, and two possible nestings of each ordering into a commutator/anti-commutator. Of these $12$ nested correlators, $6$ can be eliminated by use of generalized Jacobi relations, called \emph{sJacobi} relations in \cite{Haehl:2017qfl}.\footnote{  The generalized Jacobi relations involve constraints between $n$-point nested correlators arising from the presence of two independent brackets, the commutator and anti-commutator.} In the thermal state we have further KMS relations \eqref{eq:3okms} which amount to $4$ further relations leading to two independent 3-point functions, which we take to be $\fullcomm{123}$ and $\fullcomm{132}$ respectively. We find (modulo the caveat above):
\begin{equation}
\begin{split}    
\vev{\fullcomm{123_{_+}}}&= -\cth_3 \, \vev{\fullcomm{123} }   ,\\
\vev{\fullcomm{12_{_+} 3_{_+}}}&= -\cth_3 \cth_1 \, \vev{ \fullcomm{123}} +\cth_3\left(\cth_1+\cth_2\right)  \vev{\fullcomm{132} }  ,\\
\vev{\fullcomm{12_{_+}3}}&= \cth_1  \, \vev{\fullcomm{123}} -\left(\cth_1+\cth_2\right) 
\vev{\fullcomm{132} }     .
\end{split}
\label{eq:FD3}
\end{equation}
 Permutations of the operators $\{1,2,3\}$ give the remaining relations, upon using the sJacobi relations \cite{Haehl:2017qfl}
 \begin{equation}
 \begin{split}
 \fullcomm{312} &= - \fullcomm{132}  ,\\
 \fullcomm{231} &= -\fullcomm{123}+ \fullcomm{132}  .
 \end{split}
 \label{eq:sJac3}
 \end{equation}
to always move the operator $1$ to the left-most position (a pattern which we will repeat below).
While we have written only four of the remaining 10 nested correlators, the remaining six may be obtained by permutations of the above. 

It is also useful to record here the expression for the thermal $3$-point Wightman correlator in terms of the two independent nested basis elements identified above. To wit,
\begin{equation}\label{eq:123C}
\begin{split}
  \vev{123}   
  &= (1+\bose_{1})(1+\bose_{1,2})\, \vev{\fullcomm{123} }+ (1+\bose_{1}) \bose_{1,3}\, 
  \vev{\fullcomm{132} } .
\end{split}
\end{equation}
The remaining Wightman functions can be similarly expressed after permuting the operators on the l.h.s and using the sJacobi relation in Eq.~\eqref{eq:sJac3} to eliminate $\fullcomm{231}$. For example, we can write the following set of relations:
\begin{equation}
\begin{split}
  \vev{123}   
  &= (1+\bose_{1})(1+\bose_{1,2})\,  \vev{\fullcomm{123}} + (1+\bose_{1}) \bose_{1,3}\,  \vev{\fullcomm{132}} ,\\
  \vev{213} 
  &= \bose_{1}(1+\bose_{1,2})\,  \vev{\fullcomm{123}} + (1+\bose_1)\bose_{1,3}\,  \vev{\fullcomm{132}} ,\\
  \vev{321} 
  &=\bose_1 \bose_{1,2}\, \vev{\fullcomm{123}} +\bose_1 (1+\bose_{1,3}) \,  \vev{\fullcomm{132}} ,   
\end{split}
\end{equation}
which involves all possible positions of the operator $1$, so the remaining 3 Wightman functions are obtained by permuting $2$ and $3$, which doesn't require any use of Jacobi identities.

\paragraph{Four-point functions:} Let us now turn to $4$-point functions. There are 96 possible correlators, obtained by taking the 24 permutations of operators and inserting them into commutators and anti-commutators.  We now show that these are determined by $6$ independent structures after accounting for the sJacobi and KMS relations. 
We will choose to express all the correlators in terms of the basis generated by  $\vev{\fullcomm{1\sigma(2)\sigma(3) \sigma(4)}}$  with $\sigma \in S_3$ (we show that this is a basis in 
Appendix \ref{app:nestbasis}).\footnote{ We alert the reader that in all of this section, elements of the permutation group $S_{n-1}$ act on the set of operators $\{2,\ldots,n\}$ (instead of the more standard representation on $\{1,\ldots,n-1\}$).} Explicitly, all commutator/anti-commutator nestings can be written in terms of this basis using the KMS relations (cf.,  the caveat around \eqref{eq:FD2alt}):
\begin{align}
\vev{\fullcomm{12_{_+}3_{_+}4_{_+}}}
&
=	 -\cth_1\cth_4\frac{\cth_1 \cth_2+1}{\cth_1+\cth_2} \,\vev{\fullcomm{1234}}-
	\cth_3\cth_4\frac{\cth_1^2-1}{\cth_1+\cth_3} \, \vev{\fullcomm{1324}}
	+\cth_1\cth_4\frac{\cth_3^2-1}{\cth_3+\cth_4} \, \vev{\fullcomm{1243}}  \nonumber \\
 &\quad\;
 	+\cth_1\cth_4 \frac{\cth_2^2-1}{\cth_2+\cth_4} \vev{\fullcomm{1342} }
 	-\cth_3 \cth_4 \frac{\cth_1^2-1}{\cth_1+\cth_4} \vev{\fullcomm{1423}}
	-\cth_2\cth_4 \frac{\cth_1^2-1}{\cth_1+\cth_4}\,\vev{\fullcomm{1432} }
 	  \nonumber \\
\vev{\fullcomm{12 3_{_+}4_{_+}}}
&
=	 \frac{\cth_4}{\cth_3+\cth_4} \left( (\cth_3 \cth_4+1) \,\vev{\fullcomm{1234}} +
	(\cth_3^2-1)\, \vev{\fullcomm{1243}} \right)  \nonumber \\
\vev{ \fullcomm{12_{_+} 3 4_{_+}} } &
 = -\cth_1 \cth_4 \, 
 \, \vev{\fullcomm{1234}}+\cth_4\frac{\cth_1^2-1}{\cth_1+\cth_3}\, \vev{\fullcomm{1324}} + \cth_4 \frac{\cth_2^2-1}{\cth_2+\cth_4}  \,\vev{\fullcomm{1342}} 
\nonumber \\
 &
 \quad\;\; + \cth_4 \frac{\cth_1^2-1}{\cth_1+\cth_4}  \,\left( \vev{\fullcomm{1423}} - \vev{\fullcomm{1432}}  \right)\nonumber \\
\vev{ \fullcomm{1234_{_+}}}
&
=	 -\cth_4 \, \vev{\fullcomm{1234}}   \,, \hspace{2.1cm} 
\vev{\fullcomm{12_{_+}3_{_+} 4}}
=	- \frac{1}{\cth_4}  \vev{\fullcomm{12_{_+}3_{_+}4_{_+}}} \,,
 \nonumber \\
\vev{\fullcomm{123_{_+}4}} & 
=	 - \frac{1}{\cth_4} \,\vev{\fullcomm{123_{_+}4_{_+}}} \,, \hspace{1.7cm} 
\vev{\fullcomm{12_{_+}34}  }
 =	-\frac{1}{\cth_4}  \,\vev{\fullcomm{12_{_+}34_{_+} }}\,,
 \label{eq:FD4}
\end{align}
and the following Jacobi identities to always position the operator $1$ as the left-most one:  
 \begin{equation}
 \begin{split}
 \fullcomm{4123}& =-\fullcomm{1423} \,,\\
 \fullcomm{3412}& =-\fullcomm{1342}+\fullcomm{1432} \,, \\
\fullcomm{2341}& =-\fullcomm{1234}+\fullcomm{1324}+\fullcomm{1423}-\fullcomm{1432} \,.
\end{split}
\label{eq:sJac4}
\end{equation}
The KMS equations \eqref{eq:FD4} and all permutations of $\{1234\}$ therein, combined with the Jacobi identities \eqref{eq:sJac4} and all permutations of $\{234\}$ therein, generate the required relations to express the 96 nested correlators in terms of the basis $\vev{\fullcomm{1\sigma(2)\sigma(3) \sigma(4)}}$.
 
One can similarly  express any thermal 4-point Wightman correlator in terms of our nested commutator basis as:
\begin{equation}
\begin{split}
    \vev{1234} 
  &= (1+\bose_1) (1+\bose_{1,2}) (1+\bose_{1,2,3}) \, \vev{\fullcomm{1234}}+(1+\bose_1)(1+ \bose_{1,2})\bose_{1,2,4}\,\vev{\fullcomm{1243}} \\
   &\quad 
   +(1+\bose_1) \bose_{1,3} (1+\bose_{1,2,3}) \, \vev{\fullcomm{1324}}+ (1+\bose_{1})(1+\bose_{1,3})\bose_{1,3,4}\, \vev{\fullcomm{1342}}  \\
  &\quad 
  + (1+\bose_{1})\bose_{1,4}(1+\bose_{1,2,4})\, \vev{\fullcomm{1423}}+ (1+\bose_1)\bose_{1,4} \bose_{1,3,4}\, \vev{\fullcomm{1432}}   \,.
  \end{split}
  \label{eq:4pw}
  \end{equation}
 Taking all permutations of $(1234)$ gives the 24 Wightman correlators in terms of our six element basis of the form $\fullcomm{1\sigma(234)}$ upon using the Jacobi identities to always position the operator $1$ as the left-most one. Bose-Einstein factors can be brought to canonical form using \eqref{eq:ThermalRel}.

\paragraph{Chaos correlator:} The chaos correlator $\mathcal{C}(t)=\vev{ [\Op{V}(0),\Op{W}(t)]^2 }$ 
is an out-of-time-order correlator  as a measure of chaotic behaviour in a thermal quantum system \cite{Shenker:2013pqa,Larkin:1969aa,Maldacena:2015waa}.   We consider the following parameterization:
\begin{equation}
\begin{split}
\mathcal{C}(t_1,t_2 ,t_3\ t_4) & \equiv 
	\vev{ \comm{\Op{V}(t_1)}{ \Op{W}(t_2)} \; \comm{\Op{V}(t_3)}{\Op{W}(t_4)} } \\
 & = 
 	\frac{1}{(2\pi)^4}\int _{-\infty}^{+\infty} \prod_{k=1}^4\, d\omega_k \;
 	e^{-i \sum_k \, \omega_k\,t_k } \; 
 	\vev{ \comm{\Op{V}(\omega_1)}{ \Op{W}(\omega_2)} \; \comm{\Op{V}(\omega_3)}{\Op{W}(\omega_4)} } .
 \end{split}
\end{equation}	
Using the aforementioned results we can express this correlator  in terms of the thermal spectral functions in the Wightman basis or the nested commutator basis, respectively as:
\begin{align}
\mathcal{C}(t) &= 
 	\frac{1}{(2\pi)^4}\int _{-\infty}^{+\infty} \prod_{k=1}^4\, d\omega_k \;
 	\,e^{-i(\omega_2+\omega_4)t} 
 	\left( \vev{1234}-\vev{1243}-e^{\beta\,\omega_2}\, \vev{ 1342}+e^{\beta\,\omega_2}\, \vev{ 1432} \right) , \nonumber \\
 \mathcal{C}(t) &=
 	\frac{1}{(2\pi)^4}\int _{-\infty}^{+\infty} \prod_{k=1}^4\, d\omega_k \;
 	 \frac{e^{-i(\omega_2+\omega_4)t}}{1-e^{-\beta (\omega_1+\omega_2)}}
 	 \left( \fullcomm{1234}-\fullcomm{1243}\right) ,
\label{eq:WNchaos}
\end{align}
where $\mathcal{C}(t) = \mathcal{C}(0,t,0,t)$.
As explained in \cite{Maldacena:2015waa} it is useful to consider the regulated correlator  
\begin{equation}
 F(t_1,t_2,t_3,t_4)\equiv  \Tr{  \rhoT^{\frac{1}{4}}\; \Op{V}(t_1)\; \rhoT^{\frac{1}{4}}\; \Op{W}(t_2)\; \rhoT^{\frac{1}{4}}\; \Op{V}(t_3)\; \rhoT^{\frac{1}{4}}\; \Op{W}(t_4)} 
 = \Psi\left(\Op{V}_1 \,,\Op{W} _2\,, \Op{V}_3\,, \Op{W}_4\right) ,
\end{equation}	
which is simply the spectral Wightman 4-point function. Letting  $F(0,t,0,t)=F(t)$  
\begin{equation}
\begin{split}
F(t)&=
	\int _{-\infty}^{+\infty} \prod_{k=1}^4\, \frac{d\omega_k}{(2\pi)^4} \; 
	e^{-i(\omega_2+\omega_4)t} \; e^{\frac{\beta}{4} \sum_{j=1}^4\, j\, \omega_j}  \ 
	(1+\bose_1)  \Bigg(
	(1+\bose_{2,4})  \bose_4\, \fullcomm{1324} \,  +  \bose_{2,4}\, (1+\bose_2) \, \fullcomm{1342}
\\ 
& \qquad \qquad 
	- (1+\bose_{1,2})\left(\bose_4 \, \fullcomm{1234} + (1+\bose_3)\, \fullcomm{1243}\right)
	- \bose_{1,4}\left(\bose_3 \, \fullcomm{1423} + (1+\bose_2) \, \fullcomm{1432} \right)
	\Bigg) .
\end{split}
\label{eq:chaoscomm}
\end{equation}
 %

\subsection{Towards a causal basis}
\label{sec:nptNest}

After having seen various examples at low-point order, let us finally explain the general construction for $n$-point functions.  To understand the situation better we should first enumerate the space of nested correlators and ascertain the relations we expect amongst them. Once we do this we will be in a position to  construct a useful basis in terms of which to express the full set of correlators. As the reader might anticipate from the results in \S\ref{sec:egNest}, we will now argue that the appropriate basis is provided by a subset of fully nested commutators (which therefore pick out a basis with nice causality properties). We only give the general arguments below, and relegate most of the details to Appendices 
\ref{app:tJacobi} and \ref{app:nestbasis}, respectively.

\subsubsection{Counting}
\label{sec:countNest}

Any $n$-point nested correlator should be expressible as a linear combination of $(n-1)!$ spectral functions after using all sJacobi relations and KMS relations. 

First, recall from the beginning of this section that there are $2^{n-2}n!$ nested correlators involving (anti-)commutators \cite{Haehl:2017qfl}, but our analysis in terms of Wightman correlators reveals that these should be expressible in terms of 
$(n-1)!$ spectral functions for a thermal state.

 Of the set of nested correlators (which allow for both commutators and anti-commutators), one has  $\frac{1}{2}\, n!=\frac{1}{2}\, n\, (n-1)!$ nested commutators. It would already be sufficient to pick out from amongst these $(n-1)!$ nested commutators  and declare them to be our basis. For this to be true, we need to show that the remaining nested correlators are related to our basis choice by either sJacobi relations or through KMS relations.

 We recall that amongst the $2^{n-2}\, n!$ nested correlators, there are $(2^{n-2} -1) n!$ sJacobi relations.\footnote{  These can be thought of as generalizations of the standard $n=3$ Jacobi relation, which address both nested $n$-point commutators and anti-commutators.}
 They are operator relations in that they hold by virtue of the products of operators generating a free algebra with a graded commutator. The detailed proof of this statement and examples of relations can be found in \cite{Haehl:2017qfl}. Accounting for these sJacobi relations we would end up with a basis of $n!$ correlators which agrees with the Wightman count. However, for thermal correlators, the KMS conditions further reduce the number of independent correlators to $(n-1)!$. Thus there should be $n!-(n-1)!$ further relations. The quest now is to find a useful way to characterize the full set of relations in one go.

\subsubsection{The causal thermal commutator basis}
\label{sec:tcomm}

In \S\ref{sec:egNest} we gave various examples, expressing low-point Wightman functions and nested correlators in terms of nested commutators $\vev{ \fullcomm{1\, \sigma(2)\sigma(3)\, \cdots\,\sigma (n)} }$. We now make a completeness statement about this representation:\\

\noindent
{\bf Lemma \textlabel{1}{lemma1}:}  A useful basis of thermal $n$-point correlators is given by the nested commutators 
\begin{equation}
 \vev{ \,\fullcomm{1\, \sigma(2)\sigma(3)\, \cdots\,\sigma (n)} \,}  \qquad \text{for all }\sigma \in S_{n-1} \,.
\label{eq:causalb}
\end{equation}	
We refer to this as the {\it causal basis}.
Consequently, for a thermal density matrix, any nested correlator (and therefore any Wightman function) can always be expressed as a linear combination of causal spectral functions involving only nested commutators of the form \eqref{eq:causalb}, which we will make explicit below.\\

\noindent
 The equations expressing any thermal Wightman function in terms of the causal basis are equivalent to the {\it Fluctuation-Dissipation (FD) relations} for $n$-point functions (see below). For correlators that can be computed by the Schwinger-Keldysh path integral these were given by \cite{Wang:1998wg} using the av(erage)-dif(ference) basis of contour correlators. Since the latter can be expressed in terms of the nested correlators using the Keldysh rules, this is an equivalent presentation. We indeed check in Appendix \ref{app:thspec} that their expressions match with ours for low-point functions. Later in \S\ref{sec:toto} we will see that for $n\geq 3$ there are non-trivial KMS relations relating out-of-time order correlators which transcend the Schwinger-Keldysh ordering (cf., also \cite{Tsuji:2016kep}).
 
There are a couple of different ways to illustrate the veracity of our claims above:
\begin{itemize}
\item The most straightforward way is to explicitly demonstrate how every Wightman function can be expressed in terms of the above causal basis. This is done in \S\ref{sec:Causal} and can be seen as a generalization of FD relations. A related approach is to write any one of the $2^{n-2}n!$ nested correlators in terms of the causal basis (see \S\ref{sec:nested}).
\item Another approach is to work directly with generalized Jacobi relations to show completeness of the basis \eqref{eq:causalb}. One can first account for the sJacobi relations of \cite{Haehl:2017qfl} (which are valid in any state), and then subsequently isolate the extra relations implied by the thermal nature of the density matrix. We illustrate the latter method in \S\ref{sec:relrel}.
\item Finally one can mimic the strategy followed in \cite{Haehl:2017qfl} more closely and encapsulate the set of relations implied by the sJacobi relations and thermal KMS conditions in one swoop. This leads to the notion of {\it tJacobi} operator relations, which  are thermal deformations of Jacobi relations and their generalizations. We  explore these in Appendix \ref{app:tJacobi}. 
\end{itemize}

\subsection{Wightman functions in terms of causal basis}
\label{sec:Causal}

It is possible to generalize \eqref{eq:123C}, \eqref{eq:4pw} to $n$-point functions and give a general formula that expresses Wightman functions in terms of the causal basis of correlators $\vev{ \,\fullcomm{1\, \sigma(2)\sigma(3)\, \cdots\,\sigma (n)} \,} $. 
The result can be stated as follows (the proof can be found in Appendix \ref{app:proofThermal}):\footnote{  We thank Simon Caron-Huot for sharing valuable insights from his unpublished work which was extremely useful in obtaining this result.}\\

\noindent
{\bf Lemma \textlabel{2}{lemma2}:} An $n$-point thermal Wightman correlator can be written in terms of nested commutators as
\begin{equation} \label{eq:ShuffleResRel}
 \vev{12 \cdots n} = \sum_{\rho \in S_{n-1}} \left((1+\bose_1) \prod_{i=2}^{n-1} \Big(\tilde{s}_i^{(\rho)} + \bose_{1,\rho(2), \ldots, \rho(i)}\Big) \right) \vev{\fullcomm{1\rho(2)\cdots \rho(n)}} \,,
\end{equation}
where $\rho \equiv (\rho(2), \ldots, \rho(n))$ are understood to be permutations acting on the set $\{2,\ldots,n\}$ and the numbers $\tilde{s}_i^{(\rho)} \in \{0,1\}$ describe the {\it run structure} of the permutation $\rho$. 

We define the run structure of a permutation $\rho$ as follows. We call the index $i$ an {\it ascent} if $\rho(i) < \rho(i+1)$ (and a {\it descent} otherwise). The run structure is then the collection of ascents and descents, which culminate in peaks and valleys, respectively, along the string of elements being permuted (see \cite{Fewster:2014aa} for a recent attempt to enumerate permutations by their run structure).  

The numbers $\tilde{s}^{(\rho)}_i$ are defined in terms of the Heaviside step-function scanning the run structure of the permutation $\rho$:
\begin{equation}
  \tilde{s}_i^{(\rho)} = \Theta\big(\rho(i+1)-\rho(i)\big) \equiv \left\{ \begin{aligned}  &1 \qquad \text{if } \rho(i) < \rho(i+1)\; \text{ (ascent)} \\
  &0 \qquad \text{if } \rho(i) > \rho(i+1) \;\text{ (descent)}\end{aligned} \right.
\end{equation}
For example, the four-point function $\vev{\fullcomm{1243}}$ is characterized by the permutation $\rho = (2,4,3)$, which has $\{\tilde{s}_2=1,\tilde{s}_3=0\}$, and $\vev{\fullcomm{1432}}$ is characterized by $\rho = (4,3,2)$ with $\{\tilde{s}_2=0,\tilde{s}_3=0\}$. These examples can be used to confirm \eqref{eq:4pw}.\\

For reasons discussed around Eq.~\eqref{eq:FD2alt} one has to pay closer attention to the loci where the Bose-Einstein functions conspire to diverge, viz., in the limit $\beta\,\omega_i \to 0$. One can check that the linear map between the Wightman correlators and the nested commutator basis has an eigenspectrum  built from the set of products of the statistical factors.
More precisely, we find that the eigenvalues of the transformation \eqref{eq:ShuffleResRel} are given by\footnote{ We have inferred these eigenvalues by analyzing low-point examples (up to and including $n= 5$). In the formula below we do not give the degeneracies. Of the $(n-1)!$ eigenvalues, we empirically find 
\begin{itemize}
\item  $-\bose_{2,\ldots,n}$ has degenracy $(n-2)!$
\item  $\bose_{\sigma(2),\ldots,\sigma(n)}\, \bose_{\sigma(n)}$ has degenracy $(n-3)!$
\end{itemize}
 and so on.  We will not here attempt to prove these statements, since we are going to eschew these singular loci in our discussion. }
\begin{equation}
\label{eq:evalsM}
 -\bose_{2,\ldots,n} \quad \text{and} \quad  (-)^{N+1}\, \bose_{\sigma(2)\cdots \sigma(i_1)} \, \bose_{\sigma(i_1+1) \cdots \sigma(i_2)} \, \bose_{\sigma(i_2+1) \cdots \sigma(i_3)}\cdots \bose_{\sigma(i_N+1) \cdots \sigma(n)}
\end{equation}
for all permutations $\sigma \in S_{n-1}$ and all indices $2\leq i_1  \leq \ldots \leq i_N < n$ for any $1\leq N \leq n-2$. In words: the set of $(n-1)!$ eigenvalues is constructed by distributing all frequencies $\omega_2,\ldots,\omega_n$, arbitrarily over an arbitrary number of $\bose$-factors.
There exist divergent eigenvalues whenever any subset of the frequencies adds up to zero, i.e., whenever there exists a permutation $\sigma$ such that $\beta \sum_{i=2}^{k\leq n} \omega_{\sigma(i)} =0$.  At these points the relation given in \eqref{eq:ShuffleResRel} breaks down and one would have to account for additional contact terms on the right hand side. In the discussion below we will assume that we are always away from these singular points in frequency space and refrain from writing down the contact terms. We note however that the origin of such terms was already inferred in the analysis of \cite{Evans:1991ky} based on the analyticity domains of the Euclidean correlator.

Lemma \ref{lemma2} can be seen as one way to state $n$-point function fluctuation relations. It provides an explicit and complete implementation of KMS conditions: a trivial permutation of labels $\{2,\ldots,n\}$ on both sides of \eqref{eq:ShuffleResRel} gives a set of $(n-1)!$ relations of the same form, which then provide the matrix that transforms between Wightman functions and the causal basis (in Appendix \ref{app:proofThermal} we give an explicit formula for these permutations). Note also that this matrix is obviously invertible -- the converse problem of expressing nested commutators in terms of Wightman functions is rather trivial to solve (one simply expands out the commutators), and an explicit formula can be found in Lemma \ref{lemma5}.

\subsubsection{Application: nested correlators in terms of causal basis}
\label{sec:nested}

As an application of Lemma \ref{lemma2}, we can give the relations that give the complete set of $2^{n-2}n!$ nested correlators in terms of the causal basis. This again amounts to a complete set of generalized fluctuation-dissipation relations. To achieve this, we expand a nested correlator $\vev{\fullcomm{1\,2_{\varepsilon_2} \cdots n_{\varepsilon_n}}}$ in terms of Wightman functions, bring them to canonical form (i.e., with $\Op{O}_1$ being the left-most operator) using the KMS condition, and finally apply Lemma \ref{lemma2} to each term. The result is\\

\noindent
{\bf Lemma \textlabel{3}{lemma3}:} An $n$-point nested correlator can be written in terms of nested commutators as
\begin{equation} \label{eq:lem3}
\vev{\fullcomm{1\,2_{\varepsilon_2} \cdots n_{\varepsilon_n}}} = \sum_{\rho \in S_{n-1}} \, \mathcal{T}_{\{\varepsilon_i\}}^{(\rho)}\,  \vev{\fullcomm{1\, \rho(2) \cdots \rho(n) }}
\end{equation}
with thermal factors given by 
\begin{equation}\label{eq:NestedFactors}
 \mathcal{T}_{\{\varepsilon_i\}}^{(\rho)}= (1+\bose_1) \left(1+ \varepsilon_n e^{\beta \omega_n}\right) \!\!\!\sum_{\substack{\;\;\;\;s_2=0,1\\\;\;\;\;\smash{{\colon}} \\s_{n-1}=0,1}} \; \prod_{i=2}^{n-1} \varepsilon_i^{s_i} \, e^{\beta \, s_i \, \omega_i} \Big(s_{\rho(i+1)} (1- \tilde{s}^{(\rho)}_i) + (1-s_{\rho(i)}) \tilde{s}^{(\rho)}_i + \bose_{1,\rho(2), \ldots, \rho(i)}\Big) 
\end{equation}
As a quick consistency check, we note that $ \mathcal{T}_{\{-1,\ldots,-1\}}^{(\rho)} = \delta_{\rho,\text{id}}$. For more details on the derivation of \eqref{eq:lem3}, we refer to Appendix \ref{app:proofThermal}. Rewriting the Bose-Einstein factors in terms of $\cth_i$, one can use this formula to verify, e.g., the four-point function fluctuation relations stated in \eqref{eq:FD4}.

Lemma \ref{lemma3} can again be interpreted as solving the generalized $n$-point sJacobi relations and the KMS relations at the same time to explicitly write all nested correlators in terms of our causal basis. 
Let us also note that the Keldysh rules (and their generalization to $k$-OTO contours of \cite{Haehl:2017qfl}) can be used to write standard retarded and advanced thermal $n$-point functions in terms of the objects appearing on the left hand side of Lemma \ref{lemma3}. Therefore, \eqref{eq:lem3} allows us to express standard response functions in terms of the causal basis. We refer to Appendix \ref{app:thspec} for a more detailed consistency check with existing literature on this topic.

\subsection{Iterated KMS relations}
\label{sec:relrel}

Let us now turn to an alternative approach to verify Lemma \ref{lemma1}, which entails an implementation of KMS relations at the level of generalized Jacobi relations.

We will start with known KMS relations and employ the sJacobi relations to whittle down the space of nested correlators. We will see that a naive application leads to an over-complete set of relations, which then implies that there ought to be non-trivial identities amongst the various relations we construct. These turn out to the interesting identities amongst thermal correlators, which  is the \emph{raison d'etre} for the construction we describe herein.

The general discussion is best illustrated by an example. Consider the $n=2$ FD theorem  \eqref{eq:FD2} for two operators $\Op{X}$ and $\Op{A}$ which follows from the KMS relation, viz.,\footnote{ We emphasize again that the expressions we write down have potential ambiguities owing to the divergence of the statistics factor  $\cth$ as $\beta \omega \to 0$. See the discussion around Eq.~\eqref{eq:FD2alt} and Eq.~\eqref{eq:evalsM}.}
\begin{equation}
\vev{\fullcomm{\Op{X \, A_{_+} }}} = - \cth_{\Op{A}}\, \vev{\fullcomm{\Op{X \, A}}} .
\label{eq:FD2x}
\end{equation}	
As this holds for any $\Op{X}$ in the operator algebra it in particular holds for the choices 
$\Op{X} = \comm{\Op{C}}{\Op{B}}$ or $\Op{X} = \anti{\Op{C}}{\Op{B}}$. 
This then implies 
\begin{equation}
\vev{\fullcomm{\Op{C \, B\, A_{_+} }} } = 
- \cth_{\Op{A}} \, \vev{\fullcomm{\Op{C\, B\, A}}} \,, 
\qquad 
\vev{\fullcomm{\Op{C \, B_{_+}\, A_{_+} }} } = 
- \cth_{\Op{A}} \, \vev{\fullcomm{\Op{C\, B_{_+}\, A}}}  .
\end{equation}	
These can be further simplified using the  $n=3$ sJacobi relations \cite{Haehl:2017qfl}, 
\begin{equation}
\fullcomm{\Op{A\, B_{_+}\, C}} = 
\fullcomm{\Op{B\, C\, A_{_+}}} - 
\fullcomm{\Op{C\, A\, B_{_+}}} \,, 
\qquad 
\fullcomm{\Op{A\, B\,C}} = 
\fullcomm{\Op{B\, C\, A}} - 
\fullcomm{\Op{ C\, A\, B}} \,, 
\label{eq:sJac3rep}
\end{equation}	
leading to
\begin{equation}
\vev{\fullcomm{\Op{A\, B_{_+}\, C}} }= 
\cth_{\Op{A}} \, \cth_{\Op{C}}\, \vev{ \fullcomm{\Op{A\, B\, C}} } + 
\cth_{\Op{C}} \,\left(\cth_{\Op{A}} + \cth_{\Op{B}} \right) \vev{\fullcomm{\Op{A\, C\, B}}} .
\end{equation}	
This exercise shows that all the  $n=3$ nested correlators  can be expressed in terms of $n=2$  nested commutators and upon using the sJacobi relations, one isolates the $n=3$ FD relation.

One therefore suspects that all the $n$-point fluctuation-dissipation relations can be derived this way, i.e., using \eqref{eq:FD2}, iterating, and applying the appropriate set of sJacobi identities. This makes clear the primacy of the $n=2$ FDT \eqref{eq:FD2}, since the sJacobi identities are but operator identities owing to the presence of a graded commutator (as explained in \cite{Haehl:2017qfl} this simply follows from thinking about the operator algebra as a free Lie algebra). In particular, all the higher point relations (for example those obtained in \cite{Wang:1998wg}) are generated from the $n=2$  FDT algebraically, with no further dynamical input.

 Let us make this precise. Consider the $n=2$ FDT written in the form \eqref{eq:FD2x}. Now take 
 $\Op{X}$ to be a nested commutator involving $n-1$ operators $\Op{O}_i$ with $i \in \{1, \cdots, n-1\}$. The number of such identities we can write down is $n\, 2^{(n-1)-2}\, (n-1)! = 2^{n-3}\, n!$, the count following from the number of nested structures and cyclic permutations of the resulting $n$ objects. However, these cannot all be independent. 
 For one we have to account for sJacobi relations:  if $\Op{X} = \sum_j \, \Op{Y}_j =0$ is an sJacobi relation between some set of nested correlators $\Op{Y}_j$, then we trivially satisfy $\sum_j\, \, \vev{\Op{Y}_j\, \Op{A_{_+}}} + \cth_{\Op{A}} \, 
\vev{\Op{Y}_j\, \Op{A}} =0 $.  So this is not a new KMS relation and we should mod out by such identities which we can enumerate to be $n\, (2^{(n-1)-2} -1)(n-1)! = (2^{n-3} -1) n!$. Subtracting this from the previous count, we therefore see that the number of independent relations after accounting for the sJacobi identities is $n!$. This is however too many, since there are only $n!-(n-1)!$ independent Wightman functions after accounting for the KMS relations. Therefore among the many relations we derive by the above logic, there ought to be a further $(n-1)!$ relations among the identities obtained via use of the $n=2$ FD theorem and the sJacobis. We shall call these relations among relations, the \emph{iterated KMS relations}.

While the general construction of these $(n-1)!$ iterated KMS relations is clear, they are best understood by working them out explicitly at low orders of $n$. To write these down explicitly, we remind the reader of the notation \eqref{eq:masterFD} which emphasizes the primacy of the $n=2$ FDT, which we therefore encapsulate by the new symbol: 
\begin{equation}
\kmsfd{\Op{X}}{\Op{A}} \equiv \vev{\fullcomm{\Op{X\, A_{_+}}}} + \cth_{\Op{A}} \, \vev{
\fullcomm{\Op{X\, A}}} .
 \label{eq:masterFD2}
\end{equation}	
We can now state in terms of this master relation, the iterated KMS relations quite succinctly.

\paragraph{Two-point iterated KMS relations:} The well-known 2-point FDT in our new notation reads as
\begin{equation}
\begin{split}
\kmsfd{\Op{A}}{\Op{B}}
= 0\,.
 \end{split}
 \label{eq:2ptRofR}
\end{equation}
\paragraph{Three-point iterated KMS relations:} We claim that the $n=3$ relations are:
\begin{equation}
\begin{split}
\kmsfd{\comm{\Op{A}}{\Op{B}}}{\Op{C}}
+ \cth_{\Op{C}} \left( \kmsfd{\anti{\Op{C}}{\Op{A}}}{\Op{B}} 
- \kmsfd{\anti{\Op{C}}{\Op{B}}}{\Op{A}}\right)
&=
\kmsfd{\comm{\Op{B}}{\Op{C}}}{\Op{A}}
+ \cth_{\Op{A}} \left( \kmsfd{\anti{\Op{A}}{\Op{B}}}{\Op{C}} 
- \kmsfd{\anti{\Op{A}}{\Op{C}}}{\Op{B}}\right)
\\
&= \kmsfd{\comm{\Op{C}}{\Op{A}}}{\Op{B}}
+ \cth_{\Op{B}} \left( \kmsfd{\anti{\Op{B}}{\Op{C}}}{\Op{A}} 
- \kmsfd{\anti{\Op{B}}{\Op{A}}}{\Op{C}}\right) .
 \end{split}
 \label{eq:3ptRofR}
\end{equation}
Writing this out explicitly, we have 
\begin{equation}
\begin{split}
&
\vev{\fullcomm{\Op{A\, B\, C_{_+}}}}
 +\cth_{\Op{C}} \, \vev{\fullcomm{\Op{A\, B\, C}} }
+\cth_{\Op{C}} 
\bigg( \vev{\fullcomm{\Op{C\, A_{_+}\, B_{_+}}}}+  
\cth_{\Op{B}} \, \vev{\fullcomm{\Op{C\, A_{_+}\, B}}}
 \bigg) \\
 & \hspace{2cm}
 -\cth_{\Op{C}} \bigg(
\vev{\fullcomm{\Op{B\, C_{_+}\, A_{_+}}}} +  \cth_{\Op{A}} \, \vev{\fullcomm{\Op{B\, C_{_+}\, A}}}
 \bigg)  \\
&=
\vev{\fullcomm{\Op{B\, C\, A_{_+}}}} +\cth_{\Op{A}} \, \vev{\fullcomm{\Op{B\, C\, A}}}
+\cth_{\Op{A}} 
\bigg( \vev{\fullcomm{\Op{A\, B_{_+}\, C_{_+}}}} +
  \cth_{\Op{C}} \, \vev{\fullcomm{\Op{A\, B_{_+}\, C}}}
 \bigg)\\
 & \hspace{2cm}
 -\cth_{\Op{A}} \bigg(
\vev{\fullcomm{\Op{C\, A_{_+}\, B_{_+}}}} +  \cth_{\Op{B}} \, \vev{\fullcomm{\Op{C\, A_{_+}\, B}}}
 \bigg) 
  \\
&=\vev{\fullcomm{\Op{C\, A\, B_{_+}}}}+\cth_{\Op{B}} \, \vev{\fullcomm{\Op{C\, A\, B}}}
+\cth_{\Op{B}} 
\bigg( \vev{\fullcomm{\Op{B\, C_{_+}\, A_{_+}}}} 
+  \cth_{\Op{B}} \, \vev{\fullcomm{\Op{B\, C_{_+}\, A}}}
 \bigg)\\
 & \hspace{2cm}
 -\cth_{\Op{B}} \bigg(
\vev{\fullcomm{\Op{A\, B_{_+}\, C_{_+}}}} +  \cth_{\Op{C}} \, \vev{\fullcomm{\Op{A\, B_{_+}\, C}}}
 \bigg) .
\end{split}
\end{equation}
These relations can be proven by showing that any one  of them is equal to
\begin{equation}
\left(1+\cth_{\Op{C}} \, \cth_{\Op{A}} \right) \vev{ \Op{A\, B\, C  -C\, B\, A}} 
+\left(1+\cth_{\Op{A}}\, \cth_{\Op{B}} \right)\vev{\Op{B\, C\, A-A\, C\, B}} 
+\left(1+\cth_{\Op{B}} \, \cth_{\Op{C}}\right) \vev{\Op{C\,A\,B-B\,A\,C}}  
\end{equation}	
upon using the identity \eqref{eq:ThermalRel2}.
Since the above form is manifestly  cyclically invariant, it proves \eqref{eq:3ptRofR}. 

The iterated KMS relations can be written in a somewhat more compact form as 
\begin{equation}
\begin{split}
(\cth_1+\cth_2) \, \kmsfd{\fullcomm{12_{_+}}}{3} &= 
	\left( \cth_1 \, \kmsfd{\fullcomm{31_{_+}}}{2}
	 + \kmsfd{\fullcomm{31}}{2}  \right)  
	 +\left( \cth_2\,  \kmsfd{\fullcomm{32_{_+}}}{1} + \kmsfd{\fullcomm{32}}{1}
	  \right),\\
(\cth_1+\cth_2) \, \kmsfd{\fullcomm{12}}{3}
&=\left( \cth_1 \, \kmsfd{\fullcomm{31}}{2}
	 + \kmsfd{\fullcomm{31_{_+}}}{2}  \right)  
	 -\left( \cth_2\,  \kmsfd{\fullcomm{32}}{1} + \kmsfd{\fullcomm{32_{_+}}}{1}
	  \right).
\end{split}
\label{eq:3ptRofRalt}
\end{equation}

\paragraph{Four-point iterated KMS relations:} We find the following set of relations for $4$-point functions
\begin{subequations}
\begin{align}
(1+ \cth_1 \cth_2&+\cth_2 \cth_3+\cth_3 \cth_1) \, \kmsfd{\fullcomm{12_{_+}3_{_+}}}{4}
\nonumber \\
&=(1+\cth_1 \cth_2) \, \kmsfd{\fullcomm{12_{_+}4_{_+}}}{3}
- (\cth_1 + \cth_2) \, \kmsfd{\fullcomm{12_{_+}4}}{3}  
\nonumber \\
&
\qquad+\cth_1 
\left( \cth_3 \, \kmsfd{\fullcomm{34_{_+}1_{_+}}}{2}
- \kmsfd{\fullcomm{341_{_+}}}{2}  \right)  
+\left( \cth_3 \, \kmsfd{\fullcomm{34_{_+}1}}{2}
- \kmsfd{\fullcomm{341}}{2} \right) 
\nonumber \\
&\qquad+\cth_2
\left( \cth_3 \, \kmsfd{\fullcomm{34_{_+}2_{_+}}}{1}
- \kmsfd{\fullcomm{342_{_+}}}{1}  \right)  
+\left( \cth_3 \, \kmsfd{\fullcomm{34_{_+}2}}{1}
- \kmsfd{\fullcomm{342}}{1} \right) . 
\label{eq:4ptRofR1}
\end{align}
\begin{align}
(1+ \cth_1 \cth_2&+\cth_2 \cth_3+\cth_3 \cth_1)  \, \kmsfd{\fullcomm{12 3_{_+}}}{4}
\nonumber \\
&=(1+\cth_1 \cth_2) \, \kmsfd{\fullcomm{124_{_+}}}{3}
- (\cth_1 + \cth_2) \, \kmsfd{\fullcomm{124}}{3}  
\nonumber \\
&
\qquad+\cth_1 
\left( \cth_3 \, \kmsfd{\fullcomm{34_{_+}1}}{2}
- \kmsfd{\fullcomm{341}}{2}  \right)  
+\left( \cth_3 \, \kmsfd{\fullcomm{34_{_+}1_{_+}}}{2}
- \kmsfd{\fullcomm{341_{_+}}}{2} \right) 
\nonumber \\
&\qquad+\cth_2
\left( \cth_3 \, \kmsfd{\fullcomm{34_{_+}2}}{1}
- \kmsfd{\fullcomm{342}}{1}  \right)  
+\left( \cth_3 \, \kmsfd{\fullcomm{34_{_+}2_{_+}}}{1}
- \kmsfd{\fullcomm{342_{_+}}}{1} \right) . 
\label{eq:4ptRofR2}
\end{align}
\end{subequations}
Cyclic permutations of $123$ in the above fills out the 6 expected iterated KMS relations. In deriving these we made use of the $n=4$ identity among the thermal factors \eqref{eq:ThermalRel2}. We expect that permutations other than cyclic permutations of $123$ do not give new relations between KMS relations.  Thus Eq.~\eqref{eq:4ptRofR1} and \eqref{eq:4ptRofR2} comprise of the full set of relations between thermal 4-point functions and which from the basic FDT.

\section{Illustration: Harmonic Oscillator}
\label{sec:sho}
We now illustrate the general relations obtained hitherto with the simplest possible model one can consider -- a harmonic oscillator. Let $X(t)$ be the coordinate of the particle moving in a harmonic oscillator of frequency $\mu$. We will consider correlation functions of composite operators of the form $X^a(t)$ for $a\in \mathbb{Z}_+$. We will exhibit various KMS relations primarily  for  three point function, by computing  all the Wightman functions. In passing, we will also show how one can also obtain these via an analytic continuation from  Euclidean correlators, thus verifying explicitly some of the abstract statements encountered herein and in the literature. 

\subsection{Thermal Expectation values of Wightman correlators}

We will be interested in a thermal correlation function
\begin{equation}
\langle X^a(t_1) X^b(t_2) \cdots \rangle_\beta 
\equiv \frac{1}{\mathcal{Z}}   \, \sum_{n=1}^\infty \;
\langle  n | X^a(t_1)\, X^b(t_2) \cdots | n \rangle \; e^{-\beta \,E_n}   \,,
\label{eq:Samplecorrelator}
\end{equation}
where $ |n \rangle$ are the energy eigenstates of energy $E_{n} =  \mu \,(n + \frac{1}{2} )$ and the partition sum is $\mathcal{Z} = \sum_{n=0}^\infty e^{- \beta \mu (n + \frac{1}{2})} = \frac{1}{2 \sinh (\frac{\beta \,\mu}{ 2})}$. 

The position operator can be expanded in terms of creation and annihilation operators as (setting mass $m=1$ for simplicity)
\begin{equation}
X(t) =  \frac{1}{\sqrt{2 \, \mu }}   \left( a \, e^{- i \mu \,t} + a^\dagger \, e^{i \mu \,t}\right) .
\end{equation}
The operators act on the Hilbert space as 
\begin{equation} \label{eq:aaction}
a | n \rangle = \sqrt n | n-1 \rangle  \hspace{10mm}  
a^\dagger | n \rangle = \sqrt{ n + 1} | n +1 \rangle 
\end{equation}
and satisfy $[a , a^\dagger]  =1$. Using this, we can compute thermal expectation values and explicitly verify various identities derived in previous sections. We provide further  computational details in Appendix \ref{app:sho}.

\paragraph{Two-point functions:}  
Using the above definitions, it is easy to show that (with $t_{ij} = t_i - t_j$ henceforth to account for time translational invariance)
\begin{equation}
2 \,\mu\;  \vev{ n | X(t_1) X(t_2) |  n }  = ( e^{- i \mu \,t_{12} } + e^{  i \mu \,t_{12} }  ) \,n +   e^{- i \mu \,t_{12} } \,.
\end{equation}
Summing over $n$ with weights as in \eqref{eq:Samplecorrelator}, we can compute the thermal propagator and explicitly verify the KMS condition:\footnote{  This also gives the familiar zero temperature result 
$\vev{ \mathcal{T}X(t) X(0)   }_{\beta\rightarrow \infty} = 
\theta(t) e^{- i \mu \,t} + \theta(-t) e^{i \mu \,t}= e^{- i \mu \,|t| }$.}
\begin{equation}
2 \,\mu\,  G_M( t_{12} ) \equiv 
 2 \,\mu \,\vev{ X(t_1) \, X(t_2) }_\beta = 
 \frac{ e^{i \mu \,t_{12} } + e^{\beta \mu} e^{- i \mu \,t_{12}} }{e^{\beta \mu} - 1} 
 = 2 \,\mu  \, \vev{ X(t_2) \, X(t_1+ i \beta) }_\beta \,.
 \label{eq:propsho}
\end{equation}
In frequency space this gives the familiar form of the spectral function, for 
\begin{equation}
2 \,\mu  \,\vev{ X(\omega) X(- \omega) }_\beta = 
\frac{e^{ \beta \omega}}{e^{ \beta \omega} -1}\;  \underbrace{ \left[  \delta_{\omega-\mu} - \delta_{\omega+\mu} \right] }_{\equiv \rho(\omega)} \,.
\end{equation}

\paragraph{Three-point function:} We can compute a three-point function using similar methods. Because we are dealing with a free theory, we can express the result in terms of the propagator:
\begin{equation}
\vev{ X(t_1)  X^2(t_2) X^3(t_3) }_\beta  = 
	6\, G_M(t_{13}) \,G_M(t_{23})^2 + 3 \,G_M(t_{13}) \, G_M(0)^2 + 
	6\, G_M(t_{12}) \,G_M(t_{23}) \,G_M(0) .
\label{eq:123OrderingProp}
\end{equation}
Similarly, one can evaluate other orderings. For example, 
\begin{equation}
\vev{ X(t_1)  X^3(t_3) X^2(t_2) }_\beta  
=6\, G_M(t_{13}) \,G_M(t_{32})^2 + 3 \,G_M(t_{13}) \,G_M(0)^2 + 6 \,G_M(t_{12}) \,G_M(t_{32}) \,
G_M(0) .
\label{eq:132OrderingProp}
\end{equation}
 One can now explicitly check that the KMS conditions given in \eqref{eq:3tkms} hold. For instance,  
\begin{equation}
\begin{split}
\vev{ X(t_1)  X^2(t_2) X^3(t_3) }_\beta &= \vev{ X^2(t_2) X^3(t_3) X(t_1+ i \beta)  }_\beta  = \vev{   X^3(t_3) X(t_1+ i \beta) X^2(t_2+ i \beta)  }_\beta\,, \\
\vev{ X(t_1) X^3(t_3)  X^2(t_2)  }_\beta &= \vev{  X^3(t_3) X(t_1) X^2(t_2+ i \beta)  }_\beta  = \vev{  X(t_1) X^2(t_2+ i \beta) X^3(t_3+i \beta)   }_\beta\,.
\end{split}
\end{equation}

It is instructive to pass onto the nested correlators which (as expected) turn out to be much more simple than the Wightman correlators. For example
\begin{equation}
\begin{split}
(2 \,\mu )^3\, \vev{ \fullcomm{ X(t_1) \, X^2(t_2) \, X^3(t_3) } }_\beta &
	=  -24 \, \coth(\frac{1}{2}\,\beta\mu)  \, \sin(\mu\, t_{12}) \, \sin(\mu\, t_{23}) \,,\\
(2 \,\mu )^3\, \vev{ \fullcomm{ X(t_1) \, X^3(t_3) \,X^2(t_2) }}_\beta &= 
 24 \, \coth(\frac{1}{2}\,\beta\mu) \, \sin(\mu\, t_{13}) \,\sin(2\mu\, t_{23}).
\end{split}
\end{equation}

\paragraph{Four-point function:} Finally, let us also record a couple of four point functions
\begin{equation}
\begin{split}
(2\mu)^2\, \vev{ [X(t_1) , X(t_2)] [X(t_1) , X(t_2) ] }_\beta &= 
-4 \sin^2(\mu \, t_{12})\,,
\\
(2\mu)^4\, \vev{ [X(t_1) , X^3(t_2)] [X(t_1) , X^3(t_2) ] }_\beta  &= -108  \coth^2(\frac{1}{2}\,\beta\mu)  
\sin^2(\mu \, t_{12})
\,.
\end{split}
\end{equation}
%

\subsection{Euclidean correlators}
\label{sec:eucsho}

\paragraph{The propagator:} The Euclidean two-point function can be derived by summing the free propagator over the Matsubara modes with frequency $\omega_n = \frac{2 \pi \,n}{\beta}$, $ n \in {\mathbb Z}$ leading to the Green's function \cite{Bellac:2011kqa} (see Appendix \ref{app:sho} for details):
\begin{equation}
G_E(\tau) = 
\frac{1}{2 \mu} \left(  \frac{ e^{\beta \mu}   e^{- \mu |\tau|} +    e^{\mu |\tau|}}{ e^{\beta \mu}-1 } \right)
\end{equation}
The analytic continuation to real time can again be performed to obtain  \eqref{eq:propsho} for 
\begin{equation}
\begin{split}
2 \,\mu \,G_M(t_{12})= 2\, \mu \,\vev{ X(t_1) X(t_2)}_\beta &= 2\, \mu\,  \lim_{\epsilon_1>\epsilon_2 , \epsilon_i \to 0} \vev{ X(\tau_1=i t_1  + \epsilon_1) X(\tau_2=i t_2  + \epsilon_2) } \\
&= 2\, \mu\, \lim_{\epsilon_1>\epsilon_2 , \epsilon_i \to 0} G_E(\tau = i t_{12}+ \epsilon_1 -\epsilon_2)    \\
&= \frac{ e^{\beta \mu - i \mu \,t_{12}} +   e^{i \mu \,t_{12} }  }{  e^{\beta \mu}-1 } 
\end{split}
\end{equation}
Likewise the time ordered correlator evaluates to 
\begin{equation}
 2\mu  \vev{ {\cal T}\; X(t_1) X(t_2) }_\beta =     \frac{ e^{\beta \mu} e^{- i \mu \,|t_{12}|} +   e^{i \mu \,|t_{12}| }  }{  e^{\beta \mu}-1 } 
\end{equation}

\paragraph{Three-point function:} The Euclidean 3-point function of normal ordered operators then evaluates to
\begin{equation}
\begin{split}
 \vev{ X(\tau_1) X^2(\tau_2) X^3(\tau_3) }  
 &= 
 	 3 \vev{ X(\tau_1) X(\tau_3) } \bigg( 2 \vev{ X(\tau_2) X(\tau_3) }^2 + \vev{ X(\tau_2) X(\tau_2) }  \; \vev{ X(\tau_3) X(\tau_3) }  \bigg)\;   \\
  & 
  	+ 6 \vev{ X(\tau_1) X(\tau_2) } \vev{ X(\tau_2) X(\tau_3) } \; \vev{ X(\tau_3) X(\tau_3) } \\
   &
   =  6\, G_{E}(\tau_{13})\, G_{E}(\tau_{23})^2  
   + 3\, G_{E}(\tau_{13}) \, G_{E}(0)^2 + 6 \, G_{E}(\tau_{12}) \, G_{E}(\tau_{23})  \, G_{E}(0) 
\end{split}
\end{equation}
The real-time Wightman correlators obtained earlier can then be extracted by different analytic continuations. For instance,  \eqref{eq:123OrderingProp} follows from 
\begin{equation}
\begin{split}
 \vev{ X(t_1) X^2(t_2) X^3(t_3) } &=
 	 \lim_{\epsilon_i \to 0, \epsilon_1>\epsilon_2>\epsilon_3}  \vev{ X(\tau_1= i t_1 + \epsilon_1) X^2(\tau_2= i t_2 + \epsilon_2) X^3(\tau_3= i t_3 + \epsilon_3) }  \\
 &=  
 	6 \,G _M(t_{13}) \, G_M(t_{23})^2 + 3 \, G_M(t_{13}) \, G_M(0)^2 + 6 \, G_M(t_{12}) \, G_M(t_{23}) \, G_M(0)
 \end{split}
\end{equation}
while   \eqref{eq:132OrderingProp} results from the analytic continuation:
\begin{equation}
\begin{split}
 \vev{ X(t_1) X^3(t_3) X^2(t_2) } &= \lim_{\epsilon_i \to 0, \epsilon_1>\epsilon_3>\epsilon_2}  \vev{ X(\tau_1= i t_1 + \epsilon_1) X^2(\tau_2= i t_2 + \epsilon_2) X^3(\tau_3= i t_3 + \epsilon_3) }  \\
 &=  6 \, G_M(t_{13}) \, G_M(t_{32})^2 + 3 \, G_M(t_{13}) \, G_M(0)^2 + 6 \, G_M(t_{12}) \, G_M(t_{32}) \, G_M(0)
 \end{split}
\end{equation}

\paragraph{Chaos correlator:} Finally, let us  draw a connection between  the OTO 4-point function that provides a diagnostic for chaos \cite{Larkin:1969aa,Maldacena:2015waa} and the Wightman function evaluated at complex times. To do this note that for any operator
\begin{equation}
\begin{split}
F(t) &\equiv \frac{1}{\mathcal{Z}} \, 
\Tr{ V(0) e^{-\frac{\beta\, H}{4}} \,  W(t) e^{-\frac{\beta\, H}{4}} \, V(0) e^{-\frac{\beta\, H}{4}} \, 
W(t)  e^{-\frac{\beta\, H}{4}} }\\
&= \vev{ V(0)\,  W(t + \tfrac{ i \,\beta}{4})  \,  V(\tfrac{i \,\beta}{2})\,  W(t +   \tfrac{3\,i \,\beta}{4})\ }_\beta
\end{split}
\end{equation}
We take $V \equiv X$ and $W  \equiv X^2$ to find:
\begin{equation}
\begin{split}
F(t)& = 
	\vev{ X(0)\,  X^2(t + \tfrac{ i \,\beta}{4})  \,  X(\tfrac{i \,\beta}{2})\, 
	 X^2(t +   \tfrac{3\,i \,\beta}{4})\ }_\beta \\
&= 
6\, \bose \,  \left( e^\frac{\beta \mu}{2} ( 1 + 12\, \bose \,( 1 +\bose))  + 4\, (1 + \bose)(1 +2 \,\bose) \, \cos(2 \,\mu \,t)  \right)
\end{split}
\end{equation}
where $\bose = {1 \over e^{\beta \mu} -1 }$. The analytic continuation in $t$ of this function decays at large $\mu$ only in the strip $| \Im(t)| \le \tfrac{\beta}{2}$. This means that in a field theoretical setting with an infinite number of harmonic oscillators (with arbitrarily high frequencies) the above function exists only within this strip. We give further expressions for other 4-point functions and some tremelo 6-point functions in Appendix \ref{app:sho}.

\section{OTO classification of thermal correlators}
\label{sec:toto}

Since generic Wightman functions do not respect time-ordering, their computation necessarily involves timefolds or out-of-time order (OTO) path integral contours. While this point has been well appreciated for a long time now, only recently has there been a systematic attempt to classify the OTO correlators \cite{Haehl:2017qfl}. As discussed there, the $n!$ Wightman functions can be computed by contours with at most $\lfloor \frac{n+1}{2}\rfloor$ timefolds.  Moreover, given a particular time ordering there is a canonical proper $q$-OTO number with 
$1 \leq q\leq\lfloor \frac{n+1}{2}\rfloor $ which provides the simplest representation of the correlator (i.e., the minimum number of timefolds required to represent the correlator on a contour).

While the discussion of \cite{Haehl:2017qfl} was for generic initial density matrices, as we saw above, various Wightman correlators are related by KMS conditions.  These relations as we have described involved cyclic permutations of the operators, which do not respect the time-ordering.\footnote{  In what follows we will always make the canonical choice $t_1 > t_2 > t_3 \cdots > t_n$ for simplicity.} In fact, one can already see from the simplest example of  $3$-point functions in \eqref{eq:3tkms} that the KMS conditions do not respect OTO number:
while $\vev{123}$ is time-ordered and can be computed from a $1$-OTO contour, $\vev{231}$ necessitates a $2$-OTO contour; these two correlators (with some arguments shifted in imaginary time) are related by a KMS relation.
More generally, the proper $q$-OTO number for a thermal $n$-point Wightman function lies in the range $1\leq q \leq \lfloor \frac{n}{2}\rfloor$.

As the KMS relations map all cyclic permutations of operator insertions in a given correlator to each other, for an $n$-point function
this ends up equating correlators of different proper-OTO number. The question we propose to answer is to understand the OTO classification of the correlators in any cyclic orbit of the KMS action.  Below we add some additional structure to the classification scheme described in \cite{Haehl:2017qfl} to allow for a cleaner discussion of cyclic features inherent in the KMS relations, and show how to decompose the space of $n!$ Wightman functions into $(n-1)!$ independent spectral functions, using OTO-contours and KMS relations. Physically, it is interesting to understand these relations to ascertain which of the OTO correlators carry non-trivial and independent information in the thermal state.

\subsection{Review of OTO classification}
\label{sec:review}

Before proceeding it will be helpful to have a quick overview of the nomenclature and results of \cite{Haehl:2017qfl}, which we invoke  extensively below.

\begin{figure}[t!]
\begin{center}
\begin{tikzpicture}[scale=0.8]
\draw[thick,color=violet,dotted] (5,1.5) -- (5,-1.5);
\draw[thick,color=violet,->] (-5,3.5) -- (5,3.5) ;
\draw[thick,color=violet,->] (5,2.5) -- (-5,2.5);
\draw[thick,color=violet,->] (-5,1.5)  -- (5,1.5);
\draw[thick,color=violet,->] (5,-1.5) -- (-5,-1.5);
\draw[thick,color=violet,->] (-5,-2.5)   -- (5,-2.5);
\draw[thick,color=violet,->] (5,-3.5)   -- (-5,-3.5);
\draw[thick,color=violet,fill=violet] (-5,3.5) circle (0.75ex);
\draw[thick,color=violet,fill=violet] (-5,-3.5) circle (0.75ex);
\draw[thick,color=violet,->] (5,3.5) arc (90:-90:0.5);
\draw[thick,color=violet,->] (5,-2.5) arc (90:-90:0.5);
\draw[thick,color=violet,->] (-5,2.5) arc (90:270:0.5);
\draw[thick,color=violet,->] (-5,-1.5) arc (90:270:0.5);
\draw[thick, color=violet,dashed] (-5,-3.5) .. controls (-8.5,-1.75) and (-8.5,1.75) .. (-5,3.5);
\draw[thick, color=red]
{ (0,3.5) node [above] {\small{1R}}
(0,2.5) node [above] {\small{2L}}
(0,1.5) node [above] {\small{3R}}
(0,-1.5) node [above]{\small{3L}}
(0,-2.5) node [above] {\small{2R}}
(0,-3.5) node [above] {\small{1L}} };
\end{tikzpicture}
\caption{The k-OTO contour computing the out-of-time-ordered correlation functions encoded in the generating functional.}
\label{fig:koto}
\end{center}
\end{figure}

A $k$-OTO functional integral computes the generating function for thermal Wightman correlators and is defined to be:
\begin{equation}
\mathscr{Z}_{k-oto}[\mathcal{J}_{\alpha\skR},  \mathcal{J}_{\alpha\skL}]=
\Tr{   \cdots
U[\mathcal{J}_{3\skR}]
(U[\mathcal{J}_{2\skL}])^\dag  U[\mathcal{J}_{1\skR}] \;\rhoT \; (U[\mathcal{J}_{1\skL}])^\dag
U[\mathcal{J}_{2\skR}] (U[\mathcal{J}_{3\skL}])^\dag \cdots} \,.
\label{eq:koto}
\end{equation}
A pictorial representation of the functional integral contour in the complex time plane is given in Fig.~\ref{fig:koto}.

When computing correlators, we essentially insert operators along this contour at the appropriate time, with a contour ordering prescription that relates directly to the permutation $\sigma \in S_n$ corresponding to the Wightman correlator of interest. The contour should be viewed as an abacus, with operators free to slide about from one level to another, as long at they are unobstructed by other operators (or the density matrix). For instance:
\begin{center}
\begin{minipage}[t]{0.17\textwidth}
\begin{equation}\nonumber
\begin{split}
\quad\\
   G(t_4,t_1,t_3,t_2)\; =  \\
\quad
 \end{split}
 \end{equation}
  \end{minipage}
 \begin{minipage}[t]{0.22\textwidth}
 \begin{equation} \nonumber
 \begin{tikzpicture}[scale=0.54]
\draw[thick,color=violet,fill=violet] (-2,3.5) circle (0.75ex);
\draw[thick,color=violet,->] (-2,3.5) -- (2,3.5) ;
\draw[thick,color=red] (.7,3.5) circle (0.75ex);
\draw[thick,color=violet,->] (2,3.5) arc (90:-90:0.5);
\draw[thick,color=violet,->] (2,2.5) -- (-2,2.5);
\draw[thick,color=red] (-.7,2.5) circle (0.75ex);
\draw[thick,color=violet,->] (-2,2.5) arc (90:270:0.5);
\draw[thick,color=violet,->] (-2,1.5)  -- (2,1.5);
\draw[thick,color=violet,->] (2,1.5) arc (90:-90:0.5);
\draw[thick,color=red] (1.5,1.5) circle (0.75ex);
\draw[thick,color=violet,->] (2,0.5)  -- (-2,0.5);
\draw[thick,color=red] (-1.5,.5) circle (0.75ex);
\draw[thick,color=violet,fill=violet] (-2,0.5) circle (0.75ex);
\draw[thick, color=red]
{
(-1.5,.5) node [below] {\small{$\Op{O}_4$}}
(-.7,2.5) node [below] {\small{$\Op{O}_3$}}
(.7,3.5) node [below] {\small{$\Op{O}_2$}}
(1.5,1.5) node [below] {\small{$\Op{O}_1$}}
};
\end{tikzpicture} 
 \end{equation}
 \end{minipage}
 \begin{minipage}[t]{.02\textwidth}
 \begin{equation}\nonumber
 \begin{split}
 \quad\\
 = 
 \end{split}
 \end{equation}
 \end{minipage}
  \begin{minipage}[t]{0.22\textwidth}
 \begin{equation} \nonumber
 \begin{tikzpicture}[scale=0.54]
\draw[thick,color=violet,fill=violet] (-2,3.5) circle (0.75ex);
\draw[thick,color=violet,->] (-2,3.5) -- (2,3.5) ;
\draw[thick,color=red] (.7,2.5) circle (0.75ex);
\draw[thick,color=violet,->] (2,3.5) arc (90:-90:0.5);
\draw[thick,color=violet,->] (2,2.5) -- (-2,2.5);
\draw[thick,color=red] (-.7,2.5) circle (0.75ex);
\draw[thick,color=violet,->] (-2,2.5) arc (90:270:0.5);
\draw[thick,color=violet,->] (-2,1.5)  -- (2,1.5);
\draw[thick,color=violet,->] (2,1.5) arc (90:-90:0.5);
\draw[thick,color=red] (1.5,1.5) circle (0.75ex);
\draw[thick,color=violet,->] (2,0.5)  -- (-2,0.5);
\draw[thick,color=red] (-1.5,.5) circle (0.75ex);
\draw[thick,color=violet,fill=violet] (-2,0.5) circle (0.75ex);
\draw[thick, color=red]
{
(-1.5,.5) node [below] {\small{$\Op{O}_4$}}
(-.7,2.5) node [below] {\small{$\Op{O}_3$}}
(.7,3.5) node [below] {\small{$\Op{O}_2$}}
(1.5,1.5) node [below] {\small{$\Op{O}_1$}}
};
\end{tikzpicture} 
 \end{equation}
 \end{minipage}
  \begin{minipage}[t]{.02\textwidth}
 \begin{equation}\nonumber
 \begin{split}
 \quad\\
 = 
 \end{split}
 \end{equation}
 \end{minipage}
  \begin{minipage}[t]{0.22\textwidth}
 \begin{equation} \nonumber
 \begin{tikzpicture}[scale=0.54]
\draw[thick,color=violet,fill=violet] (-2,3.5) circle (0.75ex);
\draw[thick,color=violet,->] (-2,3.5) -- (2,3.5) ;
\draw[thick,color=red] (.7,3.5) circle (0.75ex);
\draw[thick,color=violet,->] (2,3.5) arc (90:-90:0.5);
\draw[thick,color=violet,->] (2,2.5) -- (-2,2.5);
\draw[thick,color=red] (-.7,1.5) circle (0.75ex);
\draw[thick,color=violet,->] (-2,2.5) arc (90:270:0.5);
\draw[thick,color=violet,->] (-2,1.5)  -- (2,1.5);
\draw[thick,color=violet,->] (2,1.5) arc (90:-90:0.5);
\draw[thick,color=red] (1.5,1.5) circle (0.75ex);
\draw[thick,color=violet,->] (2,0.5)  -- (-2,0.5);
\draw[thick,color=red] (-1.5,.5) circle (0.75ex);
\draw[thick,color=violet,fill=violet] (-2,0.5) circle (0.75ex);
\draw[thick, color=red]
{
(-1.5,.5) node [below] {\small{$\Op{O}_4$}}
(-.7,2.5) node [below] {\small{$\Op{O}_3$}}
(.7,3.5) node [below] {\small{$\Op{O}_2$}}
(1.5,1.5) node [below] {\small{$\Op{O}_1$}}
};
\end{tikzpicture} 
 \end{equation}
 \end{minipage}
   \begin{minipage}[t]{.03\textwidth}
 \begin{equation}\nonumber
 \begin{split}
 \quad\\
 = \,\ldots
 \end{split}
 \end{equation}
 \label{eq:g4pic}
 \end{minipage}
 \end{center}

Given a particular $\sigma\in S_n$, there is a primitive contour of minimal number of switchbacks, which is the proper-OTO contour for the said correlator.  A proper $q$-OTO Wightman function is one that can be minimally represented on a contour with  $q$ timefolds (and cannot be computed using one with less than $q$ timefolds).   For a given $n$, one can show that  proper $q$-OTO contours with $q = 1, 2, \cdots, \qmax  $ are necessary. The upper bound can be easily seen by considering the  completely oscillating or tremelo permutation, which is a sequence where insertion times alternately increase/decrease along the contour.

The proper $q$-OTO contours provide  a partitioning of the set of $n$-point Wightman functions. In particular, a $q$-OTO contour computes $g_{n,q}$ of the $n!$ time-ordering correlators, which have proper-OTO characteristic $q$. One finds:
\begin{equation}
\begin{split}
n!  &= \sum_{q=1}^{\qmax} g_{n,q} \\
g_{n,q} &= \text{Coefficient of } \mu^q  \text{ in }  \;\Bigl(2\sqrt{1-\mu}\Bigr)^{n+1} \text{Li}_{-n}\Bigl(\frac{2}{1+\sqrt{1-\mu}}-1\Bigr) \,.
\end{split}
\label{eq:gnqcount}
\end{equation}
Furthermore, each such proper $q$-OTO correlator can be computed from the generating functional \eqref{eq:koto} in a number of $h_{n,k}^{(q)}$ equivalent ways (akin to sliding relations as illustrated in the above picture). It is a non-trivial fact that this number only depends on $(n,k,q)$, but not on the specific correlator and order of insertion points. We therefore have a decomposition of the $(2k)^n$ contour $n$-point functions computed by \eqref{eq:koto} of the form 
\begin{equation}
\begin{split}
(2k)^n &=  \sum_{q=1}^{\qmax} g_{n,q} \, h_{n,k}^{(q)}  \\
h^{(q)}_{n,k}&= \text{Coefficient of } z^n t^k \text{ in } \Bigl(\frac{2z}{1-t}\Bigr)^{2q-1} \frac{t^q}{1-(z+t+zt)} \,.
\end{split}
\label{eq:hnqkcount}
\end{equation}

It is also useful in the sequel to know that for even $n$, there is a maximally out-of-time-order correlator corresponding to a tremelo permutation.\footnote{  For these permutations $g_{n,q}$ reduces to the tangent numbers.} Equivalently, the run structure of the permutation alternates between ascents and descents in a sawtooth pattern.

In the following it will be helpful to know of turning-points and turning-point operators along the OTO contour.  A future turning-point is the turning point at the right end of Fig.~\ref{fig:koto}, i.e., the junctions between
\begin{itemize}
\item 
$\{(2j-1)\,\text{R},( 2j)\, \text{L}\}$ segments, or
\item   $\{(2j)\,\text{R}, (2j-1)\, \text{L}\}$ segments, or 
\item  the $\{k\text{L},k\text{R}\}$ segments in a odd $k$-OTO contour.
\end{itemize}
Correspondingly, past turning-points are the left turning points of Fig.~\ref{fig:koto}, viz. the junctions  between
\begin{itemize}
\item 
$\{(2j)\,\text{R},( 2j+1)\, \text{L}\}$ segments , or 
\item  $\{(2j+1)\,\text{L}, (2j)\, \text{R}\}$ segments, or 
\item  the  $\{k\text{L},k\text{R}\}$ segments in a even $k$-OTO contour.
\end{itemize}
 There are $q$-future and $(q-1)$-past turning-points.  Finally, an operator inserted just before a turning-point will be referred to as a turning-point operator (an example is $\Op{O}_2$ in the pictorial representation of $G(t_4,t_1,t_3,t_2)$ depicted above).

\subsection{Generalities}
\label{sec:gen}

Let us now turn to the interplay between the OTO classification and KMS relations. Consider any Wightman correlator labeled by a permutation $\sigma \in S_n$. Having made our choice to let $t_1 > t_2> \cdots t_n$  we can let $\sigma$ be the  permutation which takes a completely time ordered correlator to the correlator in question; eg., $\vev{123\dots n}$ is mapped onto   $\vev{2134\dots n}$ by $\sigma= (12)$. 

While the physical picture should be clear from the discussion below, it is worthwhile describing our results at the outset in some formal terms. In standard combinatorics, permutations of $n$ elements can be  classified based on their run structure. Assuming the elements which are being permuted  are ordered, \emph{runs} are usually subsequences of ascent or descent as described in \S\ref{sec:Causal}. An ascent of permutation $\sigma \in S_n$ acting on the ordered set 
$\{1, 2, \cdots, n\}$ is a sub-sequence of $\sigma(1)\, \sigma(2)\, \cdots\sigma(n)$ where $\sigma(i) <\sigma(i+1)$, while a descent requires $\sigma(i) >\sigma(i+1)$. It should be transparent that the OTO classification of Wightman correlators is isomorphic to the run structure of the permutation; each ascent corresponds to a forward evolution, while a descent to the backward evolution, and the entire sequence can be decomposed into a set of ascents and descents. The discussion of \cite{Haehl:2017qfl} should be then interpreted as an analysis of \emph{linear run structure}, since one is only interested in the absolute ordering of the operator insertions. 

The KMS conditions on the other hand, involve cyclic relations, which therefore implies that we should look to classify the \emph{cyclic run structure} of permutations.  This can be understood as follows: in the OTO contour of 
\cite{Haehl:2017qfl} (see also \cite{Haehl:2016pec}), it was noted that the future and past turning-points of the contour should be interpreted as the insertion of an identity matrix, since the gluing conditions on the contour require that the future-most end of a forward directed segment of the contour, gets mapped back without change to the future-most end of a past directed segment. For generic density matrices, once the forward/backward evolution is complete, one takes the trace. 

However, in the case of thermal density matrices, one can view the past-most point on the initial and final legs of the contour (denoted 1R and 1L in Fig.~\ref{fig:koto}) as also being joined together through an imaginary time evolution. In other, words a thermal $k$-OTO contour is a thermal circle, with forward/backward excursions in real time.\footnote{  This is for example quite clear in the discussion of \cite{Maldacena:2015waa}, see their Fig.~1.}  Operators inserted on the contour can then be slid around, as long as they do not encounter another operator blocking their way. Sliding an operator through the density matrix (i.e., along the dashed line of Fig.~\ref{fig:koto}) will result in a shift of the operator argument by $-i\,\beta$. In keeping with our conventions, we will only allow sliding counter-clockwise through the density matrix. 

In the thermal state in addition to the future/past turning-points, we now also have the \emph{density matrix turning-point}. Note that in the infinite temperature limit $\rhoT$ reduces to the maximally mixed state given by the identity operator, which results in the density matrix turning-point becoming isomorphic (modulo normalization) to a past turning-point. The cyclic run structure is now transparent, since we are required to decompose $n!$ ordering into sequences that can be nicely arranged on the thermal circle, with real-time forward/backward evolutions.

\begin{figure}[t!]
\begin{center}
\begin{tikzpicture}[scale=0.57]
\foreach \x in {0,7,14}
\foreach \y in {0}
{
\draw[thick,color=violet,fill=violet] (\x-2,\y+3.5) circle (0.75ex);
\draw[thick,color=violet,->] (\x-2,\y+3.5) -- (\x+2,\y+3.5) ;
\draw[thick,color=violet,->] (\x+2,\y+3.5) arc (90:-90:0.5);
\draw[thick,color=violet,->] (\x+2,\y+2.5) -- (\x-2,\y+2.5);
\draw[thick,color=violet,dotted,->] (\x-2,\y+2.5) arc (90:270:0.5);
\draw[thick,color=violet,dotted,->] (\x-2,\y+1.5)  -- (\x+2,\y+1.5);
\draw[thick,color=violet,dotted,->] (\x+2,\y+1.5) arc (90:-90:0.5);
\draw[thick,color=violet,dotted,->] (\x+2,\y+0.5)  -- (\x-2,\y+0.5);
\draw[thick,color=violet,fill=violet] (\x-2,\y+0.5) circle (0.75ex);
\draw[thin,color=violet,dashed] (\x-2,\y+0.5) .. controls (\x-3.5,\y+1.5) and (\x-3.5,\y+2.5) .. (\x-2,\y+3.5);
}
\foreach \x in {0,7,14,21}
\foreach \z in {5}
{
\draw[thick,color=violet,fill=violet] (\x-2,\z+3.5) circle (0.75ex);
\draw[thick,color=violet,->] (\x-2,\z+3.5) -- (\x+2,\z+3.5) ;
\draw[thick,color=violet,->] (\x+2,\z+3.5) arc (90:-90:0.5);
\draw[thick,color=violet,->] (\x+2,\z+2.5) -- (\x-2,\z+2.5);
\draw[thick,color=violet,dotted,->] (\x-2,\z+2.5) arc (90:270:0.5);
\draw[thick,color=violet,dotted,->] (\x-2,\z+1.5)  -- (\x+2,\z+1.5);
\draw[thick,color=violet,dotted,->] (\x+2,\z+1.5) arc (90:-90:0.5);
\draw[thick,color=violet,dotted,->] (\x+2,\z+0.5)  -- (\x-2,\z+0.5);
\draw[thick,color=violet,fill=violet] (\x-2,\z+0.5) circle (0.75ex);
\draw[thin,color=violet,dashed] (\x-2,\z+0.5) .. controls (\x-3.5,\z+1.5) and (\x-3.5,\z+2.5) .. (\x-2,\z+3.5);
}
\foreach \x in {-1.5,-0.5,0.5,1.5}
{\draw[thick,color=red] (\x,8.5) circle (0.75ex);}
\foreach \x in {-1,1,6,8,13,15,20,22}
{\draw[thick,color=red] (\x,7.5) circle (0.75ex);}
\foreach \x in {-1.5,-0.5,0.5,1.5, 5.5,6.5,7.5,8.5, 12.5,13.5,14.5,15.5}
{\draw[thick,color=red] (\x,1.5) circle (0.75ex);}
 \foreach \x in {5.5, 12.5, 19.5}
{\draw[thick,color=red] (\x,6.5) circle (0.75ex);}
 \foreach \position in {(6.5,8.5), (7.5,8.5),(8.5,8.5),
(13.5,6.5), (14.5,8.5),(15.5,8.5), 
(20.5,6.5), (21.5,6.5),(22.5,8.5),
(20-7,2.5),(22-7,2.5),(15-7,0.5),(13-7,2.5),(8,0.5),(-1,0.5),(1,0.5)
 }
{\draw[thick,color=red] \position circle (0.75ex);}
 \draw[thick,color=black,->] (3.5,2) -- (2.8,2);
 \draw[thick,color=black,->] (10.5,2) -- (9.8,2);
 \draw[thick,color=black,->] (2.8,7) -- (3.5,7);
 \draw[thick,color=black,->] (9.8,7) -- (10.5,7);
 \draw[thick,color=black,->] (16.8,7) -- (17.5,7);
 \draw[thick,color=black,->] (20.6,4.6) arc (0:-90:2.4);
\end{tikzpicture} 
\caption{The KMS orbit of a six-point Wightman function. The starting permutation is a proper 1-OTO which we have redundantly represented as a $2$-OTO for ease of visualization of the sliding. The subsequent steps represent sliding of successive operators counter-clockwise through the density matrix to complete the KMS orbit. The last picture of the sequence is the same correlator as the first picture (but all time arguments have been shifted by $-i\beta$). The $q$-list for this example is $\{1,2,2,2,2,1\}$ (which gives the proper-OTO numbers if the figure is read from the end to the beginning in reversed order). The $\delta$-list is $\{0,0,0,0,0,1\}$.}
\label{fig:kmsoto}
\end{center}
\end{figure}
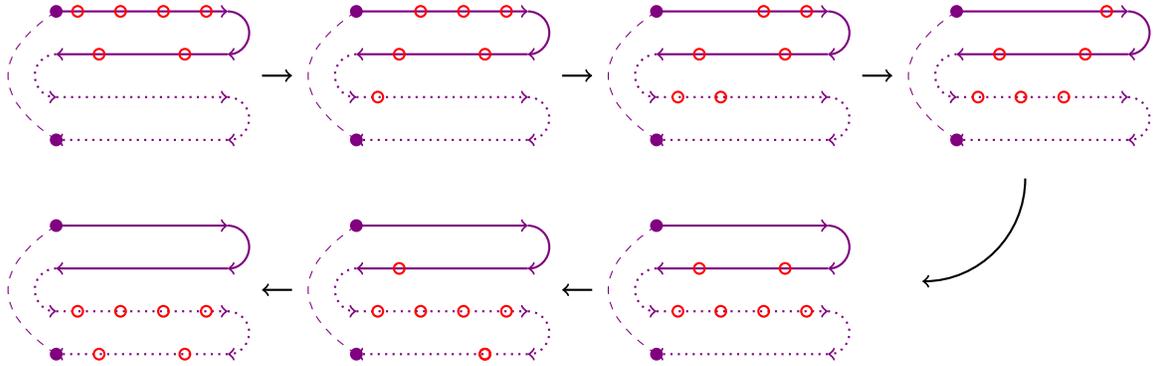

The KMS relations can then be simply understood in terms of sliding operators on the closed thermal OTO contour. Algorithmically we proceed as follows:
\begin{itemize}
\item Begin with a proper $q$-OTO representation of the $n$-point Wightman function associated with the permutation $\sigma$. 
\item Embed the $q$-OTO contour into a redundant $(q+1)$-OTO contour by appending one further timefolds at the bottom of the contour, i.e., inserted just before the density matrix.\footnote{  For ease of visualization it is in fact worthwhile being even more redundant and embedding the given $q$-OTO into a $2q$ -OTO by adding  empty timefolds at the bottom (which  helps to keep the temporal  flow  along contours intact).} 
\item Starting  at the very past of the  $1{\rm R}$  contour, take operators sequentially through the density matrix, traversing counter-clockwise. The operators should be brought to the first empty contour having the same direction of time evolution as the original  contour from which they were taken. 
\item Each operator transition through the density matrix gives a KMS related correlator. At the end of the sliding, we canonicalize the OTO by erasing the redundant legs (adjacent forward/backward evolution).
\end{itemize}
We illustrate the above algorithm pictorially in Figure~\ref{fig:kmsoto}. An immediate consequence of the construction is that it makes manifest the fact that starting with a Wightman correlator with proper OTO number being $q$, in the cyclic orbit of $\sigma$, we are guaranteed to find proper OTO numbers being either $q$ or $q+1$ and none other.

\subsection{OTO sliding and KMS relations}
\label{sec:otoslide}

Let us now turn to giving a more concrete picture which will result in a decomposition of $n$-point Wightman functions into equivalence classes under KMS relations.   To this end we will introduce some basic objects, $q$- and $\delta$-lists which will be suitable sequences of integers that capture the OTO structure of a permutation and its cyclic cousins. The analysis below will complement the general picture above and result in an explicit formula for the count  of  KMS relations.

\subsubsection{$\delta$ and $q$ lists} 

Given a permutation $\sigma\in S_n$, we associate with it an ordered set of proper OTO numbers $\{q_1(\sigma),q_2(\sigma),\ldots,q_n(\sigma)\}$ which we will call the $q$-list. These are  the proper OTO numbers of Wightman correlators associated with $\sigma$ and its cyclic cousins. This  list suffices to determine the OTO structure of KMS relations, for $q_j(\sigma)$ is the proper-OTO number when density matrix is sandwiched between the $j^{\rm th}$ and $(j -1)^{\rm st}$ operators.

It is actually more efficient to consider a more primitive object which characterizes the OTO structure of the Wightman correlator. For a given  permutation $\sigma$, we begin by defining a  binary sequence of $n$ numbers which we refer to as the \emph{$\delta$-list}:
$$\{\dptp_1(\sigma),\dptp_2(\sigma),\ldots,\dptp_n(\sigma)  \}  \,, \quad\text{with} \;\; 
\dptp_i(\sigma) \in \{0,1\} \,.$$ 
We take $\dptp_i(\sigma) =1$ if the $i^{\rm th}$ operator in the $\sigma$ Wightman correlator is a past turning-point operator.\footnote{  We remind the reader that the past turning-point operator lies to the past of both its neighbours \cite{Haehl:2017qfl}.} 
Otherwise, we take $\dptp_i(\sigma) =0$.
Note that for $i=1$ and $i=n$, we check the past turning-point by cyclically moving them away from the edges (since by definition, edge operators can never be a past turning-point operator).

The $\delta$-list determines the $q$-list, since the total number of (non-edge) past turning-point operators in a correlator should be one less than its OTO number. Hence,  
\begin{equation}
 q_j(\sigma)=1+\sum_{i=1}^n\dptp_i(\sigma)-\left[\dptp_{j-1}(\sigma)+\dptp_j(\sigma)\right]= q_0-\left[\dptp_{j-1}(\sigma)+\dptp_j(\sigma)\right]\  
 \label{eq:deltatoq}
\end{equation} 
 where we take $\dptp_0(\sigma) = \dptp_n(\sigma)$ and have defined $q_0\equiv 1+\sum_{i=1}^n\dptp_i(\sigma)$. 

\begin{table}[htb!]
\centering
\begin{tabular}{|c|c|c| }
  \hline
$\{\dptp_1,\dptp_2\}$ &$\{q_1,q_2\}$ &$\vev{ \sigma(1) \sigma(2) }$  \\
  \hline
 $\{0,1\} $& $\{1,1\} $&$\vev{ 12 }$ \\
  $\{1,0\} $& $\{1,1\} $&$\vev{  21}$ \\
 \hline\hline
\end{tabular}
\caption{ $q$-lists and $\delta$-lists for  two-point functions  }
\label{tab:2ptq}
\end{table}

\begin{table}[htb!]
\centering
\begin{tabular}{|c|c|c|}
  \hline
$\{\dptp_1,\dptp_2,\dptp_3\}$ &$\{q_1,q_2,q_3\}$ &$ \vev{ \sigma(1) \sigma(2) \sigma(3) } $ \\
  \hline\hline
 $\{1,0,0\} $&  $\{1,1,2\} $&$\vev{ 3 1 2 },\vev{ 321} $ \\
 $\{0,0,1\} $& $\{1,2,1\} $&$\vev{ 1 2 3 },\vev{ 213 } $ \\
  $\{0,1,0\} $& $\{2,1,1\} $&$\vev{ 231  },\vev{  1 3 2 } $ \\
\hline
\end{tabular}
\caption{$q$-lists and $\delta$-lists for  three-point functions}
\label{tab:3ptq}
\end{table}

\begin{table}[htb!]
\centering
\begin{tabular}{|c|c|c|}
  \hline
$\{ \dptp_1,\dptp_2, \dptp_3, \dptp_4 \}$ & $\{q_1,q_2,q_3,q_4\}$ &$\vev{ \sigma(1) \sigma(2) \sigma(3) \sigma(4)  }$ \\
  \hline\hline
 $\{1,0,0,0 \} $ & $\{1,1,2,2\} $&$\vev{ 4 1 2 3 },\vev{ 4 2 1 3 }, \vev{ 4 3 1 2 },\vev{ 4 3 2 1 }$  \\
 $\{0,0,0,1 \} $ &   $\{1,2,2,1\} $&$\vev{ 12 3 4 },\vev{ 2 1 3 4 }, \vev{ 3 1 2 4 },\vev{ 3 2 1 4 }$  \\
 $\{0,0,1,0 \} $ &  $\{2,2,1,1\} $&$\vev{ 1 2 4 3 },\vev{ 1 3 4 2 }, \vev{ 2 1 4 3 },\vev{ 2 3 4 1 }$ \\  
 $\{0,1,0,0 \} $ &    $\{2,1,1,2\} $&$\vev{ 14 3 2 },\vev{ 2 4 3 1 }, \vev{ 3 4 1 2 },\vev{ 3 4 2 1 }$  \\  
    \hline
 $\{0,1,0,1 \} $ &    $\{2,2,2,2\} $&$\vev{ 1 3 2 4 },\vev{ 1 4 2 3 }, \vev{ 2 3 1 4 },\vev{ 2 4 1 3 },$   \\   
 $\{1,0,1,0 \} $    & &$\vev{ 3 1 4 2 },\vev{ 3 2 4 1 }, \vev{ 4 1 3 2 },\vev{ 4 2 3 1 }$    \\
\hline
\end{tabular}
\caption{$q$-lists and $\delta$-lists for four-point functions}
\label{tab:4ptq}
\end{table}

 The relation  \eqref{eq:deltatoq} can then be used to compute the \emph{$q$-list} defined as  the sequence $\{q_1(\sigma),q_2(\sigma),\ldots,q_n(\sigma)\}$ for a given permutation $\sigma$. This analysis gives a simple way to obtain the proper OTO numbers associated with all correlators related to a given permutation $\sigma$ by KMS condition. We illustrate this construction with some low point examples $n\leq 4$ in Tables~\ref{tab:2ptq}, \ref{tab:3ptq}, and \ref{tab:4ptq} respectively. Figure \ref{fig:kmsoto} provides another detailed example. 

 We are interested in enumerating how many KMS relations can be represented on a proper $q$-OTO contour. For example for the three-point functions, KMS conditions relate the first entry in the third columns of Table~\ref{tab:3ptq} with each other, i.e., they relate the two 1-OTOs (e.g.,  $\vev{ 312 }, \vev{ 123 }$) with a 2-OTO ($\vev{ 231 }$). Similarly KMS conditions also relate the  second entry in the third columns of the tables with each other, viz., they relate the two 1-OTOs (i.e.,  $\vev{ 321 }, \vev{ 213 }$) with a 2-OTO ($\vev{ 132 }$). Thus by knowing the $q$-list we will have identified how the KMS relations cycle through $q(\sigma)$-OTOs and $(q(\sigma)+1)$-OTOs.

\subsubsection{Properties of $q$ and $\delta$-necklaces.}
\label{sec:dqneck}

Given the $q$- and $\delta$-lists, we can now define the notion of $q$- and $\delta$-necklaces. All the $q$ (or $\delta$) lists obtained by a cyclic permutation of the operators in the correlator is collectively termed as $q$ (or $\delta$)-necklace.  As noted these are equivalence classes of lists once we account for the KMS relations in thermal states. We now proceed to give a concise way of thinking about these objects and also note some interesting and useful properties of these necklaces. 

Any $\delta$-necklace must obey the property that two successive entries can never both be $1$ (because two successive operators cannot both be in the past of each other). In fact, if the total number of $1$'s in a $\delta$-necklace is  $p$, then  by performing cyclic shifts, one can always bring the $\delta$ to the form 
\begin{equation} 
\{\dptp_1,\dptp_2,\ldots,\dptp_n\} = \{1,\underbrace{0,\ldots,0}_{m_1\ \text{times}},1,\underbrace{0,\ldots,0}_{m_2\ \text{times}}, \ldots, 1,\underbrace{0,\ldots,0}_{m_p\ \text{times}} \} 
\label{eq:delneck}
\end{equation}
We have then $\sum_{i=1}^p \, m_i = n-p$  and the necklace can simply be labeled by $\{m_1,m_2,\ldots,m_p\}$ up to cyclicity. It is also clear that all possible $\delta$-lists can be obtained from a 
$\delta$-necklace by just cutting the necklace given in \eqref{eq:delneck} at an arbitrary location.

From the $\delta$-necklace we can construct the corresponding (unique) $q$-necklace  using \eqref{eq:deltatoq}:
\begin{equation} 
\{q_1,q_2,\ldots,q_N\} = \{p,p,\underbrace{ p+1,\ldots p+1}_{m_1-1\ \text{times}}, p,p,\underbrace{ p+1,\ldots p+1}_{m_2-1\ \text{times}} \dots p,p,\underbrace{ p+1,\ldots p+1}_{m_p-1\ \text{times}}   \}
\label{eq:qneck}
\end{equation} 
This shows that a $q$-necklace (again up to cyclicity) can be represented by $\{m_1-1,m_2-1,\ldots m_p-1 \}$. Physically, all that matters in constructing the  necklaces is the knowledge of the past turning-point operators. We give some examples of q-necklaces and the number of times the particular necklace appears in any given $n$ point functions in Table \ref{tab:qneck}. 

\paragraph{$q$-necklaces and KMS relations:} 
These abstract considerations can now be put to use to understand some features of the KMS relations. A couple of obvious properties of $q$-necklaces are useful in this regard:
\begin{itemize}
\item Since KMS conditions relate correlators within the  $q$-necklace, it is now apparent that they relate a proper $p$-OTO correlator  either to $p$-OTO or to $(p+1)$-OTO correlators as promised.
\item In any $q$-necklace, $p$-OTO correlators appear $2p$ times and $(p+1)$-OTO correlators appear $n-2p$ times.  
\item For $n=2p$, there are special necklaces where all correlators related through the KMS conditions are proper $p$-OTO. These correspond to cyclic orbits of tremelo permutations where every alternate operator is a past turning-point operator, cf., last row of Table~\ref{tab:4ptq}.
\end{itemize}

These considerations make clear that the $\delta$- and $q$-necklaces of a permutation are sufficient to encode the KMS relations of interest.

\subsection{Necklace degeneracies }

Having identified the utility of working with the necklaces, we now turn to computing the number of necklaces we can form for a given $n$. The count can be efficiently organized in terms of ascertaining the number of potential necklaces that are allowed for a given $n$, and fixed proper-OTO number $p$, and thence determining the degeneracy with which each is encountered. We now give the essential ideas behind these counts, noting that the general picture should be such that at the end of the day we get a partitioning of  $(n-1)!$ (the number of independent thermal Wightman functions) into necklace sets.

\paragraph{Enumerating the necklace types:} In order to enumerate $\delta$-necklaces, it suffices  to count configurations with $p$ $1$'s. This is a standard problem in Polya theory. We need the count  $\mathcal{P}(n,p)$, corresponding to the partition of $n$ into $p$ objects, with each object bigger than $1$, but ordered up to cyclic permutations. This counting also gives the possible independent $q$-necklaces since the latter is inferred from the $\delta$-necklace directly.



\paragraph{Necklace degeneracies:} Given a $\delta$-necklace $\{m_1,m_2,\ldots,m_p\}$ we will assume that we have somehow chosen some canonical representative for each   $\{m_1,m_2,\ldots,m_p\}$ which we 
denote as $\{m_1,m_2,\ldots,m_p\}_c$. We now enumerate the number of such necklaces, which will give a complete decomposition of the $n!$ Wightman functions in terms of KMS cognizant parts.

Consider then one of the elements among the  $d^{n,p}_{\{m_1,m_2,\ldots,m_p\}_c}$ permutations of the type $\{m_1,m_2,\ldots,m_p\}_c$. Let us remove the past-most operator (or equivalently the deepest valley) of this  permutation. This converts one of the triples (say $j^{\rm th}$ triple) $010$ in the   $\delta$-necklace to either $01$ or $10$ or $00$. Consequently, it converts the pair $(m_{j-1}, m_j)$  in the $n$-necklace to $(m_{j-1}, m_j-1)$ or $(m_{j-1}-1, m_j)$ or shortens the necklace by replacing the pair with a single number $m_{j-1}+m_j$. So we can write  a recursion relation
\begin{equation}\label{eq:rec_d}
\begin{split}
 d^{n,p}_{\{m_1,m_2,\ldots,m_p\}_c} &= 
 \sum_{j=1}^p 
 	\Bigl[ d^{n-1,p}_{\{m_1,m_2,\ldots,m_{j-1},m_j-1,\ldots m_p\}_c}
 	+ d^{n-1,p}_{\{m_1,m_2,\ldots,m_{j-1}-1,m_j,\ldots  m_p\}_c}
  \\
  & \hspace{2cm}	+d^{n-1,p-1}_{\{m_1,m_2,\ldots,m_{j-1}+m_j,\ldots m_p\}_c} \Bigr] \\
 &= 
 	\sum_{j=1}^p \Bigl[ 2d^{n-1,p}_{\{m_1,m_2,\ldots,m_j-1,\ldots m_p\}_c}
 	+d^{n-1,p-1}_{\{m_1,m_2,\ldots,m_{j-1}+m_j,\ldots m_p\}_c} \Bigr] \\
 \end{split}
\end{equation}
This recursion relation can solved  with the boundary condition that 
\begin{equation}\label{eq:bc_d}
 d^{n,p}_{\{m_1,m_2,\ldots,m_p\}_c} = 0 , \quad\mbox{if }\; m_i = 0 \mbox{ for any }i
\end{equation}
Using these recursion relations, one can see that 
  \begin{equation}
\begin{split}
d^{n,1}_{\{n-1\}_c}&= 2^{n-2}\\
d^{n,2}_{\{p,n-2-p\}_c}&= 2^{n-2} \left[\binom{n-2}{p}-1\ \right]\\
d^{n,3}_{\{1,p,n-4-p\}_c} &= 2^{n-3} \left[p\binom{n-1}{p+2}+\binom{n-2}{p+1}-2\binom{n-2}{p}-\binom{n-4}{1}\ \right]
 \end{split}
\end{equation}
These degeneracy factors encode the basic information about how many times (modulo symmetry factors, see below) a canonical $\delta$-necklace shows up within a KMS orbit. 

\paragraph{Symmetry factors:} We are almost done with the count, but for accounting of a simple symmetry factor. A given necklace $\{m_1,m_2,\cdots m_p\}$ may end up being invariant under cyclic translations and we need to account for this additional wrinkle. The symmetry factor $S_{\{m_1,m_2,\cdots m_p\}}$ is defined to be the smallest  cyclic translation which leaves  the necklace sequence invariant. In other words, a cyclic shuffle by 
$$\frac{p}{S_{\{m_1,m_2,\cdots m_p\}}}$$ 
leaves the necklace invariant.

\paragraph{A resolution of Wightman basis:} Finally, we can put all of the above together to find an explicit decomposition of the $(n-1)!$ independent thermal Wightman functions. We can write:
\begin{equation}
(n-1)!  = \sum_{p=1}^n \; \frac{ d^{n,p}_{\{m_1,m_2,\ldots,m_p\}_c}}{S_{\{m_1,m_2,\cdots m_p\}}}
\label{eq:}
\end{equation}	

\begin{table}[htb!]
\centering
\begin{tabular}{|r|c|c|c|}
  \hline
  &   $q$ necklace & $\delta$ necklace & \# of occurrences: $\frac{ d^{n,p}_{\{m_1,m_2,\ldots,m_p\}_c}}{S_{\{m_1,m_2,\cdots m_p\}}}$  \\
  \hline
$n=2$ &  $\{1,1\}$ & $\{1,0\}$ & $1$ \\ \hline 
$n=3$ &  $\{1,1,2\}$ & $\{1,0,0\}$ & $2$ \\ \hline 
$n=4$ &   $\{1,1,2,2\}$ & $\{1,0,0,0\}$ & $2^2$ \\
             &   $\{2,2,2,2\}$ & $\{1,0,1,0\}$  & $2$ \\ \hline 
$n=5$ &   $\{1,1,2,2,2\}$ & $\{1,0,0,0,0\}$ & $2^3$ \\
             &   $\{2,2,2,2,3\}$ & $\{1,0,1,0,0\}$ & $2^3\times 2$ \\    \hline             
$n=6$ &   $\{1,1,2,2,2,2\}$ & $\{1,0,0,0,0,0\}$  & $2^4$ \\
              & $\{2,2,2,2,3,3\}$ & $\{1,0,1,0,0,0\}$  & $2^4\times 3$ \\        
              & $\{2,2,3,2,2,3\}$ & $\{1,0,0,1,0,0\}$  & $2^3\times 5$ \\      
              &  $\{3,3,3,3,3,3\}$ & $\{1,0,1,0,1,0\}$ & $2^4$ \\   \hline
$n=7$ &   $\{1,1,2,2,2,2,2\}$& $\{1,0,0,0,0,0,0\}$  & $2^5$ \\
             &  $\{2,2,2,2,3,3,3\}$& $\{1,0,1,0,0,0,0\}$ & $2^5\times 4$ \\        
              &   $\{2,2,3,2,2,3,3\}$& $\{1,0,0,1,0,0,0\}$  & $2^5\times 9$ \\      
              &   $\{3,3,3,3,3,3,4\}$& $\{1,0,1,0,1,0,0\}$ & $2^4\times 17$ \\   
 \hline
$n=8$ &  $\{1,1,2,2,2,2,2,2\}$&$\{1,0,0,0,0,0,0,0\}$  & $2^6$ \\
             &   $\{2,2,2,2,3,3,3,3\}$&  $\{1,0,1,0,0,0,0,0\}$ & $2^5\times 10$\\ 
              &   $\{2,2,3,3,2,2,3,3\}$& $\{1,0,0,0,1,0,0,0\}$ & $2^5\times 19$ \\        
              &   $\{2,2,3,2,2,3,3,3\}$& $\{1,0,0,1,0,0,0,0\}$ & $2^5\times 28$ \\      
             &  $\{3,3,3,3,3,3,4,4\}$& $\{1,0,1,0,1,0,0,0\}$ &  $2^6\times 17$ \\    
              &   $\{3,3,3,3,4,3,3,4\}$& $\{1,0,1,0,0,1,0,0\}$ & $2^8\times 7$ \\   
              &   $\{4,4,4,4,4,4,4,4\}$& $\{1,0,1,0,1,0,1,0\}$ & $2^4\times 17$\\                                    
 \hline
 $n=9$  & $\{1,1,2,2,2,2,2,2,2\}$ & $\{1,0,0,0,0,0,0,0,0\}$ & $2^7$\\
             &  $\{2,2,2,2,3,3,3,3,3\}$& $\{1,0,1,0,0,0,0,0,0\}$ & $2^7\times 6$ \\        
              &   $\{2,2,3,2,2,3,3,3,3\}$& $\{1,0,0,1,0,0,0,0,0\}$ & $2^7\times 20$ \\      
              &   $\{2,2,3,3,2,2,3,3,3\}$& $\{1,0,0,0,1,0,0,0,0\}$ & $2^7\times 34$ \\ 
              &  $\{3,3,3,3,3,3,4,4,4\}$& $\{1,0,1,0,1,0,0,0,0\}$ & $2^7\times 29 $ \\    
              &  $\{3,3,3,3,4,3,3,4,4\}$& $\{1,0,1,0,0,1,0,0,0\}$  & $2^{13}$\\  
              &  $\{3,3,3,3,4,4,3,3,4\}$& $\{1,0,1,0,0,0,1,0,0\}$ & $2^{13}$ \\  
              &  $\{3,3,4,3,3,4,3,3,4\}$& $\{1,0,0,1,0,0,1,0,0\}$ & $2^7\times 35$ \\ 
              &  $\{4,4,4,4,4,4,4,4,5\}$& $\{1,0,1,0,1,0,1,0,0\}$ & $2^8\times 31$ \\                                    
 \hline
\end{tabular}
\caption{The list of allowed $q$- and $\delta$-necklaces and their degeneracies (accounting for the symmetry factors). One can check that the sum of the entries in the last column for a given $n$ is $(n-1)!$, which confirms that the necklaces provide an OTO classification of thermal $n$-point Wightman functions according to KMS orbits.}
\label{tab:qneck}
\end{table}

In summary, we have provided an explicit decomposition of the $n!$ Wightman functions into equivalence classes under KMS relations. This was aided by encoding the OTO structure into the $\delta$- and $q$-necklaces. The key point to note is the degeneracies with which these necklaces appear, and understanding them gives a complete decomposition of the Wightman functions into KMS equivalence classes, as has been explicitly enumerated for some low-point functions in Table~\ref{tab:qneck}. 
We see explicitly from there that for each $n$-point function, there are $q$-necklaces of all proper $p$-OTO types with $p\leq \lfloor\frac{n+1}{2}\rfloor$. Further, each necklace contains a set of OTO numbers differing by at most one unit. However, for odd $n$, the highest proper-OTO number $p = \frac{n+1}{2} $ lies in the KMS orbit of $p = \frac{n-1}{2}$, thereby informing us that the KMS relations are making the higher OTO correlator redundant in the thermal state.

\section{Discussion}
\label{sec:discuss}

We have primarily focused on synthesizing known features of thermal correlation functions and arguing that they are best understood in the  space of out-of-time-order (OTO) observables.  While traditional presentations avoid the OTO correlators by explicitly convolving the KMS condition with time reversal or a CPT transformation to restore operator ordering, the generalization as discussed herein allows for a simpler interpretation. 
The KMS condition acts by cyclically permuting operators in a given correlator. Among other things we have shown that the proper-OTO numbers of all correlators in the KMS orbit of a given Wightman function will at most differ by unity. We also `solved' the KMS relations explicitly, by constructing a causal basis for the independent thermal correlation functions in terms of the fully nested commutators (with one operator held in fixed position). We have given explicit formulae that express various complete classes of correlation functions in terms of this causal basis; these can be interpreted as the full set of generalized $n$-point fluctuation-dissipation relations. 

There are several interesting avenues that are ripe for further exploration. For instance, one could attempt to understand the structures herein in kinetic theory and in explicit models. However, there are a set of conceptual questions that we have not addressed herein which are interesting to contemplate further:
\begin{itemize}
\item Thermalization of chaotic quantum systems is a subtle process. One relevant observable in the context of quantum chaos is the 4-point tremolo correlator (see below). We can ask: are there finer-grained  or perhaps novel features of thermalization and chaotic behaviour, that are captured excluvively by higher $2k$-point tremelo correlators? If so, are these features captured more simply in terms of the causal basis of nested correlators?
\item How does the notion of thermal equivariance introduced in the context of hydrodynamic effective field theories \cite{Haehl:2016uah}, which are constructed in the Schwinger-Keldysh context, generalize to higher-point functions?
\item What is the implication of 2-OTO thermal correlators which are related to Schwinger-Keldysh correlators 
in terms of transport? 
\end{itemize} 

The fact that KMS relations lead to relations between observables of different proper-OTO number can be interpreted to imply that non-trivial information in thermal correlators only appears at even-point functions. For instance, while there are 2-OTO 3-point functions, they lie in the KMS orbit of a 1-OTO correlator and their physics should thus not belong to the class of 2-OTO observables. The first place where we encounter non-trivial 2-OTO observables is in a 4-point function, where one can canonically choose it to be the chaos correlator \cite{Shenker:2013pqa,Larkin:1969aa,Maldacena:2015waa}. The natural observable to pick, is the tremelo correlator, where every alternate operator is a past turning point operator. Equivalently the $\delta$-list would comprise of $\frac{n}{2}$ alternating 1s, e.g., the last two roles of Table~\ref{tab:4ptq}. These are distinguished by the fact that every correlator in their KMS orbit has maximal allowed OTO-number for the given number of operators.

Given that this observable is cognizant of the detailed dynamics of thermalization, one may wonder if higher even-point functions provide further detailed signatures. In particular, could one view the higher-point functions as higher moments of some distributions that captures equilibration? One useful avenue to examine is to explore how operator scrambling interplays with higher-OTOs. Similar considerations have inspired explorations of $k$-designs in chaotic quantum channels \cite{Roberts:2016hpo,Cotler:2017jue}. 

It is also worthwhile to explore in more detail the physical implication of 2-OTO thermal correlators which lie in the KMS orbit of a Schwinger-Keldysh correlator. Since a natural subset of the latter forms the basis of transport theory, it would be interesting to understand the OTO fluctuations in terms of more natural response coefficients.  These considerations  could provide useful generalizations of the Jarzynski relation \cite{Jarzynski:1997aa}, which we hope to explore elsewhere.

Any perturbation theoretic/path-integral approach to these correlators needs a way of effectively re-packaging the contour ordered correlators that arise from generalized Schwinger-Keldysh contours.

This would mean extending the existing results on Schwinger-Keldysh contour thermal correlators \cite{Hou:1998yc,Carrington:1996rx}  whereby these correlators have been written as outer-products of certain 2-component column vectors thus bringing out their KMS structure. Like their Schwinger-Keldysh analogues, this form could be very useful for summering the non-linear fluctuation-dissipation relations, for establishing spectral representations and sum rules, for efficient hard thermal loop (HTL) approximations and for deriving generalized Kubo formulae.

Another interesting extension of the work here is to set up the analogue of Brownian motion which keeps track of higher OTO correlations of the bath. This is a natural higher OTO extension of the famous work by Feynman-Vernon \cite{Feynman:1963fq} and Caldeira-Leggett \cite{Caldeira:1982iu}.

Finally, while we have focused exclusively on thermal density matrices, it should be possible to generalize the above discussion to a general initial state $\rhoi$. Working with the modular Hamiltonian, $K = -\log(\rhoi)$ we could effectively run the same arguments for the modular evolved operators (see also the discussion section of \cite{Haehl:2016pec}). Sliding an operator through the density matrix as here would now result in the modular evolution of the operator. Various authors have noted the similarity between thermal and modular evolution, and in the holographic context, the modular evolution plays a role in construction of local bulk observables in the entanglement wedge \cite{Jafferis:2015del,Faulkner:2017vdd,Cotler:2017erl}. It would be interesting to examine whether the modular KMS relations which one can derive by sliding operators through the density matrix have any useful information to impart for these considerations.

\acknowledgments

It is a great pleasure to thank Simon Caron-Huot, Michael Geracie, Veronika Hubeny,  David Ramirez, Jaroslav Trnka, and Nicole Yunger Halpern for useful discussions. 
FH gratefully acknowledges support through a fellowship by the Simons Collaboration `It from Qubit' and hospitality at QMAP, UC Davis, and  McGill University during the course of this project. RL gratefully acknowledges support from International Centre for Theoretical Sciences (ICTS), Tata Institute of Fundamental Research, Bengaluru and Ramanujan fellowship from Govt.\ of India.  MR would like to thank ICTP and  Weizmann Institute for hospitality during the course of this project. MR was supported in part by U.S.\ Department of Energy grant DE-SC0009999 and by the University of California.
AAN, PN and RL would  also like to acknowledge their debt to the people of India for their steady and generous support to research in the basic sciences. 

\appendix
\section{Thermal Schwinger-Keldysh correlators}
\label{app:thspec}

Motivated by developing a nonlinear generalization of the Fluctuation-Dissipation (FD) theorem for higher-point response functions, \cite{Wang:1998wg} used the Schwinger-Keldysh formalism as the appropriate framework for studying nonlinear response theory near thermal equilibrium. They worked with Schwinger-Keldysh contour correlators and used the KMS relations to derive the generalized FD relations for real time (1-OTO) thermal Green functions. The FD relations furnish a spectral representation for  retarded and advanced thermal Green functions.
 
In this appendix we will compare the results of our paper with those of \cite{Wang:1998wg}. In the process, we will derive expressions for 3 and 4 point (1-OTO) thermal contour correlators in terms of our commutator basis.\footnote{ Similar expressions are derived in \cite{Carrington:1996rx,Carrington:2006xj}.} 

The thermal $n$-point Green function in \cite{Wang:1998wg} is defined as 
\be 
G_{rrara....}\equiv (-i)^{n-1} 2^{n_r-1} \vev{ \mathcal{T}_{SK} \, 1_r \, 2_r \,3_a\,4_r\,5_a .......}
\ee
where $i_r$ stands for $(A_{iR}(t_i)+A_{iL}(t_i))/2$ (average basis), $i_a$ stands for $A_{iR}(t_i)-A_{iL}(t_i)$ (difference basis), and ${\cal T}_{SK}$ denotes Schwinger-Keldysh (i.e., 1-OTO) contour ordering.\footnote{ The factors of $2$ are chosen so that they cancel on using the Keldysh rules to move to the nested commutator basis: $n_r$ is the number of average operators in the correlator}

\paragraph{Three-point spectral function relations:}
Consider 1-OTO 3-point contour correlators:
\begin{equation}
\begin{split}
G_{raa} & \equiv (-i)^2 \vev{ \mathcal{T}_{SK} \, 1_r \, 2_a \,3_a }  \\
G_{rra} & \equiv 2(-i)^2\vev{ \mathcal{T}_{SK} \, 1_r\,  2_r\,  3_a } \\
G_{rrr} & \equiv 2^2(-i)^2\vev{ \mathcal{T}_{SK} \, 1_r \,2_r \, 3_r } 
\end{split}
\end{equation}
We can use Keldysh rules to write the contour correlators as nested correlators:
\begin{equation}
\begin{split}
 G_{raa}&= -\Theta_{123 } \,  
 \vev{ \fullcomm{123}}  - \Theta_{132}\,\vev{\fullcomm{132}}  \\
G_{rra} &= 
-\Theta_{123} \,  \vev{ \fullcomm{12_{_+}3}}-\Theta_{213}\,   \vev{ \fullcomm{21_{_+}3}}
-\Theta_{132} \,  \vev{ \fullcomm{13_{_+}2}}-\Theta_{231} \,  \vev{ \fullcomm{23_{_+}1}}  \\
G_{rrr}  &= 
-(\Theta_{123}+\Theta_{213}) \, \vev{ \fullcomm{12_{_+}3_{_+}}}
-(\Theta_{231}+\Theta_{321})\,  \vev{ \fullcomm{23_{_+}1_{_+}}}  
-(\Theta_{312}+\Theta_{132}) \, \vev{ \fullcomm{31_{_+}2_{_+}}}
\end{split}
\end{equation}
where we have introduced the time-ordering step-function: 
$\Theta_{ijk} \equiv \Theta(t_i > t_j > t_k)$.  Passing onto our causal basis for nested correlators we can write expressions for these contour correlators in terms of nested commutators: 
\begin{equation}
\begin{split}
G_{raa}&=- \Theta_{123} \, \vev{\fullcomm{123}}- \Theta_{132} \,\vev{\fullcomm{132}}  \\
G_{rra}&= -\cth_1\,  \Theta_{23} \, \vev{\fullcomm{123}}+(\cth_1 \,\Theta_{23}+\cth_2 \,\Theta_{13})
\vev{\fullcomm{132}} \\
G_{rrr}&= (\cth_1\,\cth_3+\Theta_{132}+\Theta_{312}) \vev{\fullcomm{123}}+
(\cth_1\,\cth_2+\Theta_{123}+\Theta_{213}) \, \vev{\fullcomm{132}} 
\end{split}
\end{equation}
This gives, for a thermal density matrix, an expression for 1-OTO 3-point contour correlators in terms of two spectral functions. This expression, in a different notation, was given in Eq.~(31) of \cite{Carrington:1996rx}.

Wang-Heinz \cite{Wang:1998wg} write the FD relations for  3-point 1-OTO contour correlators in the Av-Dif basis as 
\begin{equation}
\begin{split}
G_{rra}&=\cth_1\, (G^*_{aar}-G_{ara})+\cth_2\, (G^*_{aar}-G_{raa})  \\
G_{rrr}&=G^*_{raa} + G^*_{ara} + G^*_{aar} + \cth_2 \,\cth_3\, (G_{raa} + G^*_{raa}) 
+\cth_3 \,\cth_1\,(G_{ara} + G^*_{ara}) + \cth_1 \,\cth_2\,(G_{aar} + G^*_{aar})
\end{split}
\label{eq:whfdt3}
\end{equation}
As above, we can evaluate all these contour correlators in our commutator basis (noting that complex conjugation reverses the time ordering in Theta functions), to find:
\begin{equation}
\begin{split}
G_{raa}&=-\Theta_{123} \, \vev{ \fullcomm{123} }- \Theta_{132} \, \vev{\fullcomm{132} }  \\
G^*_{raa}&=-\Theta_{321}\, \vev{ \fullcomm{123}}- \Theta_{231}\, \vev{\fullcomm{132} } .
\end{split}
\end{equation}
 Similar expressions for $G_{aar},\,G_{ara},\,G^*_{aar},\,G^*_{ara}$ obtained by cyclic permutations of $\{1,2,3\}$ in the above (and using $\fullcomm{231}=-\fullcomm{123}+ \fullcomm{132}$).
With all these expressions in hand we can compare both sides of \eqref{eq:whfdt3}  and confirm that they indeed hold.

\paragraph{Four-point spectral function relations:}
We now move to the 4-point function case. Wang-Heinz \cite{Wang:1998wg} write the FD relations for  4-point 1-OTO contour correlators in the Av-Dif basis in the following form: 
\begin{equation}
\begin{split}
G_{rrrr} &=
	  -\cth_2\, \cth_3\,\cth_4\, G_{raaa} +(\cth_2\cth_3\cth_4 + \cth_2 + \cth_3 + \cth_4)
	  \,G^*_{raaa} +\cth_1\cth_3\cth_4\, G_{araa}  \\
&+	\cth_2(\cth^{(14)}_{(23)} +\cth_4^2 \cth^{(13)}_{(24)})\, G^*_{araa} 
	+\cth_1\cth_2\cth_4\, G_{aara} +\cth_3(\cth^{(12)}_{(34)} 
	+\cth_2^2\cth^{(14)}_{(23)})\, G^*_{aara}   \\
&	+\cth_1\cth_2\cth_3\, G_{aaar} + \cth_4(\cth^{(13)}_{(24)} 
	+ \cth_3^2\cth^{(12)}_{(34)})\, G^*_{aaar} + \cth_1\cth_4\, G_{arra} 
	+ \cth_2\cth_3\cth^{(14)}_{(23)}\, G^*_{arra}  \\
&	+\cth_1\cth_3\, G_{arar} + \cth_2\cth_4\cth^{(13)}_{(24)}\, G^*_{arar} 
	+ \cth_1\cth_2\, G_{aarr} + \cth_3\cth_4\cth^{(12)}_{(34)}\, G^*_{aarr} ,   \\
G_{rrra}&=  
	\cth_2\cth_3\, G_{raaa}- \cth_2\cth_4\cth^{(13)}_{(24)}\, G^*_{araa} 
	-\cth_3\cth_4\cth^{(12)}_{(34)}\, G^*_{aara}-(\cth^{(13)}_{(24)} 
	+\cth_3^2\cth^{(12)}_{(34)})\, G^*_{aaar} \\
&
-\cth_1\, G_{arra} -\cth_2\cth^{(13)}_{(24)} \, G^*_{arar} 
-\cth_3\cth^{(12)}_{(34)}\, G^*_{aarr},  \\
 G_{rraa}&=
 	 -\cth_2\, G_{raaa} -\cth_1\, G_{araa} + \cth_4\cth^{(12)}_{(34)}\, G^*_{aara} 
 	 + \cth_3\cth^{(12)}_{(34)}\, G^*_{aaar} + \cth^{(12)}_{(34)}\, G^*_{aarr}   
 \end{split}
 \label{eq:whfdt4a}
\end{equation} 

We use the Keldysh rules to get the following contour correlators in nested form:
\begin{align}
G_{rrrr}& 
	=(-i)^3 2^3\sum_{\sigma} \Theta_{\sigma(1234)}\,
	\vev{
	\fullcomm{\sigma(1)\,\sigma(2)_{_+} \, \sigma(3)_{_+}\, \sigma(4)_{_+} } }  \nonumber \\
G_{rrra}& =(-i)^3 2^2 \bigg(
	\Theta_{1234} \,\vev{ \fullcomm{1\,2_{_+}\,3_{_+}\,4 } }
	 +\Theta_{1243}\, \vev{\fullcomm{1\,2_{_+}\,4\,3_{_+}} }
	 +\Theta_{1423}\, \vev{\fullcomm{1\,4\,2_{_+}\,3_{_+} } }\bigg)+ (123)\text{sym} 
	  \nonumber \\
G_{rraa}& =(-i)^3 2\bigg( \Theta_{1234}\vev{\fullcomm{1\,2_{_+}\,3\,4 }}
	+\Theta_{1324}\vev{\fullcomm{1\,3\,2_{_+}\,4 }}
	+\Theta_{1342}\vev{\fullcomm{1\,3\,4\,2_{_+}} }\bigg)
	+ (12)(34)\text{sym}  \nonumber \\
G_{raaa}& =(-i)^3\sum_{\sigma} \, \Theta_{1\sigma(234)} \, 
	\vev{\fullcomm{1\,\sigma(2)\,\sigma(3)\,\sigma(4)} } 
\label{eq:whfdt4b}
\end{align}	
where the sums run over permutations of labels as indicated.

Using \eqref{eq:FD4} (or more generally \eqref{eq:lem3}) we can change all nestings to commutators and go to our causal basis:
\begin{align}
G_{rrrr} &= -(-i)^3 2^3
\sum_{\sigma}  \Theta_{\sigma(1234)} \left\{ \cth_{\sigma_1}\cth_{\sigma_4}\frac{\cth_{\sigma_1} \cth_{\sigma_2}+1}{\cth_{\sigma_1}+\cth_{\sigma_2}} \,\vev{\fullcomm{\sigma(1234)}}+\cth_{\sigma_3}\cth_{\sigma_4}\frac{\cth_{\sigma_1}^2-1}{\cth_{\sigma_1}+\cth_{\sigma_3}} \, \vev{\fullcomm{\sigma(1324)}}  \right.   \nonumber \\
 &\qquad\qquad\quad \left.  -\cth_{\sigma_1}\cth_{\sigma_4}\frac{\cth_{\sigma_3}^2-1}{\cth_{\sigma_3}+\cth_{\sigma_4}} \, \vev{\fullcomm{\sigma(1243)}} -\cth_{\sigma_1}\cth_{{\sigma_4}} \frac{\cth_{\sigma_2}^2-1}{\cth_{\sigma_2}+\cth_{\sigma_4}} \vev{\fullcomm{\sigma(1342)} }    \right.   \nonumber \\
 &\qquad\qquad\quad \left. +\cth_{\sigma_3} \cth_{\sigma_4} \frac{\cth_{\sigma_1}^2-1}{\cth_{\sigma_1}+\cth_{\sigma_4}} \vev{\fullcomm{\sigma(1423)}} + \cth_{\sigma_2}\cth_{\sigma_4} \frac{\cth_{\sigma_1}^2-1}{\cth_{\sigma_1}+\cth_{\sigma_4}}\,\vev{\fullcomm{\sigma(1432)} }
\right\}   \nonumber \\
G_{rrra} &=(-i)^3 2^2 \bigg\{ \frac{\cth_1 \left(\cth_1 \cth_2+1\right) }{\cth_1+\cth_2}\,\Theta _{1234}\vev{\fullcomm{1234}}+\frac{\left(\cth_1^2-1\right) \cth_3}{\cth_1+\cth_3}\, \left(\Theta _{1234}+\Theta _{1243}\right)\vev{\fullcomm{1324}}\nonumber \\
 &\qquad\qquad\quad- \left( \frac{\cth_1 (\cth_2^2-1)}{\cth_2+\cth_4}  \,\Theta_{1234} + \frac{(\cth_1^2-1) \cth_3}{\cth_1 + \cth_3}\, \Theta_{1243}  \right)\vev{\fullcomm{1342}}   \nonumber \\
  &\qquad\qquad\quad+ \bigg(  \frac{\cth_3\left(\cth_1^2-1\right)}{\cth_1+\cth_4}(\Theta _{1234}+ \Theta _{1243})  - \frac{\cth_3 (\cth_1 \cth_4+1)}{\cth_1+\cth_4}  \Theta _{1423} \bigg) \vev{\fullcomm{1423}}   \nonumber \\
 &\qquad\qquad\quad+ \left( \frac{\cth_2(\cth_1^2-1)}{\cth_1+\cth_4} \, \Theta_{1234} + \frac{\cth_3(\cth_2^2-1)}{\cth_2+\cth_3} \,( \Theta_{1243} + \Theta_{1423} ) \right)\vev{\fullcomm{1432} }  \nonumber \\
   &\qquad\qquad\quad -\cth_1 \bigg( \frac{\cth_3^2-1}{\cth_3+\cth_4} \, \Theta _{1234}+ \cth_3 \Theta _{1243}\bigg)\vev{\fullcomm{1243}} \bigg\}+ (123)\text{sym}   \nonumber \\
G_{rraa} &=(-i)^32 \bigg\{ \cth_1 \Theta _{1234} \vev{\fullcomm{1234}}+\frac{\cth_1^2-1}{\cth_1+\cth_4}\Theta _{1234}\left(\vev{\fullcomm{1432}}-\vev{\fullcomm{1423}}\right)\nonumber \\
 &\qquad\qquad\quad- \bigg( \frac{\cth_2^2-1}{\cth_2+\cth_4} \,(\Theta _{1234}+ \Theta _{1324})+\cth_2  \Theta _{1342} \bigg) \vev{\fullcomm{1342}} \nonumber \\
 &\qquad\qquad\quad-\left( \frac{\cth_1^2-1}{\cth_1+\cth_3} \, \Theta _{1234}-  \frac{\cth_1 \cth_3+1}{\cth_1+\cth_3} \, \Theta _{1324}\right) \vev{\fullcomm{1324}} \bigg\} + (12)(34)\text{sym}  \nonumber \\
G_{raaa} &=(-i)^3 \sum_{\sigma} \Theta_{1\sigma(234)}\vev{\fullcomm{1\sigma(234)} }
\label{eq:whfdt4c}
\end{align}	
These equations (together with the Jacobi relations of the kind in \eqref{eq:sJac4}) give the SK correlators $G_{rrrr},G_{rrra},G_{rraa},G_{raaa}$ in our causal commutator basis $\vev{\fullcomm{1\sigma(234)}}$. Note that for any fixed time ordering the above expressions simplify significantly.

We can similarly get expressions for all the $G$'s in \eqref{eq:whfdt4a} (the complex conjugation again reverses the time ordering in Theta functions) and explicitly check that it is true.

\section{Nested Thermal brackets and tJacobi relations}
\label{app:tJacobi}
In this appendix we illustrate the use of the thermal nested brackets defined in \eqref{eq:tac} for deriving FD relations for nested correlators. We will consider nested thermal (anti-)commutators and present various {\it tJacobi} operator relations between them. These are analogous to the sJacobi relations discussed in \cite{Haehl:2017qfl} and in fact comprise a one-parameter deformation of those identities.
An important tool for the following considerations is\\

\noindent
{\bf Lemma \textlabel{4}{lemma4}:}  Given the graded commutator \eqref{eq:gradedcomm} and the thermally deformed graded commutator  \eqref{eq:tac}, the following 
\emph{thermal involution identity} holds:
\begin{equation}
\vev{
[{\cdots}{
	\Ecomm{
	\Ecomm{
	\Ecomm{\Op{X}}{\Op{C}}{\Op{C}}
	}{\Op{B}}{\Op{B}}
	}
	{\Op{A}}{\Op{A}}
	 {\cdots}{}} ]
}
= 
\vev{
	\Op{X}\ 
	\tcomm{\Op{C}}{
	\tcomm{\Op{B}}{
	\tcomm{\Op{A}}{\cdots \Op{1}}{\Op{A}}
	}{\Op{B}}
	}{\Op{C}}
}
\label{eq:tInv}
\end{equation}	
where $\Op{1}$ is the identity operator.\\  

\noindent
The terminology thermal involution refers to the fact that the right hand side of \eqref{eq:tInv} is a nested thermally deformed correlator, where the nesting is inverted compared to the left hand side. 

{\it Proof}: We will prove Lemma \ref{lemma4} inductively, illustrating the basic ideas and using the KMS relations in appropriate steps. Firstly, we have 
\begin{equation}
\begin{split}
\mathcal{G}_2(\Op{X}) &\equiv 
	\vev{
	\Ecomm{
	\Ecomm{\Op{X}}{\Op{B}}{\Op{B}} 
	}{\Op{A}}{\Op{A}} }\\
&= \vev{
	\Op{X \,B \,A} + \varepsilon_{_\Op{B}} \, e^{\beta \omega_{_\Op{B}}} \Op{X\,A\,B}  +  \varepsilon_{_\Op{A}} \,  e^{\beta \omega_{_\Op{A}}} \,\Op{X\,B\,A} 
	+  \varepsilon_{_\Op{A}}    \varepsilon_{_\Op{B}} \,  e^{\beta(\omega_{_\Op{A}}+ \omega_{_\Op{B}})}\, \Op{X\,A\,B} 
}\\
&=(1 + \varepsilon_{_\Op{A}} \, e^{\beta \omega_{_\Op{A}}}) \vev{
	\Op{X}\; \tcomm{\Op{B}}{\Op{A}}{\Op{B}}
}
= \vev{
	\Op{X}\; \tcomm{\Op{B}}{\tcomm{\Op{A}}{\Op{1}}{\Op{A}} }{\Op{B}}
}
\end{split}
\end{equation}
where we have used KMS relations in the first equality above to bring $\Op{X}$ to the left. These manipulations can be recursively applied to pass to higher orders. For instance replacing 
$\Op{X}$ by $\Ecomm{\Op{X}}{\Op{C}}{\Op{C}}$ in the above equation we get:
\begin{equation}
\mathcal{G}_3(\Op{X})  \equiv \vev{
	\Ecomm{
	\Ecomm{
	\Ecomm{\Op{X}}{\Op{C}}{\Op{C}}}
	{\Op{B}}{\Op{B}}}{\Op{A}}{\Op{A}}
}
\end{equation}	
But since  $\vev{\Ecomm{\Op{X}}{\Op{C}}{\Op{C}}\, \Op{O}}= \vev{\Op{X\,C\,O}}
+\varepsilon_{_\Op{C}}\, e^{\beta \omega_{_\Op{C}}}\, \vev{\Op{X\,O\,C}}
= \vev{\Op{X}\, \tcomm{\Op{C}}{\Op{O}}{\Op{C}}}$, we thus  obtain

\begin{equation}
 \mathcal{G}_3(\Op{X})  = 
\vev{
	\Op{X}\; 
	\tcomm{\Op{C}}{
	\tcomm{\Op{B}}{\tcomm{\Op{A}}{\Op{1}}{\Op{A}} }{\Op{B}}
	}{\Op{C}}
}
\end{equation}	
and so on recursively. This process makes clear that the relations in \eqref{eq:tInv} hold at arbitrary level.\\

Having established \eqref{eq:tInv}, let us do a counting of the  number of correlators contained therein, and of the number of relations between those.
There are $2^n\, n!$ thermal nested $n$-point structures of the form
\begin{equation}\label{eq:thermalX}
\mathcal{X}_{\sigma,\varepsilon_\sigma} = \tcomm{\Op{A}_{\sigma(n)}}
	{\tcomm{\Op{A}_{\sigma(n-1)}}{\cdots\tcomm{\Op{A}_{\sigma(1)}}{\Op{1}}{{\sigma(1)}}\cdots}{\sigma(n-1)}}{\sigma(n)} \,.
\end{equation}
These correlators should be expanded in a Wightman basis of $n!$ structures. 
Thus there ought to be  $(2^n -1)n!$  relations between them. We refer to these as {\it tJacobi relations}. Of these, $(2^{n-1}-1)n!$ encode the information of $(n+1)$-point sJacobi relations, which are valid for any density matrix \cite{Haehl:2017qfl}. The remaining (improper) ones are more directly related to a basic fluctuation-dissipation theorem. More precisely, we can understand all tJacobi relations according to the following two complementary observations.
\begin{itemize}
\item {\bf Improper tJacobi relations:} These are the $2^{n-1}n!$ tJacobi relations, which can be described as a simple nesting of the basic $n=1$ ``seed'' of fluctuation-dissipation relations inside thermally deformed (anti-)commutators. The $n=1$ FD seed is the trivial identity
\begin{equation}
\ta{\Op{A}}{\Op{1}} + \cth_{\Op{A}}\; \tc{\Op{A}}{\Op{1}} =0\,.
\label{eq:1tJ}
\end{equation}	
This relation captures the basic FD theorem, since we can multiply by $\Op{B}$ from the left and upon taking expectation value infer using \eqref{eq:tInv} that the FD theorem \eqref{eq:FD2} holds (see also below for illustration). Nesting this relation with $(n-1)$ further operators inside $(n-1)$ thermal brackets and considering all operator permutations gives $2^{n-1}n!$ of the tJacobi relations.
\item {\bf Relations between $n$-point tJacobis and $(n+1)$-point sJacobis:} 
To understand the remaining $(2^{n-1}-1)n!$ tJacobi relations, we start with the observation that $(2^{n-1}-1)n!$ is precisely the number of $(n+1)$-point sJacobi relations, where the $(n+1)^{\rm st}$ operator is being given a fixed position (i.e., only considering permutations of the remaining $n$ operators).

This is not coincidental: the thermal involution relation \eqref{eq:tInv} allows us to relate any given $n$-point tJacobi relation to a standard sJacobi-type relation for $(n+1)$-point functions. To see this, let us assume we have a tJacobi relation of the form $ \sum_{\sigma,\varepsilon_\sigma}  \, \mathcal{X}_{\sigma,\varepsilon_\sigma} = 0$ with $\mathcal{X}_{\sigma,\varepsilon_\sigma}$ as in \eqref{eq:thermalX}. We can now multiply with the $(n+1)^{\rm st}$ operator and find, using \eqref{eq:tInv}:
\begin{equation}\label{eq:ExpsJac}
\begin{split}
0 = \vev{\Op{A}_{n+1}\, \sum_{\sigma,\varepsilon_\sigma}  \, \mathcal{X}_{\sigma,\varepsilon_\sigma}}  = \vev{\sum_{\sigma,\varepsilon_\sigma}  \,
	\Ecomm{\cdots 
	\Ecomm{
	\Ecomm{\Op{A}_{n+1}}{\Op{A}_{\sigma(n)}}{\sigma(n)}
	}{\Op{A}_{\sigma(n-1)}}{\sigma(n-1)}
	}
	{\cdots \Op{A}_{\sigma(1)}}{\sigma(1)}
	  }  \,.
\end{split}
\end{equation}
The right hand side is of the form of an $(n+1)$-point sJacobi relation.

\end{itemize} 
Note that one the one hand it is justified to call the relations \eqref{eq:ExpsJac} as sJacobi identities: all thermal factors occurring on the left hand side have been removed on the right hand side and the relations thus do not rely on thermality! On the other hand, we should note that the right hand side of \eqref{eq:ExpsJac} is not written in the standard form of sJacobi relations explored in \cite{Haehl:2017qfl}, since a particular operator $\Op{A}_{n+1}$ is singled out to be innermost in all terms. We conjecture that the mapping described above is nevertheless one-to-one.
\paragraph{Conjecture:} Every $n$-point tJacobi relation is of one of two types: either it is one of the $2^{n-1}n!$ improper ones (as described above), or it can be derived as descending from an $(n+1)$-point sJacobi relation via \eqref{eq:ExpsJac} (now reading that equation from right to left).\\

This conjecture nicely unifies the ideas of generalized Jacobi relations and KMS condition. A simple counting argument lends support to the conjecture: by permuting $\Op{A}_{n+1}$ in \eqref{eq:ExpsJac}, we obtain $(n+1)$ such sJacobi-type relations for each of the remaining $(2^{n-1}-1)n!$ tJacobi relations $ \sum_{\sigma,\varepsilon_\sigma}  \, \mathcal{X}_{\sigma,\varepsilon_\sigma} = 0$. Note that $(n+1) \times (2^{n-1}-1)n! = (2^{(n+1)-2}-1)(n+1)!$ is the correct number of $(n+1)$-point sJacobi relations.\\

Let us now illustrate these general constructions for small values of $n$.
\paragraph{$n =1$ tJacobi:} This is the simplest relation since we have a single operator. From \eqref{eq:tInv} we find:
\begin{equation}
\ta{\Op{A}}{\Op{1}} + \cth_{\Op{A}}\; \tc{\Op{A}}{\Op{1}} =0
\label{eq:1tJ}
\end{equation}	
This relation captures the basic FD theorem, since we can multiply by $\Op{B}$ on the left and upon taking expectation value infer using \eqref{eq:tInv} that \eqref{eq:FD2} holds.

\paragraph{$n=2$ tJacobi:} The tJacobi relations for $n=2$ split into proper and improper relations. The improper tJacobi relations are obtained by nesting in the $n=1$ tJacobi relation with permutations, viz., 
\begin{equation}
\tcomm{\Op{B}}{
\ta{\Op{A}}{\Op{1}} + \cth_{\Op{A}}  \, \tc{\Op{A}}{\Op{1}}	
}{\Op{B}} =0 \,, \qquad
\tcomm{\Op{A}}{
\ta{\Op{B}}{\Op{1}} + \cth_{\Op{B}}  \, \tc{\Op{B}}{\Op{1}}	
}{\Op{A}} =0 
\label{eq:n2tjI}
\end{equation}	
On the other hand, the $n=2$ proper tJacobi relations arise as new identities involving two operators. These can be brought to the form:
\begin{equation}
\begin{split}
\ta{\Op{A}}{\ta{\Op{B}}{\Op{1}}}  - \tc{\Op{A}}{\tc{\Op{B}}{\Op{1}}} 
& =
	\ta{\Op{B}}{\ta{\Op{A}}{\Op{1}}} - \tc{\Op{B}}{\tc{\Op{A}}{\Op{1}}} \\
\ta{\Op{A}}{\tc{\Op{B}}{\Op{1}}}  - \tc{\Op{A}}{\ta{\Op{B}}{\Op{1}}} 
& =
	\tc{\Op{B}}{\ta{\Op{A}}{\Op{1}}} -  \ta{\Op{B}}{\tc{\Op{A}}{\Op{1}}} 
\end{split}
\label{eq:n2tjp}
\end{equation}	

As before multiplying the tJacobi relations by a new operator $\Op{C}$ on the left, taking expectation values, and using the thermal involution identity \eqref{eq:tInv} we obtain
\begin{subequations}
\begin{align}
\eqref{eq:n2tjI} & 
\;\; \Longrightarrow \;\;
\begin{cases}
& 
	\vev{\fullcomm{\Op{B \, C\,  A_{_+}} }}
	+ \cth_{\Op{A}} \, \vev{\fullcomm{\Op{B \, C\,  A} } } =0  \\
& 
	\vev{\fullcomm{\Op{B \, C_{_+}\,  A_{_+}} }}
	+ \cth_{\Op{A}} \, \vev{\fullcomm{\Op{B \, C_{_+}\,  A} } } =0 \\
\end{cases}
\label{eq:FDTI3}
\end{align}
\begin{align}
\eqref{eq:n2tjp}  & 
\;\; \Longrightarrow \;\;
\begin{cases}
& 
	\vev{\fullcomm{\Op{C \, A_{_+}\,  B_{_+}} }}
	- \vev{\fullcomm{\Op{C\, A\, B}}}
	 =
	 \vev{\fullcomm{\Op{C\, B_{_+} \, \Op{A}_{_+} }}} - 
	 \vev{\fullcomm{\Op{C\, B\, A}}}
	   \\
& 
	\vev{\fullcomm{\Op{C \, A_{_+}\,  B} }}
	- \vev{\fullcomm{\Op{C\, A\, B_{_+}}}}
	 =
	 \vev{\fullcomm{\Op{C\, B_{_+} \, \Op{A} }}} - 
	 \vev{\fullcomm{\Op{C\, B\, A_{_+}}}} \\
\end{cases} 
\nonumber \\
& \;\; \Longrightarrow \;\;
\vev{\fullcomm{\Op{C, B_{_+} A}}} = \cth_{\Op{C}}\;  \vev{\fullcomm{\Op{C\, B \, A}}} - 
\left(\cth_{\Op{B}} + \cth_{\Op{C}} \right)\; \vev{\fullcomm{\Op{C\, A \, B}}}
\label{eq:FDTP3}
\end{align}
\end{subequations}
The relations \eqref{eq:FDTI3} and \eqref{eq:FDTP3} and their cyclic permutations are the $n=3$ fluctuation-dissipation relations. 

One can similarly write down the tJacobi relations for higher values of $n$; explicit expressions for $n=3$ tJacobi relations are given below.

\paragraph{$n=3$ tJacobi:} The $n=3$ tJacobi relations following from KMS relations are:
\begin{equation}
\begin{split}
\tcomm{\Op{A}}{
\tcomm{\Op{B}}{
\ta{\Op{C}}{\Op{1}} + \cth_{\Op{C}}  \, \tc{\Op{C}}{\Op{1}}	
}{\Op{B}}
}{\Op{A}}&=0\\
\tcomm{\Op{B}}{
\tcomm{\Op{C}}{
\ta{\Op{A}}{\Op{1}} + \cth_{\Op{A}}  \, \tc{\Op{A}}{\Op{1}}	
}{\Op{C}}
}{\Op{B}}&=0\\
\tcomm{\Op{C}}{
\tcomm{\Op{A}}{
\ta{\Op{B}}{\Op{1}} + \cth_{\Op{B}}  \, \tc{\Op{B}}{\Op{1}}	
}{\Op{A}}
}{\Op{C}}&=0\\
\tcomm{\Op{A}}{
\tcomm{\Op{C}}{
\ta{\Op{B}}{\Op{1}} + \cth_{\Op{B}}  \, \tc{\Op{B}}{\Op{1}}	
}{\Op{C}}
}{\Op{A}}&=0\\
\tcomm{\Op{C}}{
\tcomm{\Op{B}}{
\ta{\Op{A}}{\Op{1}} + \cth_{\Op{A}}  \, \tc{\Op{A}}{\Op{1}}	
}{\Op{B}}
}{\Op{C}}&=0\\
\tcomm{\Op{B}}{
\tcomm{\Op{A}}{
\ta{\Op{C}}{\Op{1}} + \cth_{\Op{C}}  \, \tc{\Op{C}}{\Op{1}}	
}{\Op{A}}
}{\Op{B}}&=0
\end{split}
\label{eq:tJac3-24}
\end{equation}
The choice of signs give $2^{n-1} n! = 24$ of the tJacobi relations at $n=3$. One can check explicitly that these relations are all linearly independent.
\begin{equation}
\begin{split}
\tcomm{\Op C}{   \ta{\Op A }{\ta{\Op B }{\Op 1}}-\tc{\Op A }{\tc{\Op B }{\Op 1}}        }{\Op C }     & = \tcomm{\Op C}{   \ta{\Op B }{\ta{\Op A }{\Op 1}}-\tc{\Op B }{\tc{\Op A }{\Op 1}}        }{\Op C }     \\
\tcomm{\Op C}{   \ta{\Op A }{\tc{\Op B }{\Op 1}}-\tc{\Op A }{\ta{\Op B }{\Op 1}}        }{\Op C }     & = \tcomm{\Op C}{   \tc{\Op B }{\ta{\Op A }{\Op 1}}-\ta{\Op B }{\tc{\Op A }{\Op 1}}        }{\Op C }     \\
\tcomm{\Op B}{   \ta{\Op C }{\ta{\Op A }{\Op 1}}-\tc{\Op C }{\tc{\Op A }{\Op 1}}        }{\Op B }     & = \tcomm{\Op B}{   \ta{\Op A }{\ta{\Op C }{\Op 1}}-\tc{\Op A }{\tc{\Op C }{\Op 1}}        }{\Op B }     \\
\tcomm{\Op B}{   \ta{\Op C }{\tc{\Op A }{\Op 1}}-\tc{\Op C }{\ta{\Op A }{\Op 1}}        }{\Op B }     & = \tcomm{\Op B}{   \tc{\Op A }{\ta{\Op C }{\Op 1}}-\ta{\Op A }{\tc{\Op C }{\Op 1}}        }{\Op B }     \\
\tcomm{\Op A}{   \ta{\Op B }{\ta{\Op C }{\Op 1}}-\tc{\Op B }{\tc{\Op C }{\Op 1}}        }{\Op A }     & = \tcomm{\Op A}{   \ta{\Op C }{\ta{\Op B }{\Op 1}}-\tc{\Op C }{\tc{\Op B }{\Op 1}}        }{\Op A }     \\
\tcomm{\Op A}{   \ta{\Op B }{\tc{\Op C }{\Op 1}}-\tc{\Op B }{\ta{\Op C }{\Op 1}}        }{\Op A }     & = \tcomm{\Op A}{   \tc{\Op C }{\ta{\Op B }{\Op 1}}-\ta{\Op C }{\tc{\Op B }{\Op 1}}        }{\Op A } 
\end{split}
\label{eq:tJac3-12}
\end{equation}
With choice of signs this accounts for $12$ of the tJacobi relations in $n=3$.

The total number of tJacobi relations is the total number of t-nested correlators ($2^n n!= 48$ for $n=3$) minus the total number of Wightman 
correlators ($n!=6$). We have obtained above $24+12= 36$ of the $42$ tJacobi relations. 
The remaining six relations are given by 
\begin{equation}
\begin{split}
\ta{\Op{A}}{
\ta{\Op{B}}{ 
\tc{\Op{C}}{
\Op{1}
}}}
&- \cth_{\Op{A},\Op{C}}\;
\tc{\Op{A}}{
\ta{\Op{B}}{ 
\tc{\Op{C}}{
\Op{1}
}}}\\
&=
\tc{\Op{A}}{
\tc{\Op{B}}{ 
\tc{\Op{C}}{
\Op{1}
}}}
- \cth_{\Op{A},\Op{C}}\;
\ta{\Op{A}}{
\ta{\Op{B}}{ 
\tc{\Op{C}}{
\Op{1}
}}}\\
&\qquad +2 \frac{(1+\bose_{\Op{A}}) \bose_{\Op{A},\Op{C}}}{\bose_{\Op{C}}} \Bigl(
\tc{\Op{B}}{
\ta{\Op{C}}{ 
\tc{\Op{A}}{
\Op{1}
}}}+
\ta{\Op{B}}{
\tc{\Op{C}}{ 
\tc{\Op{A}}{
\Op{1}
}}}
\Bigr)
\end{split}
\label{eq:tJac3rem}
\end{equation}
 and their six permutations.  We believe that the relations in \eqref{eq:tJac3rem} are linearly independent, and together with \eqref{eq:tJac3-24} and \eqref{eq:tJac3-12} give all the 42 tJacobi relations which whittle down the space of 4-point functions to 6.

\section{Nested correlator bases: derivations and proofs}
\label{app:nestbasis}

In this appendix, we first derive a basis of $n!$ nested correlators, which span the space of all $n$-point functions for any initial density matrix. In the second part, we prove that a canonical choice of $(n-1)!$ of these serve as a basis, if in addition we assume a thermal initial state with KMS condition (Lemma \ref{lemma2}).

\subsection{A nested basis for $n$-point functions (in generic states)}
\label{app:nestedGeneric}

In the formulation of $k$-oto $n$-point correlators in terms of nested commutators and anti-commutators, there are $2^{n-2} n!$ basic correlators
\begin{equation}\label{eq:generalForm}
\vev{[\,\cdots
\Ecomm{\Ecomm{\Ecomm{\OpH{O}_{\sigma(1)}}{\OpH{O}_{\sigma(2)}}{1}}{\OpH{O}_{\sigma(3)}}{2}}{\cdots\,}{n-1}}  \qquad \text{for}
\qquad \varepsilon_i \in \{+,-\} \,,\; \sigma\in S_n^+ \,,
\end{equation}
where $S_n^+$ denotes the group of even permutations of $n$ objects. We know that there are only $n!$ independent Wightman functions, which means that the representation \eqref{eq:generalForm} is highly redundant. We will now describe a canonical choice of $n!$ basis elements of the above form. All other nested correlators (and hence all Wightman functions) can then be expressed in terms of these using sJacobi identities described in \cite{Haehl:2017qfl}. We do not assume the KMS condition in this subsection.

We will construct this basis recursively as follows, to allow for ease of visualization of the process. At the end of the day we will prove that \eqref{eq:causalb} is a suitable basis for our considerations in thermal field theories. 
\begin{itemize}

\item {\bf$\mathbf{n=2}$:} For $2$-point function the basis is trivial to state:
\begin{equation}
{\mathfrak{B}}_2 = \left\{ \; \mathfrak{b}^{(2)}_1 = \fullcomm{12} \;,\;\mathfrak{b}^{(2)}_2 = \fullcomm{12_{_+}} \; \right\} \,.
\end{equation}

\item {\bf$\mathbf{n=3}$:} For $3$-point functions, we nest the $2$-point function basis from above inside a commutator or anti-commutator and consider all $3$ choices for the outermost operator:
\begin{equation}
\begin{split}
{\mathfrak{B}}_3 &= \Big\{ \; 
	\mathfrak{b}^{(3)}_1 =  \fullcomm{132}  \,,\; \mathfrak{b}_2^{(3)}=  \fullcomm{13_{_+}2}   \,,\\
 &\qquad\; 
	\mathfrak{b}_3^{(3)} =  \fullcomm{123}  \,, \; \mathfrak{b}_4^{(3)}= \fullcomm{12_{_+}3} \,, \\
&\qquad\; 
	 \mathfrak{b}_5^{(3)} =  \fullcomm{231_{_+}}  \,,\; \mathfrak{b}_6^{(3)}= \fullcomm{23_{_+}1_{_+}}  \; \Big\} \,.
\end{split}
\end{equation}
Here we picked an anti-commutator for the outermost nesting only in the case where the operator with smallest index $\OpH{O}_1$ is outermost. For later convenience, we think of each row as a block labeled by the index of the outermost operator $\in\{1,2,3\}$. Each such block has exactly $(n-1)!$ elements induced by the $2$-point function basis. 

Then the remaining nested operators are given by the linear combinations
\begin{equation}
\begin{split}
\fullcomm{12_{_+}3_{_+}} &= -\mathfrak{b}^{(3)}_1 + \mathfrak{b}^{(3)}_6 \,,\quad \;\;\;
\fullcomm{13_{_+}2_{_+}}  = -\mathfrak{b}^{(3)}_3 + \mathfrak{b}^{(3)}_6 \,, \\
\fullcomm{23_{_+}1}  &= -\mathfrak{b}^{(3)}_2-\mathfrak{b}^{(3)}_4 \,,\qquad\;\, 
\fullcomm{123_{_+}} = \mathfrak{b}^{(3)}_2+\mathfrak{b}^{(3)}_5 \,,\\
\fullcomm{132_{_+}} &= \mathfrak{b}^{(3)}_4-\mathfrak{b}^{(3)}_5 \,,\qquad\quad\;\;\;
\fullcomm{231}  = \mathfrak{b}^{(3)}_1 - \mathfrak{b}^{(3)}_3 \,.
\end{split}
\end{equation}
\end{itemize}

Let us now consider arbitrary values of $n$. Let $\mathfrak{B}_{n-1}$ denote the basis of $(n-1)$-point nested correlators. Then the basis $\mathfrak{B}_n$ is constructed as follows.
We simply nest the $(n-1)!$ objects of the $\mathfrak{B}_{n-1}$ basis inside a commutator or anti-commutator and consider all $n$ possibilities for the outermost operator (which then accounts for $n \times (n-1)! = n!$ choices). We can choose to always pick a commutator for this nesting, except for one case, say, when the outermost operator is $\OpH{O}_1$ whence we pick an anti-commutator.\footnote{  There is some freedom here. Another possible choice would be to choose all the outermost brackets to be anti-commutators except for the case when the outermost operator is $\OpH{O}_1$ and $n$ is odd. Various other prescriptions are possible, but we choose a particularly simple one here.} Explicitly, let us denote the $n!$ elements of $\mathfrak{B}_{n}$ as $\{\mathfrak{b}^{(n)}_i\}_{i=1,\ldots,n!}$ and construct them recursively as follows:
\begin{equation}\label{eq:basisGeneral}
\begin{split}
\mathfrak{b}^{(n)}_{(j-2)(n-1)!+i} &= \left\{\begin{aligned}
&  [ \mathfrak{b}^{(n-1)}_i(1,\ldots,j-1,j+1,\ldots,n) \,,\, j ]  \qquad \text{for } j=2,\ldots,n\\
&  \{ \mathfrak{b}^{(n-1)}_i(2,\ldots,n) \,,\, 1 \}  \qquad\qquad\qquad\qquad\;\; \text{for } j=n+1
\end{aligned}\right.
\end{split}
\end{equation}
where $i = 1,\ldots,(n-1)!$. The index $j$ labels what we called as ``blocks'' above. 
On the right hand side, we use round brackets to show the operator numbers on which the respective basis element is evaluated on. For example, if $\mathfrak{b}_1^{(2)} = \fullcomm{12} \equiv [ \OpH{O}_1, \OpH{O}_2]$, then by $\mathfrak{b}_1^{(2)}(2,3)$ we mean the same object evaluated for operators $\OpH{O}_2$ and $\OpH{O}_3$, viz.,  $\mathfrak{b}_1^{(2)}(2,3)= \fullcomm{23}$.

{\it Proof:} To prove that the construction \eqref{eq:basisGeneral} gives a basis of $n!$ $n$-point functions, we use induction. For small values of $n$, we have given explicit constructions in the main text, so it remains to show that assuming $\mathfrak{B}_{n-1}$ forms a basis of $(n-1)$-point functions, then also $\mathfrak{B}_{n}$ forms a basis of $n$-point functions. We show this by explicitly constructing any given element of the Wightman basis in terms of $\mathfrak{B}_n$. 

Consider an arbitrary $n$-point Wightman function, which we can characterize by a permutation $\sigma \in S_{n-1}$ as follows: 
\begin{equation}
  G_{\sigma,k}^{(n)} = \vev{ \sigma_2 \cdots \sigma_k \, 1 \, \sigma_{k+1} \cdots \sigma_{n} }
\end{equation}
where we use the usual shortcut $\sigma_i \equiv \OpH{O}_{\sigma_i}(t_{\sigma_i})$ and $1\equiv \OpH{O}_1(t_1)$. Note that $\sigma$ here acts as a permutation on the set $\{2,\ldots,n\}$. The index $k$ indicates the position of operator $\OpH{O}_1$. We now give an explicit linear combination of commutators and anti-commutators, which reproduces the Wightman correlator $G_{\sigma,k}^{(n)}$: 
\begin{equation}\label{eq:proofeq1}
\begin{split}
 2\,G_{\sigma,k}^{(n)} &= \Big\langle
 	 \comm{\sigma_2 \cdots \sigma_k \, 1 \, \sigma_{k+1} \cdots \sigma_{n-1} }{ \sigma_{n} } + 
 	\comm{\sigma_{n}\sigma_2 \cdots \sigma_k \, 1 \, \sigma_{k+1} \cdots \sigma_{n-2}}{\sigma_{n-1} } \\
 &\qquad 
 	+\; \ldots \ldots +\comm{\sigma_{k+2} \cdots \sigma_{n} \sigma_2 \cdots \sigma_{k}\,1 }{\sigma_{k+1} } 
 	+ \anti{ \sigma_{k+1} \cdots \sigma_{n} \sigma_2 \cdots \sigma_k }{ 1 } \\
 &\qquad 
	 - \comm{1\, \sigma_{k+1} \cdots \sigma_{n} \sigma_2 \cdots \sigma_{k-1}}{\sigma_{k} } - \ldots \ldots \\
& \qquad 
- \comm{\sigma_4 \cdots \sigma_k \, 1 \, \sigma_{k+1} \cdots \sigma_{n} \sigma_2}{\sigma_3 } - \comm{\sigma_3 \cdots \sigma_k \, 1 \, \sigma_{k+1} \cdots \sigma_{n} }{\sigma_2 } \Big\rangle \,.
\end{split}
\end{equation}
It is straightforward to check this equation. We will now argue that every line of this equation is a combination of basis elements of $\mathfrak{B}_n$. This involves two steps. First, it follows by induction that any string of $n-1$ operators appearing as the first entry of any of the (anti-)commutators above is a combination of the basis elements appearing in the basis $\mathfrak{B}_{n-1}$. This requires a different labeling of operators for each line, but the important part is that this can be done, simply because the first entry of each (anti-)commutator is an $(n-1)$-point Wightman function. Secondly, we realize that these (relabeled permutations of) the basis elements of $\mathfrak{B}_{n-1}$ are precisely the objects that we nest inside (anti-)commutators in order to construct $\mathfrak{B}_n$. Further, according to the construction \eqref{eq:basisGeneral}, we always nest inside commutators, except when the outermost operator is $1$, in which case we use an anti-commutator. This is exactly the same structure as that appearing in the linear combination \eqref{eq:proofeq1}. More explicitly, the first line of \eqref{eq:proofeq1} is a linear combination of the $(\sigma_{n})$-th block of $\mathfrak{B}_n$, the second line is a linear combination of the elements of the $(\sigma_{n-1})$-th block, and so on. 

This completes the proof. As a corollary, we can immediately see from the explicit construction \eqref{eq:proofeq1} that the linear combination that expresses any Wightman correlator in terms of the basis $\mathfrak{B}_n$ contains every single one of the $n!$ elements of $\mathfrak{B}_n$ and all coefficients have the same absolute value. That is, we can write any Wightman $n$-point function as 
\begin{equation}
  \vev{\OpH{O}_{\sigma_1}(t_{\sigma_1}) \cdots \OpH{O}_{\sigma_n}(t_{\sigma_n})  } = \sum_{i=1}^{n!} \frac{s_i}{2^{n-1}} \, \mathfrak{b}^{(n)}_i \,,
\end{equation}
where $\sigma \in S_n$ and $s_i = \pm 1$. 
In this sense the basis $\mathfrak{B}_n$ is very democratic: no particular Wightman function is any simpler or any more complicated than any other one, when expressed in this basis.

\subsection{A causal basis for thermal correlators}
\label{app:proofThermal}

In the previous subsection we constructed a basis $\mathfrak{B}_n$ of $n!$ nested correlators, which form a basis of $n$-point functions in generic states. Let us now assume that the initial state is thermal, so we can use the KMS condition to further reduce the basis to $(n-1)!$ elements. 

One can easily see from the previous subsection that $\mathfrak{B}_n$ contains precisely $(n-1)!$ nested commutators (i.e., nested correlators which involve no anti-commutators). These are all of the form $\fullcomm{1\sigma(2)\cdots \sigma(n)}$, i.e., the operator $1$ is innermost and all others occur in all possible orders. We need to prove that all remaining elements of $\mathfrak{B}_n$ can be expressed in terms of the ``causal basis'' $\fullcomm{1\sigma(2)\cdots \sigma(n)}$ using the KMS condition. Alternatively, we can proceed by showing that every Wightman function can be expressed in terms of the causal basis. Since the KMS condition acts cyclically on Wightman functions (as in \eqref{eq:kmsfr}), we can restrict to Wightman functions of the form $\vev{1\sigma(2)\cdots \sigma(n)}$ and show that they are in one-to-one correspondence with the causal correlators $\vev{\fullcomm{1\sigma(2)\cdots \sigma(n)}}$. 

We will proceed in two steps: first, we give a general expansion for the causal basis correlators in terms of Wightman functions. While it is obvious that this exists, it will be useful to write it out more explicitly. Secondly, we will show the converse: every Wightman function of the form $\vev{1\sigma(2)\cdots \sigma(n)}$ can be written in terms of $\vev{\fullcomm{1\sigma(2)\cdots \sigma(n)}}$. Finally, we will also prove Lemma \ref{lemma3} to express generic nested correlators in terms of the causal basis, thus completing the arc reaching from the nested  correlators to the basis $\mathfrak{B}_n$, and its reduction to the causal basis in thermal states. 

\paragraph{1. Causal basis in terms of Wightman correlators:} Any element of the causal basis can be expressed in terms of Wightman functions of the form $  \vev{1\sigma_2\cdots \sigma_n}$ (where $\sigma_i \equiv \sigma(i)$). While this is obvious from the involution relation \eqref{eq:tInv}, we wish to make it more explicit:\\

\noindent
{\bf Lemma \textlabel{5}{lemma5}:}  Expanding out nested commutators and using the KMS condition leads to the following expression in terms of thermal Wightman functions:
\begin{equation}
\begin{split}
\vev{\fullcomm{1\sigma_2\cdots \sigma_n}}
&=- \bose_{\sigma_n}^{-1} \,\sum_{s_2 = 0,1} \cdots\!\!\!\! \sum_{s_{n-1}=0,1} (-)^{\sum_i s_i}  \;  e^{ \beta \sum_{i=2}^{n-1} s_i \,\omega_{\sigma_i} }  \; \vev{ 1 \, (\sigma \circ \pi_{^{\{s_i\}}})_2 \cdots (\sigma \circ \pi_{^{\{s_i\}}})_n } \\
 \text{where } \pi_{^{\{s_i\}}} &=  ((n-1)n)^{s_{n-1}} \circ ((n-2)(n-1)n)^{s_{n-2}} \circ \ldots \circ (3\cdots n)^{s_3}\circ (2\cdots n)^{s_2}\,.
\end{split}
\label{eq:CausalWight}
\end{equation}	
The representation of the permutations $\pi_{^{\{s_i\}}}$ is a cycle decomposition of the action on the set $\{2,\ldots,n\}$.\footnote{  We wish to alert the reader that we use two different (but standard) notations for permutations. To indicate operator insertions inside correlators, we use the `one-line notation' $\sigma \equiv (\sigma_2,\ldots,\sigma_n)$, giving an ordered list of the images of $\{2,\ldots,n\}$ under $\sigma$. In \eqref{eq:CausalWight}, we employ instead the `cycle notation', indicating the action of individual cycles on the same set of labels.}

{\it Proof:} Proving the identity \eqref{eq:CausalWight} boils down to using the involution relation \eqref{eq:tInv} and carefully reading off what terms can appear and what their thermal weight is. To guide the eye, let us demonstrate how the first few steps of the expansion work:
\begin{align}
\vev{\fullcomm{1\sigma_2\cdots \sigma_n}} &\equiv  \vev{
\comm{{\cdots}{
	\comm{
	\comm{1}{\sigma_2}{}
	}{\sigma_3}{}
	 {}}}{\cdots\sigma_n}{}
}
\nonumber \\
& = 
\vev{
	1\ 
	{}_-\comm{\sigma_2}{
	{}_-\comm{\sigma_3}{
	\cdots {}_-\comm{\sigma_n}{\Op{1}}\cdots
	}
	}
}
\nonumber \\
& = -\bose_{\sigma_n}^{-1} \vev{ 
1\ {}_-\comm{\sigma_2}{\cdots {}_-\comm{\sigma_{n-2}}{ \left(\sigma_{n-1}\sigma_n - e^{\beta\omega_{\sigma_{n-1}}} \sigma_n \sigma_{n-1}\right)}\cdots}
} \nonumber \\
& = -\bose_{\sigma_n}^{-1} \Big\langle
 1\ {}_- [\sigma_2,\cdots {}_- [\sigma_{n-3}, \sigma_{n-2} \left(\sigma_{n-1}\sigma_n - e^{\beta\omega_{\sigma_{n-1}}} \sigma_n \sigma_{n-1}\right)  \nonumber \\
 &\qquad\qquad\qquad\qquad\qquad\qquad -e^{\beta\omega_{\sigma_{n-2}}} \left(\sigma_{n-1}\sigma_n - e^{\beta\omega_{\sigma_{n-1}}} \sigma_n \sigma_{n-1}\right) \sigma_{n-2}]\cdots ]
\Big\rangle \nonumber \\
& = -\bose_{\sigma_n}^{-1} \Big\langle
 1\ {}_- [\sigma_2,\cdots {}_- [\sigma_{n-4}, \sigma_{n-3} \left( \,\star\, \right) - e^{\beta\omega_{\sigma_{n-3}}} \left( \,\star\,\right) \sigma_{n-3} ]\cdots ]
\Big\rangle \nonumber \\
& = \ldots  \,,
\label{eq:tInv2}
\end{align}	
where we used \eqref{eq:tInv} to get the second line and we abbreviated 
\begin{equation}
(\,\star\,) \equiv \sigma_{n-2} \left(\sigma_{n-1}\sigma_n - e^{\beta\omega_{\sigma_{n-1}}} \sigma_n \sigma_{n-1}\right)-e^{\beta\omega_{\sigma_{n-2}}} \left(\sigma_{n-1}\sigma_n - e^{\beta\omega_{\sigma_{n-1}}} \sigma_n \sigma_{n-1}\right) \sigma_{n-2} \,.
\end{equation}
By inspection of how this expansion proceeds from one line to the next by commuting blocks of operator insertions in each step, one quickly realizes that the Wightman functions and thermal factors occurring are precisely as claimed in \eqref{eq:CausalWight}. The characteristic numbers $s_i$ in that formula encode whether ($s_i=1$) or not ($s_i=0$) the operator $\sigma_i$ has been commuted past the block of operators $\sigma_{i+1},\ldots,\sigma_n$ further inside the nesting structure on the right hand side of \eqref{eq:tInv2}: every time we do permute an operator $\sigma_i$ past its adjacent block, we pick up a thermal factor $(-e^{\beta\omega_{\sigma_i}})$. The permutation $\pi_{^{\{s_i\}}}$ simply implements the corresponding permutation of $\sigma_i$ past said block for a given signature $\{s_i\}_{i=2,\ldots,n-1}$.


\paragraph{2. Wightman correlators in terms of causal basis (proof of Lemma \ref{lemma2}):}

We now need to show that the relation \eqref{eq:CausalWight} can be inverted to express a given Wightman function $\vev{1\sigma_2 \cdots \sigma_n}$ in terms of causal nested correlators $\vev{\fullcomm{1\rho_2\cdots \rho_n}}$. The result reads
\begin{equation} \label{eq:ShuffleRes}
 \vev{1\,\sigma_2 \cdots \sigma_n} = \sum_{\rho \in S_{n-1}} \left((1+\bose_1) \prod_{i=2}^{n-1} (\tilde{s}_i^{(\rho)} + \bose_{1,(\sigma\circ\rho)_2, \ldots, (\sigma\circ\rho)_i}) \right) \vev{\fullcomm{1(\sigma\circ\rho)_2\cdots (\sigma\circ\rho)_n}} \,.
\end{equation}
which was given for $\sigma = \text{id}$ in Lemma \ref{lemma2} (we refer to \S\ref{sec:Causal} for more explanation). In the following, we also restrict to $\sigma = \text{id}$ to simplify notation.

{\it Proof:} To prove \eqref{eq:ShuffleRes}, we need to show that it inverts \eqref{eq:CausalWight}. To this end, let us define the following thermal factor associated with two permutations $\rho$ and $\sigma$ of $\{2,\ldots,n\}$:  
\begin{equation}
\mathcal{T}^{(\rho,\sigma)}_{\{-\}} \equiv (1+\bose_1) (-\bose_{\sigma_n}^{-1})\!\!\!\sum_{\substack{\;\;\;\;s_2=0,1\\\;\;\;\;\smash{{\colon}} \\s_{n-1}=0,1}} \; \prod_{i=2}^{n-1} (-)^{s_i} \, e^{\beta \, s_i \, \omega_{\sigma_i}} \Big(\tilde{s}_i^{(\pi^{-1}_{^{\{s_i\}}}\rho)}+ \bose_{1,(\sigma \rho)_2, \ldots, (\sigma\rho)_i}\Big) \,.
\end{equation}
A slight variation of this object was encountered in Lemma \ref{lemma3}.
We claim that $\mathcal{T}^{(\rho,\sigma)}_{\{-\}} = \delta_{\rho ,\text{id}}$, i.e., it is unity if $\rho=\text{id}$, and vanishes otherwise. We will prove this statement below. First, however, let us note that \eqref{eq:ShuffleRes} immediately follows once we show this. To see this, let us plug the equation $\mathcal{T}^{(\rho,\sigma)}_{\{-\}} = \delta_{\rho ,\text{id}}$ into a sum over permutations and exchange the order of summation: 
\small
\begin{equation}
\begin{split}
&  \vev{\fullcomm{1\sigma_2\cdots \sigma_n}} \\
&\quad=\sum_{\rho \in S_{n-1}} \mathcal{T}^{(\rho,\sigma)}_{\{-\}} \, \vev{\fullcomm{1\, (\sigma \rho)_2 \cdots (\sigma \rho)_n }} \\
  &\quad=   (-\bose_{\sigma_n}^{-1}) \!\!\!\sum_{\substack{\;\;\;\;s_2=0,1\\\;\;\;\;\smash{{\colon}} \\s_{n-1}=0,1}}   \sum_{\rho \in S_{n-1}}(1+\bose_1)\left( \prod_{i=2}^{n-1} (-)^{s_i} \, e^{\beta \, s_i \, \omega_{\sigma_i}}\, \Big(\tilde{s}_i^{(\pi^{-1}_{^{\{s_i\}}} \rho)} + \bose_{1,(\sigma \rho)_2, \ldots, (\sigma \rho)_i}\Big) \right)\vev{\fullcomm{1\, (\sigma \rho)_2 \cdots (\sigma \rho)_n }}\\
  &\quad=  (-\bose_{\sigma_n}^{-1}) \!\!\!\sum_{\substack{\;\;\;\;s_2=0,1\\\;\;\;\;\smash{{\colon}} \\s_{n-1}=0,1}} (-)^{\sum_{i=2}^{n-1} s_i} \, e^{\beta \, \sum_{i=2}^{n-1} s_i \, \omega_{\sigma_i}} \\
  &\qquad\qquad\qquad\quad \times \left(\sum_{\rho \in S_{n-1}}  (1+\bose_1) \prod_{i=2}^{n-1}  \Big(\tilde{s}_i^{(\rho)} + \bose_{1,(\sigma \pi_{^{\{s_i\}}} \rho)_2, \ldots, (\sigma \pi_{^{\{s_i\}}} \rho)_i}\Big) \vev{\fullcomm{1\, (\sigma \pi_{^{\{s_i\}}} \rho)_2 \cdots (\sigma \pi_{^{\{s_i\}}} \rho)_n }} \right)
\end{split}
\end{equation}
\normalsize
where we redefined $\rho \rightarrow\pi_{^{\{s_i\}}}\circ \rho$ in the last step. The statement \eqref{eq:ShuffleRes} now follows immediately from comparison with the already established formula \eqref{eq:CausalWight}.

Let us now prove  $\mathcal{T}^{(\rho,\sigma)}_{\{-\}} = \delta_{\rho ,\text{id}}$. For simplicity, let us restrict to $\sigma = \text{id}$ and show $\mathcal{T}^{(\rho,\text{id})}_{\{-\}} = \delta_{\rho ,\text{id}}$ (the general case works analogously). Let us start with analyzing the permutation $\pi_{^{\{s_i\}}}^{-1}$, which takes the explicit form
\begin{equation}
 (\pi_{^{\{s_i\}}}^{-1})_2 = 2+ (n-2) s_2 \,, \qquad (\pi_{^{\{s_i\}}}^{-1})_j = j + (n-j) s_j - \sum_{k=2}^{j-1} s_k  \quad (j=3,\ldots,n)\,.
\end{equation}
From this one can infer the following:
\begin{equation} \label{eq:stildesim}
\tilde{s}_\ell^{(\pi^{-1}_{^{\{s_i\}}}\circ\rho)} = s_{\rho_{\ell+1}} \, \big(1- \tilde{s}^{(\rho)}_\ell\big) + \big(1-s_{\rho_\ell} \big)\, \tilde{s}^{(\rho)}_\ell  \,.
\end{equation}
This simplifies the expression for $\mathcal{T}^{(\rho,\sigma)}_{\{-\}}$:
\begin{equation}\label{eq:Ttemp}
\mathcal{T}^{(\rho,\text{id})}_{\{-\}} = (1+\bose_1) (-\bose_{n}^{-1})\!\!\!\sum_{\substack{\;\;\;\;s_2=0,1\\\;\;\;\;\smash{{\colon}} \\s_{n-1}=0,1}} \; \prod_{i=2}^{n-1} (-)^{s_i} \, e^{\beta \, s_i \, \omega_{i}} \Big(s_{\rho_{i+1}} \, \big(1- \tilde{s}^{(\rho)}_i \big) + \big(1-s_{\rho_i} \big)\, \tilde{s}^{(\rho)}_i+ \bose_{1,\rho_2, \ldots, \rho_i}\Big) \,.
\end{equation}
We will now distinguish between $\rho = \text{id}$ and $\rho \neq \text{id}$.

Case 1: If $\rho = \text{id}$, it means that $\tilde{s}_i^{(\rho)} = 1$ for all $i$. Therefore, the terms in the product decouple and we simply find a telescopic product:
\begin{equation}
\begin{split}
\mathcal{T}^{(\text{id},\text{id})}_{\{-\}} &= (1+\bose_1) (-\bose_{n}^{-1})\!\!\!\sum_{\substack{\;\;\;\;s_2=0,1\\\;\;\;\;\smash{{\colon}} \\s_{n-1}=0,1}} \; \prod_{i=2}^{n-1} (-)^{s_i} \, e^{\beta \, s_i \, \omega_{i}} \Big( 1-s_{i} + \bose_{1,2, \ldots,i}\Big) \\
&= (1+\bose_1) (-\bose_{n}^{-1}) \, \prod_{i=2}^{n-1} \left[ \sum_{s_i=0,1} (-)^{s_i} \, e^{\beta \, s_i \, \omega_{i}} \Big( 1-s_{i} + \bose_{1,2, \ldots,i}\Big) \right] \\
&= (1+\bose_1) (-\bose_{n}^{-1}) \prod_{i=2}^{n-1} e^{\beta\, \omega_i} \, (- \bose_{1,\ldots,i}) (-\bose^{-1}_{1,\ldots,i-1}) \\
&= 1 \,.
\end{split}
\end{equation}

Case 2: If $\rho \neq \text{id}$, there exists at least one index $i$ where $\tilde{s}_i^{(\rho)} = 0$, i.e., where the permutation $\rho$ has a descent. Let $j \in \{2,\ldots, n-1\}$ be the largest index for which this is the case. From \eqref{eq:stildesim} we conclude that $\tilde{s}_j^{(\pi^{-1}_{^{\{s_i\}}}\circ\rho)} = s_{\rho_{j+1}}$. Apart from the $j^{\rm th}$ one, the only other Bose-Einstein factor in \eqref{eq:Ttemp}, which is sensitive to the summation over $s_{\rho_{j+1}}$ is the one with index $i=j+1$ for which we have $\tilde{s}_{j+1}^{(\pi^{-1}_{^{\{s_i\}}}\circ\rho)} = 1-s_{\rho_{j+1}}$ (due to our assumption that $j$ is the largest index with a descent, so $j+1$ cannot be a descent as well). The sum over $s_{\rho_{j+1}}$ therefore takes the form\footnote{ Note that in the case $j=n-1$ (which happens, for example, for $\rho=(n,n-1,\ldots,2)$), this equation looks somewhat different since there is no index $j+1$ in the product: the last factor in \eqref{eq:proofcalc} will be absent. Since in this case $\bose_{1,\rho_2,\ldots,\rho_j} = -1-\bose_{\rho_n}$, one can check that the sum over $s_{\rho_{j+1}} \equiv s_{\rho_n}$ vanishes nevertheless.}
\begin{equation}
\begin{split}
\sum_{s_{\rho_{j+1}}=0,1} (-)^{s_{\rho_{j+1}}} \, e^{\beta\, s_{\rho_{j+1}} \, \omega_{\rho_{j+1}}}\, \left( s_{\rho_{j+1}} + \bose_{1,\rho_2,\ldots, \rho_j} \right) \left( 1 - s_{\rho_{j+1}} + \bose_{1 , \rho_2 , \ldots, \rho_{j+1}} \right) = 0 \,.
\end{split}
\label{eq:proofcalc}
\end{equation}
This shows that there is a vanishing factor in \eqref{eq:Ttemp}, so $\mathcal{T}^{(\rho\neq \text{id},\text{id})}_{\{-\}} = 0$. This completes the proof.

\paragraph{3. Nested correlators in terms of causal basis (proof of Lemma \ref{lemma3}):} 
We now wish to use the results given hitherto and apply them to derive the formula for nested correlators:
\begin{equation} \label{eq:lem3rep}
\vev{\fullcomm{1\,2_{\varepsilon_2} \cdots n_{\varepsilon_n}}} = \sum_{\rho \in S_{n-1}} \, \mathcal{T}_{\{\varepsilon_i\}}^{(\rho)}\,  \vev{\fullcomm{1\, \rho(2) \cdots \rho(n) }}
\end{equation}
with thermal factors given in \eqref{eq:NestedFactors}. 
We start with an immediate generalization of \eqref{eq:CausalWight} to the case where we allow for both commutators and anti-commutators (and we fix $\sigma = \text{id}$ in \eqref{eq:CausalWight} to declutter notation):
\begin{equation}
\begin{split}
\vev{\fullcomm{1\,2_{\varepsilon_2} \cdots n_{\varepsilon_n}}}
&= (1+\varepsilon_n e^{\beta\omega_n})\!\! \sum_{\substack{\;\;\;\;s_2=0,1\\\;\;\;\;\smash{{\colon}} \\s_{n-1}=0,1}} \left( \prod_{i=2}^{n-1} \varepsilon_i^{s_i}  \;  e^{ \beta \, s_i \,\omega_{i} } \right) \, \vev{ 1 \, ( \pi_{^{\{s_i\}}})_2 \cdots (\pi_{^{\{s_i\}}})_n } \,,
\end{split}
\label{eq:CausalWight3}
\end{equation}	
which again follows from expanding \eqref{eq:tInv}. Next, we write the Wightman functions in this expression in terms of nested commutators, using \eqref{eq:ShuffleRes}:
\begin{equation}
\begin{split}
&\vev{\fullcomm{1\,2_{\varepsilon_2} \cdots n_{\varepsilon_n}}}\\
&\quad= (1+\varepsilon_n e^{\beta\omega_n})\!\! \sum_{\substack{\;\;\;\;s_2=0,1\\\;\;\;\;\smash{{\colon}} \\s_{n-1}=0,1}} \left( \prod_{i=2}^{n-1} \varepsilon_i^{s_i}  \;  e^{ \beta \, s_i \,\omega_{i} } \right)  \\
&\qquad\qquad \times  \sum_{\rho \in S_{n-1}} \left((1+\bose_1) \prod_{\ell=2}^{n-1} (\tilde{s}_\ell^{(\rho)} + \bose_{1,(\pi_{^{\{s_i\}}}\circ \rho)_2, \ldots, (\pi_{^{\{s_i\}}}\circ\rho)_\ell}) \right) \vev{\fullcomm{1(\pi_{^{\{s_i\}}}\circ\rho)_2\cdots (\pi_{^{\{s_i\}}}\circ \rho)_n}} \\
&\quad= \sum_{\rho \in S_{n-1}} (1+\bose_1) (1+\varepsilon_n e^{\beta\omega_n})\sum_{\substack{\;\;\;\;s_2=0,1\\\;\;\;\;\smash{{\colon}} \\s_{n-1}=0,1}}    \left( \prod_{\ell=2}^{n-1}\varepsilon_\ell^{s_\ell}  \;  e^{ \beta \, s_\ell \,\omega_{\ell} }  \Big(\tilde{s}_\ell^{(\pi^{-1}_{^{\{s_i\}}}\circ\rho)} + \bose_{1,\rho_2, \ldots, \rho_\ell}\Big) \right) \vev{\fullcomm{1\rho_2\cdots \rho_n}} \,,
\end{split}
\label{eq:CausalWight4}
\end{equation}	
where we exchanged the order of summation in the second step by relabelling $\rho \rightarrow \pi^{-1}_{^{\{s_i\}}} \circ \rho$. Formula \eqref{eq:NestedFactors} for the thermal factors now follows from the observation \eqref{eq:stildesim}.

\section{Details on the harmonic oscillator}
\label{app:sho}

We provide here some details on the harmonic oscillator example that we skipped in \S\ref{sec:sho}.

 Since $[a , a^\dagger]  =1$, one can get the various powers of $X(t)$, for example,
\begin{equation}
\begin{split}
(2 \mu )\, X^2(t) =& a^2 e^{-2 i \mu \,t} + {a^\dagger}^2 e^{2 i \mu \,t} + 2 a^\dagger a + 1  \,,\\
(2 \mu )^{3 \over 2}\, X^3(t) =& a^3 e^{-3 i \mu \,t} + {a^\dagger}^3 e^{3 i \mu \,t} + 3 ( a^\dagger a^2 e^{-i \mu \,t} + {a^\dagger}^2 a e^{i \mu \,t} ) + 3  (   a e^{-i \mu \,t} + a^\dagger e^{i \mu \,t}) \,.
\end{split}
\end{equation}
From the basic action on the Hilbert space, \eqref{eq:aaction}, we find 
\begin{equation}
\begin{split}
a^m | n \rangle &= \Theta(n-m) \,  \sqrt{n! \over (n-m)! }\; | n-m \rangle 
 \hspace{10mm} 
  {a^\dagger}^m | n \rangle = \sqrt{ (n + m)! \over n!}\; | n +m \rangle \\
& {a^\dagger}^{m_1} a^{m_2} | n \rangle = {\sqrt{n! (n-m_2+m_1)!} \over (n-m_2)!} \; | n-m_2 + m_1 \rangle  \\
& \langle n | {a^\dagger}^{m_1} a^{m_2} | n \rangle = \delta_{m_1,m_2} { n!  \over (n-m_2)!} 
\end{split}
\end{equation}
where  $\Theta(n-m)$ is the Heaviside step-function.

\paragraph{Three-point functions:} For three-point functions we find 
\begin{equation}
\begin{split}
(2 \mu)^{3} \langle X(t_1) & X^2(t_2) X^3(t_3) \rangle_\beta   = {3 \over (e^{\beta \mu}-1)^3 } \times  \\ 
& 2 e^{ i \mu \,(3t_{13} - 2 t_{12}) } + 2 e^{3 \beta \mu}  e^{- i \mu \,(3t_{13} - 2 t_{12}) }   \\
& + e^{ i \mu \,t_{13}} \left[3 + 8 e^{\beta \mu} + e^{2 \beta \mu } + e^{ -2 i \mu \,t_{12}  } e^{\beta \mu}  (4 + 2 e^{\beta \mu} )\right] \\
& + e^{- i \mu \,t_{13}} e^{\beta \mu } \left[1 + 8 e^{\beta \mu} +3 e^{2 \beta \mu } + e^{ 2 i \mu \,t_{12}  }  (2 + 4 e^{\beta \mu} )\right] 
\end{split}
\label{eq:123Ordering}
\end{equation}

\paragraph{Four-point functions:} To derive the chaos correlator given in the main text, it is useful to note that: 
\begin{equation}
\begin{split}
(2\mu)^2\, \vev{ [X(t_1) , X(t_2)] [X(t_1) , X(t_2) ] }_\beta & = 
- 4 \sin^2(\mu\, t_{12}) \, \\ 
(2\mu)^4\vev{ [X(t_1) , X^3(t_2)] [X(t_1) , X^3(t_2) ] }_\beta  &
	=  -108 \coth^2(\frac{1}{2}\,\beta\mu) \sin^2(\mu\, t_{12}) .
\end{split}
\end{equation}

\paragraph{Tremelo commutators:} To exemplify some of the statements in the text, it is also useful to record  a  particular set of tremelo commutators all the way up to 6-point functions . 
\begin{equation}
\begin{split}
(2\mu)^2 \, \vev{ [X(0) , X^3(t)]   }_\beta 
& =  
  6\, i\,  \coth(\frac{1}{2} \, \beta\mu) \sin(\mu \,t) \,, \\
(2\mu)^4 \, \vev{ [X(0) , X^3(t)]^2   }_\beta  &=  -108 \coth^2(\frac{1}{2}\, \beta\mu) \sin^2(\mu \,t)
  \,, \\
(2\mu)^6\, \vev{ [X(0) , X^3(t)]^3   }_\beta &=  -
3240 i \coth^3(\frac{1}{2}\, \beta\mu) \sin^3(\mu \,t)
	\, . 
\end{split}
\end{equation}

For completeness, we also give the regulated version (spectral functions) corresponding to the above commutators
\begin{subequations}
\begin{align}
(2 \mu)^2 \, \vev{ e^{-\frac{\beta}{2}\,H} X(0) e^{-\frac{\beta}{2}\,H} X(t) } 
& = (2 \mu)^2 \vev{ X(0)\, X^3(t+{i \beta \over 2})  } 
 = { 3 \coth{ \beta \mu \over 2} \over \sinh{\beta \mu \over 2} }  \cos(t \mu ) \nonumber \\
& = 6 \sqrt{\bose (1+\bose)} \; (1+ 2\, \bose) \cos(t \mu) 
\end{align}
\begin{align}
(2 \mu)^4 \, & \vev{ e^{-\frac{\beta}{4}\, H} \, X(0)  \,e^{-\frac{\beta}{4}\, H} \, X(t)\, e^{-\frac{\beta}{4}\, H} \, X(0) \,e^{-\frac{\beta}{4}\, H} \, X(t) }  \nonumber \\
& = 
	(2 \mu)^4 \, \vev{ X(0) X^3(t+ i\,\tfrac{\beta}{4})  X(i \tfrac{\beta}{2}) X^3(t+i \tfrac{3 \beta}{4}) } \nonumber \\
& = 
	\frac{3}{2 \,\sinh^4(\frac{\beta \mu }{ 2})} 
	\left[ 4 (7 + 3 \cosh \tfrac{\beta \mu }{ 2} ) + 3 \cos(2\,\mu\,t) \, \cosh\tfrac{\beta \mu }{ 2} 
	(9 + \cosh \beta \mu )  \right]  \nonumber \\
&= 
	6 \, \bose\,  \left[  8 (1+ \bose) (3 + 20 \,\bose +  20 \, \bose^2) + 3\, \sqrt{1 + \bose^{-1}} \,
	(1 + 2\, \bose ) (1 + 20\,\bose + 20\, \bose^2 )\,  \cos(2 t \mu) \right] 
\end{align}
\end{subequations}

\paragraph{Euclidean correlators:}
 Working with the analytically continued Euclidean oscillator, we can check that the propagator in frequency space takes the familiar form:
\begin{equation}
 G_E(\omega_1,\omega_2) \equiv  \vev{ x(\omega_1) x(\omega_2) } = {\delta(\omega_1+\omega_2) \over  (\omega_1^2 + \mu^2)} \equiv  \delta(\omega_1 + \omega_2) G_E(\omega_1)
\end{equation}
Working first at zero temperature, we can Fourier transform back to the time domain to find
 \begin{equation}
G_E(\omega) = {1 \over \omega^2 + \mu^2} \implies G_{E}(\tau) =\int {d\omega\over 2\pi} {e^{-i \omega \tau} \over \omega^2 + \mu^2 }  = {1 \over 2 \mu} \left[  \theta(\tau) e^{-\mu \tau} + \theta(-\tau) e^{\mu \tau} \right] = {1 \over 2 \mu}  e^{-\mu |\tau|}
\end{equation}

To obtain correlators in Minkowski space, one needs to analytically continue $\tau = i t$. Standard manipulations leads to 
\begin{equation}
\begin{split}
\vev{ X(t_1) X(t_2)} =& \lim_{\epsilon_1>\epsilon_2 , \epsilon_i \to 0} G_E(\tau_1= i t_1  + \epsilon_1,\tau_2= i t_2  + \epsilon_2) \\
=& \lim_{\epsilon_1>\epsilon_2 , \epsilon_i \to 0} G_E(\tau = i t_{12}+ \epsilon_1 -\epsilon_2)   = {1 \over 2 \mu} e^{-i \mu \,t_{12}}
\end{split}
\end{equation} 
which of course agrees with \eqref{eq:propsho} at zero temperature. The time-ordered correlator can be extracted from the above equation (or equivalently obtained via an $i \epsilon$ prescription).

At finite temperature, the only difference is that the frequencies are not continuous, but discrete Matsubara modes, $\omega_n = {2 \pi n \over \beta}$, $ n \in {\mathbb Z}$. Working out the Fourier series we have (cf., \cite{Bellac:2011kqa})
\begin{equation}
G_E(\tau) = {1 \over \beta} \sum_{n = -\infty}^\infty {e^{ - i \omega_n \tau} \over \omega_n^2 + \mu^2} = {1 \over 2 \mu} \left[   { e^{\beta \mu}   e^{- \mu |\tau|} +    e^{\mu |\tau|}    \over  e^{\beta \mu}-1 } \right]
\end{equation}

\newpage
\bibliographystyle{JHEP}

\providecommand{\href}[2]{#2}\begingroup\raggedright\endgroup

\end{document}